# Galactic Kinematics Towards the South Galactic Pole. First Results from the Yale-San Juan Southern Proper-Motion Program

René A. Méndez

*Cerro Tololo Inter-American Observatory, Casilla 603, La Serena, Chile.*
*E-mail: rmendez@noao.edu*

Imants Platais[1], Terrence M. Girard, Vera Kozhurina-Platais, William F. van Altena

*Department of Astronomy, Yale University, P. O. Box 208101, New Haven, CT 06520-8101.*
*E-mail: imants, vera, girard, vanalten @astro.yale.edu*

## ABSTRACT

The predictions from a Galactic Structure and Kinematic model code are compared to the color counts and absolute proper-motions derived from the Southern Proper-Motion survey covering more than 700 deg$^2$ toward the South Galactic Pole in the range $9 < B_{\rm J} \leq 19$. The theoretical assumptions and associated computational procedures, the geometry for the kinematic model, and the adopted parameters are presented in detail and compared to other Galactic Kinematic models of its kind.

The data to which the model is compared consists of more than 30,000 randomly selected stars, and it is best fit by models with a solar peculiar motion of $+5$ km s$^{-1}$ in the V-component (pointing in the direction of Galactic rotation), a large LSR speed of 270 km s$^{-1}$, and a (disk) velocity ellipsoid that always points towards the Galactic center. The absolute proper-motions in the U-component indicate a solar peculiar motion of $11.0 \pm 1.5$ km s$^{-1}$, with no need for a local expansion or contraction term.

The fainter absolute motions show an indication that the thick-disk must exhibit a rather steep velocity gradient of about -36 km s$^{-1}$ kpc$^{-1}$ with respect to the LSR. We are not able to set constraints on the overall rotation for the halo, nor on the thick-disk or halo velocity dispersions. Some substructure in the U & V proper-motions could be present in the brighter bins $10 < B_{\rm J} < 13$, and it might be indicative of (disk) moving groups.

*Subject headings:* Astrometry: stellar dynamics – stars: kinematics – stars: fundamental parameters – Galaxy: fundamental parameters – Galaxy: kinematics and dynamics – Galaxy: structure

---

[1]Visiting Astronomer, Cerro Tololo Inter-American Observatory. CTIO is operated by AURA, Inc. under contract to the National Science Foundation.

## 1. Introduction

Proper-motions are one of the fundamental observational quantities in astronomy, as they provide an estimate of the distribution of stellar velocities within a few kiloparsecs from the Sun. This information provides constraints on the various theories of the structure and dynamics of the Galaxy, and hence it is an important observational quantity.

One of the limitations of the available proper-motions for the study of the large-scale kinematic properties of stars in our Galaxy has been the relative lack of *absolute* proper-motions, as opposed to the more common *relative* motions giving only the transverse component of the motion with respect to, e.g., the mean motion of a group of stars. The solution to the problem of calculating absolute proper-motions came with the possibility of using galaxies to define an inertial (i.e., *non-rotating*) reference frame for the evaluation of the proper-motions of stars. The concept is based on the fact that, since galaxies are so far away, their proper-motions are extremely small, and could be considered zero. Even if the transverse velocities of galaxies were comparable to their radial velocities, this would amount to proper-motions of only 0.02 mas yr$^{-1}$ (where 1 mas = 1 milli arc-second) for a Hubble constant of 50 km s$^{-1}$ Mpc$^{-1}$.

The idea of using galaxies to determine the zero-point of the motions probably first originated at Lick Observatory, with W. H. Wright in about 1919 (Klemola et al. 1987), before the extragalactic nature of these nebulae was widely accepted. Wright realized that a wide-field telescope was required since a large number of galaxies would be needed to establish the inertial frame to sufficient accuracy. In 1947 the Lick Northern Proper-Motion program was started. By 1986 all the second-epoch plates had been taken, and the first interpretative results were published soon thereafter (Hanson 1987).

A similar program (the Southern proper-motion program), aimed at determining proper-motions for stars in the Southern hemisphere was started jointly by Yale & Columbia University in 1965 in Argentina, and the first results have been recently released (Platais et al. 1998).

At the same time, our knowledge of stellar populations and Galactic structure in the Milky Way has increased enormously in the last few decades (for a review see Majewski 1993). In particular, Galactic structure models have been developed that allow the prediction of the *observed* kinematic properties of stars by making a number of assumptions based on our current knowledge about the local stellar system. The predictions of these models can then be directly compared to observations, from which broad properties of the stellar velocity distribution can indeed be inferred (Méndez and van Altena 1996). The use of models is necessary because the observed proper-motion distribution is, in itself, the product of the convolution of stars located at different distances from the Sun having different tangential velocities, and representing different Galactic components: The disentanglement of these various contributions when no additional information is available must be done, then, in a statistical fashion.

In this paper we present the first systematic analysis using a Galactic Kinematic model of the data gathered in the course of the SPM survey in a large area (more than 700 deg$^2$) towards the South Galactic Pole, down to an apparent magnitude of $B_{\rm J} \sim 19$. A number of issues related to the velocity distribution function of stars from the disk, thick-disk, and halo are addressed.

The arrangement of this paper is the following: In Section 2 we present an overview of the SPM photometric and proper-motion data and their errors. In Section 3 the model used to compare the observed proper-motion distributions with those predicted is described in detail, while Sections 4 and 5 present the model predictions as compared to the magnitude/color and proper-motion data respectively. Finally, Section 6 presents the main conclusions of the paper.

## 2. The Southern Proper-Motion kinematic and color catalog

We analyze the kinematic and color data gathered in the context of a massive proper-motion survey, the Southern Proper-Motion Program (SPM hereafter), which is the southern-skies counterpart of the Lick Observatory Northern Proper-Motion (NPM hereafter) program (Klemola et al. 1987). The main goal of both programs is the measurement of absolute proper-motions relative to external galaxies. A description of the scientific motivation for the SPM can be found in Wesselink (1974) and van Altena et al. (1990). When finished, the SPM will produce absolute proper-motions, positions, and $B, V$ photographic photometry for approximately one million stars south of $\delta = -17°$.

In this paper we present an analysis of the SPM Catalog 1.1, an improved version of the SPM Catalog 1.0 as described by Platais et al. (1998). This catalog provides positions, absolute proper-motions, and $BV$ photometry for 58,887 objects at the South Galactic Pole. The sky coverage of the SPM Catalog 1.0 is about 720 deg$^2$ in the magnitude range $5 < V < 18.5$. Boundaries of the SGP area are indicated in Fig. 1 of Platais et al. (1998, see also http://www.astro.yale.edu/astrom/). In this paper we utilize only about 31,000 stars which have been randomly chosen. The accuracy of individual absolute proper-motions is 3-8 mas yr$^{-1}$ depending on the star's magnitude. The mean motion as a function of magnitude has random errors below 1 mas yr$^{-1}$, and various comparisons with Hipparcos motions at the bright end, and other independent measurements at the faint, end indicate systematic errors also smaller than 1 mas yr$^{-1}$ (see Section 5.1). A great effort has been put into correcting positions and proper-motions for magnitude-dependent systematic errors (see Section 5.1). For further details about the catalogue structure, contents, plate measurement and other astrometric/photometric details, the reader is referred to Girard et al. (1998, Paper I) and Platais et al. (1998, Paper II).

In the SGP region the mean number of stars per field down to the plate limit $B_J \approx 19$ is about 35,000. Due to the scan-time limitations imposed by the plate measuring machine, we could measure only ∼10% of that number of stars per field. Therefore, an effort was made to predict a minimal number of anonymous, randomly selected stars which had to be measured in order to ensure statistically well-sampled components of the Galactic disk, thick-disk, and halo, over the $B_J$-magnitude range from 9 to 19. This was done using an earlier version of our Galactic Structure and Kinematic Model code (cf. Méndez et al. 1993) which also allowed for the incorporation of the expected accuracy of the SPM proper-motions. For instance, the thick-disk stars can be sampled properly starting from $B_J = 14$, whereas halo stars appear in statistically significant numbers only at $B_J > 17$ (see Section 4.1). The initial working numbers of how many stars had to be measured in each magnitude bin were given in van Altena et al. (1994). Later, more stars with $V > 15$ were added in order to increase the density of a secondary reference frame, thus on average at the SGP there are 46 anonymous stars per square degree.

## 2.1. The randomly selected sample

In this section we provide details of the basic observational material analyzed in this paper. We focus on the randomly selected sample mentioned in the previous section. Out of a total 33,498 randomly selected stars, 31,023 have complete $B - V$ color and absolute proper motion information - and we only utilize these. Most stars missing colors are extremely faint $(B, V > 19)$ and could not be measured in one bandpass.

The random sample has been chosen such that, at a given $B_J$ magnitude interval, a fixed number of randomly selected stars were extracted from the COSMOS/UKST Object Catalog (Yentis et al. 1992). We note that the COSMOS catalog is known to be complete to magnitudes much fainter ($B_J \sim 22$) than the limit imposed by our astrometric plates. Even though the SPM catalog presents calibrated B & V photographic photometry, the initial selection of random objects was performed using the COSMOS $B_J$ magnitudes. Therefore, for the random SPM sample, one-magnitude intervals have been chosen in the range $9 \leq B_J \leq 19$, where we have adopted the relation $B_J = B - 0.28(B-V)$ from Blair and Gilmore (1982), valid for the UKST plate passband (see also Bertin and Dennefeld 1997). Correspondingly, the analysis below partitions the whole catalogue into these magnitude bins. Indeed, lumping the SPM-SGP data into two or more $B_J$ magnitude bins will lead to selectively incomplete samples - proper statistical corrections must be applied when doing this. However, at a given magnitude interval, the complete sample and the randomly selected sample differ only by a scale factor, because the observed sample has been drawn randomly from all the stars available in that particular magnitude bin. Therefore, color and kinematic properties binned in the proper magnitude intervals should be the same as those derived from a complete sample, except for the increased uncertainty on the derived values because of the smaller sample - an effect that is fully taken into account in the analysis below.

Similarly to Méndez and van Altena (1996), the kinematic comparisons to the model predictions take the form of histograms of proper-motions along the Galactocentric direction, and along Galactic rotation. Both, the observed and the model histograms are fully convolved with observational errors, so that the model comparisons can be carried out directly. Rather than making the comparisons graphically, we have used statistical descriptors of the proper-motion distribution; the median proper-motion and the proper-motion dispersion (however, the full shape of the proper-motion histograms is also used for an assessment of the *overall* fit to the model predictions, see Section 5.3).

Table 1 presents the basic data concerning uncertainties in the photographic colors as a function of $B_J$ magnitude derived from the catalog itself. The second and third columns indicate the mean and median value of the error in color respectively, while the fourth and fifth columns indicate the actual number of stars used in the computations and the total number available in that particular magnitude range. We have used a iterative procedure to determine robust estimates of the values quoted using the procedure presented in Méndez and van Altena (1996). Basically, the method trims outliers in an iterative fashion by computing preliminary values for the median and dispersion. Then, a window of semi-width three times the dispersion centered on the median is used to recompute the median and dispersion, until convergence. This method is similar to using the robust technique of probability plots (Daniel and Wood 1980) to estimate dispersions and properly account for outliers (Lutz and Upgren 1980), and indeed from Table 1 we can see that, in all cases, the number of stars excluded amounts to less than 5%, i.e., smaller than the fraction of excluded objects which are known to bias the derived values for the dispersion ($\pm 10\%$ at both extremes of the distribution of values). The same method has been applied to obtain mean, median, and dispersion values for the absolute proper motions.

Tables 2 and 3 indicate the mean, median and uncertainties of the mean for the absolute proper-motions in the U and V Galactic directions respectively. In this paper, U is oriented toward the Galactic center, while V points in the direction of Galactic rotation. Since we are looking down to the SGP, the derived tangential motions decouple nicely into these two physically meaningful quantities that are easier to model and interpret. From the observed absolute $\mu_\alpha$ and $\mu_\delta$ proper motions, we derived $\mu_l$ and $\mu_b$, proper-motions in Galactic longitude and latitude respectively. Then, these motions are projected into $\mu_U$ and $\mu_V$. The first conversion is straight forward. The last step, however, deserves further comments, because of the lack of radial velocities. Basically, to convert from the observed $\mu_l$ and $\mu_b$ to $\mu_U$ and $\mu_V$ we would need to apply the following equation:

$$\mu_U = -\mu_b \sin b \cos l - \mu_l \sin l + \frac{V_r}{r} \cos b \cos l \quad (1)$$

$$\mu_V = -\mu_b \sin b \sin l + \mu_l \cos l + \frac{V_r}{r} \cos b \sin l \quad (2)$$

As it can be seen form the above equations, the transformation necessarily involves the ratio $V_r/r$ between each star's radial velocity and its heliocentric distance, but we have neither. We bypass this by just dropping this last term from Eqs. (1) and (2). Obviously, a similar procedure should be adopted in the model computations, this is further discussed in Section 3.2 in this context. The correction term is, however, small since, for these data, $\cos b \sim 0$. Errors have also been properly propagated from $\sigma_{\mu_\alpha}$ and $\sigma_{\mu_\delta}$ to $\sigma_{\mu_U}$ and $\sigma_{\mu_V}$ using equations similar to Eqs 1 and 2. Notice that, near the SGP, the value of $l$ can take any value from $0^o$ to $360^o$, and Eqs (1) and (2) imply that $\mu_U$ and $\mu_V$ have a contribution from both $\mu_l$ and $\mu_b$ and, in turn, from $\mu_\alpha$ and $\mu_\delta$. Therefore, it is unlikely that, e.g, a systematic effect on the proper-motions in $\mu_\delta$ would propagate to affect *only* $\mu_V$. We have, however, tested the effects of any remaining systematic effect on our equatorial absolute proper-motions upon the derived motions along the U and V Galactic components, see Section 5.1.

Table 4 presents the values for the proper-motion dispersions and its corresponding uncertainty for U and V as a function of apparent magnitude. It is evident that the dispersions are *not* the same along U and V. This fact reveals the intrinsically different kinematic behavior of stars along the Galactocentric and the Galactic rotation direction, a property fully accounted for by our kinematic model (see Section 5).

Tables 5 and 6 indicate the mean and median proper motion errors, along with the dispersion (standard deviation) on the mean. These values, along with those presented in Table 1, will be used for convolving the model predictions with the appropriate errors in both color and proper-motion (see Section 4).

## 3. The starcounts and kinematic model

In the following analysis we model the starcounts concurrently with the kinematics by using the model presented by Méndez and van Altena (1996). The starcounts model employed here has been tested under many different circumstances, and has proved to be able to predict starcounts that match the observed magnitude and color counts (in both shape *and* number) to better than 10%, and in many cases to better than 1%. In particular, the model has been recently shown to provide an excellent match to the overall magnitude and color counts at the SGP and other lines-of-sight from the deep ESO Imaging Survey (da Costa et al. 1998, see http://www.eso.org/science/eis/ for the current status). The kinematic model presented by Méndez and van Altena (1996) has been also shown to be able to reproduce the kinematics of disk stars in two intermediate Galactic latitude fields, providing for the first time constraints on the expected run of velocity dispersion for disk stars as a a function of distance from the Galactic plane, in agreement with theoretical (dynamical) expectations (Fuchs and Wielen 1987, Kuijken and Gilmore 1989).

Even though we adopt the model described by Méndez and van Altena (1996), the fainter magnitude limits now available from the SPM data allows us, for the first time, to study the contribution of other Galactic components. This makes it necessary to provide an outline of the kinematic model parameters, as the starcounts parameters have already been described extensively elsewhere (Méndez & van Altena 1996, Méndez et al. 1996, Méndez & Guzmán 1998)

### 3.1. Basic Assumptions

The magnitude and color model is based on the *fundamental equation of stellar statistics* (Trumpler and Weaver 1962, Mihalas and Binney 1981). This equation can be easily extended to include kinematics. If we call $N_\mathrm{j}$ the number of stars per unit of solid angle, per unit of apparent magnitude, and per unit of apparent color for Galactic component j, then, the number of stars per unit of velocity, per unit of solid angle, per unit of apparent magnitude, and per unit of apparent color for component j at position $\vec{r}$, and velocity $\vec{V}$ is given by:

$$\frac{dN_\mathrm{j}}{d^3\vec{V}} = N_\mathrm{j} \cdot f_\mathrm{j}(\vec{r}, \vec{V}) \qquad (3)$$

where $N_\mathrm{j}$ is given by the fundamental equation of stellar statistics, and $f_\mathrm{j}(\vec{r}, \vec{V})$ is the velocity distribution function for component j at position $\vec{r}$ and velocity $\vec{V}$. For the velocity distribution we have adopted generalized Schwarzschild velocity ellipsoids (1907, 1908), represented by orthogonal three-dimensional Gaussian functions with (in general) different velocity dispersions along the principal axes of the velocity ellipsoid. For a detailed justification on the use of this function, the reader is referred to Méndez and van Altena (1996).

The velocity distribution functions adopted in the model are completely specified by three velocity dispersions along the principal axes of the velocity ellipsoid, as well as by the orientation of the velocity ellipsoid in a given coordinate system. If $\vec{V} = (U', V', W')$ are the velocities along the principal axes of the velocity ellipsoid relative to an inertial reference frame moving with the instantaneous mean speed for Galactic component j, then the (normalized) function is given by:

$$f_\mathrm{j}(\vec{r}, \vec{V}) = f_\mathrm{j_o}(\vec{r}) e^{-\frac{1}{2}\left[\left(\frac{U'}{\Sigma_{U_\mathrm{j}}(\vec{r})}\right)^2 + \left(\frac{V'}{\Sigma_{V_\mathrm{j}}(\vec{r})}\right)^2 + \left(\frac{W'}{\Sigma_{W_\mathrm{j}}(\vec{r})}\right)^2\right]} \quad (4)$$

$$f_\mathrm{j_o} = \frac{1}{(2\pi)^{3/2} \cdot \Sigma_{U_\mathrm{j}}(\vec{r}) \cdot \Sigma_{V_\mathrm{j}}(\vec{r}) \cdot \Sigma_{W_\mathrm{j}}(\vec{r})} \quad (5)$$

where $\Sigma_{U_\mathrm{j}}(\vec{r})$, $\Sigma_{V_\mathrm{j}}(\vec{r})$, and $\Sigma_{W_\mathrm{j}}(\vec{r})$ are the velocity dispersions for component j, evaluated at position $\vec{r}$.

It is possible to show that, for an axisymmetric system in steady-state, if the velocity distribution function is of the type given in Equation (4), then one of the axes of the velocity ellipsoid *must* point toward the direction of Galactic rotation, and another axis *must* be oriented toward the axis of rotation of the Galaxy (Fricke 1952, King 1990). This prevents the velocity ellipsoids from having any vertex-deviation, the cause of which is still not well understood. It probably reflects the initial conditions present at the star formation site (Mihalas and Binney 1981) and so it is a consequence of departures from steady state, from axial symmetry, or a combination of both (King 1990). In the current model implementation, we have neglected any vertex-deviation.

Similarly, it is possible to show that, at the Galactic plane, one of the axes will point toward the galactic center, and the other axis will be perpendicular to the plane, however for locations far from the plane, these axes may change their orientation (King 1990). Indeed, numerical computations of star orbits using a realistic Galactic potential (Carlberg and Innanen 1987) show that the envelopes of these orbits have a tendency to tilt toward the Galactic center, providing thus an estimate of the velocity ellipsoid orientation far from the plane (Gilmore 1990). Additionally,

Statler (1989a, b) has shown that the spatial variation of the velocity ellipsoid dispersions is still unknown, and it can be determined only by velocity observations at intermediate latitudes (King 1993), which are not yet available (although see Méndez & van Altena 1996 for a discussion of this in the context of proper-motions).

The velocity ellipsoid yields the relative frequency of stars for a given Galactic component as a function of velocity with respect to the *mean rotational motion* of the stars *at a given Galactic position* (note however that this is *different* from the motion of the Local Standard of Rest, LSR). The $(U', V', W')$ velocities in Equation (4) can be thus viewed as the *peculiar velocities* of the stars considered. Consequently, the velocity ellipsoids are centered at zero velocity, unless there is some kind of streaming motion present. Streaming motions, which can be recognized as *moving groups*, have not been included in the code so far due to their considerable complexity (e.g., Dehnen 1998). The model peculiar velocities are converted to Heliocentric velocities as described below, and the relative fraction of stars for a range of velocities is accumulated (via Equation 3) onto the corresponding Heliocentric velocities to be compared with the observations. In this way, marginal distributions (histograms) of proper-motion and/or radial velocity can be output, subject to any kind of restrictions on the observables implemented in the model.

As mentioned before, in order to convert Equation (4) into a function that describes the distribution of velocities for a Heliocentric observer, it is necessary to know the orientation of the $(U', V', W')$ system relative to a (fixed) set of axes. In what follows we thus describe the geometry employed to evaluate the velocity distribution function in the observable space of Heliocentric velocities, as well as the assumptions concerning the velocity dispersions and velocity lags of the different Galactic components included in the code. We will also compare, whenever possible, our assumptions with those of the only two existing *global* models that include kinematics, namely, the model described by Ratnatunga et al. (1989, RBC89 hereafter), based on the Bahcall and Soneira (1980) starcounts model, and the model by Robin and Oblak (1987, RO87 hereafter), based on the model of synthesis of stellar populations by Robin and Créze (1986, RC86 hereafter). RBC89's kinematic model has been compared to a sample of stars from the Bright Star Catalogue (BSC hereafter, Hof-fleit 1982), while RO87's model has been compared with Chiu's (1980) proper-motion survey toward Selected Areas SA 51 ($l = 189^o, b = +21^o$), SA 57 ($l = 69^o, b = +85^o$), and SA 68 ($l = 111^o, b = -46^o$), and with Murrays's (1986) proper-motion survey toward the South Galactic Pole. Both models have been, broadly speaking, successful in predicting the observed distribution of stars as a function of proper-motion.

### 3.2. Geometry of the kinematics: Heliocentric velocities

The observed Heliocentric velocity of a star at distance R from the Galactic center (as measured on the Galactic plane) and distance Z from the Galactic plane can be computed from:

$$\vec{V}_{\text{Hel}} = \vec{\bar{V}}(R,Z) - \vec{V}_{\text{LSR}}(R_\odot) + \vec{V}_{\text{Pec}} - \vec{V}_\odot \quad (6)$$

where $\vec{\bar{V}}(R, Z)$ is the mean rotational velocity for a star located at distance $R$ from the Galactic center, and distance $Z$ from the Galactic plane, and $\vec{V}_{\text{LSR}}(R_\odot)$ is the Solar LSR velocity (which it is useful to distinguish from the LSR speed at any other location in the Galaxy, see Section 3.4), while $\vec{V}_{\text{Pec}}$ is the peculiar velocity of the object considered with respect to its own mean Galactic rotational speed, and $\vec{V}_\odot$ is the Solar peculiar velocity with respect to the Solar LSR. If we use a right-handed cylindrical coordinate system, oriented so that one of the axes points toward the Galactic center (U-axis), another axis points toward Galactic rotation (V-axis), and another axis points toward the North Galactic Pole (W-axis), then $\vec{V}_{\text{Hel}} = (U_{\text{Hel}}, V_{\text{Hel}}, W_{\text{Hel}})$, and the different terms in Equation (6) are given by the following expressions:

$$\vec{\bar{V}}(R,Z) = \begin{pmatrix} \bar{V}(R,Z) \cdot \frac{r}{R} \cdot \cos b \sin l \\ \bar{V}(R,Z) \cdot \frac{R_\odot - r \cos b \cos l}{R} \\ 0 \end{pmatrix} \quad (7)$$

$$\vec{V}_{\text{LSR}}(R_\odot) = \begin{pmatrix} 0 \\ V_{\text{LSR}}(R_\odot) \\ 0 \end{pmatrix} \quad (8)$$

$$\vec{V}_{\text{Pec}} = \begin{pmatrix} U' \cos\beta \cos\alpha + V' \sin\alpha + W' \sin\beta \cos\alpha \\ -U' \cos\beta \sin\alpha + V' \cos\alpha - W' \sin\beta \sin\alpha \\ -U' \sin\beta + W' \cos\beta \end{pmatrix} (9)$$

$$\vec{V}_\odot = \begin{pmatrix} U_\odot \\ V_\odot \\ W_\odot \end{pmatrix} \quad (10)$$

where $(R, Z)$ are the distances on the plane of the Galaxy from the Galactic center and perpendicular to it for an object located at Heliocentric distance $r$, and Galactic latitude and longitude $(l, b)$ respectively, $R_\odot$ being the Solar Galactocentric (cylindrical) distance. $\vec{\bar{V}}$ is the mean rotation velocity for the particular component in question at distance $R$, and height $Z$ (in general, different from the rotation curve, specially for large radial-velocity dispersion systems, see Méndez and van Altena (1996) and Section 3.4 below), while $V_{\rm LSR}(R_\odot)$ is the LSR velocity of the stars in the solar neighborhood (the Solar LSR). The velocities $(U', V', W')$ are the peculiar velocities with respect to those oriented-along the principal axes of the velocity ellipsoid for that particular component (Equation (4). The angles $(\alpha, \beta)$ correct for the tilt of the velocity ellipsoid with respect to the local $(U, V, W)$ system. Finally, $(U_\odot, V_\odot, W_\odot)$ are the components of the Solar peculiar velocity with respect to the Solar LSR. The angles $(\alpha, \beta)$ are given by:

$$\sin \alpha = \frac{r}{R} \cos b \sin l \quad (11a)$$
$$\cos \alpha = \frac{R_\odot - r \cos b \cos l}{R} \quad (11b)$$
$$\sin \beta = \frac{r \sin b}{\sqrt{R^2 + Z^2}} \quad (11c)$$
$$\cos \beta = \frac{R}{\sqrt{R^2 + Z^2}} \quad (11d)$$

The Heliocentric proper-motions along the $(U, V)$ axes would be given by:

$$\begin{pmatrix} \mu_{\rm U} \\ \mu_{\rm V} \end{pmatrix} = \frac{\rm K}{r} \begin{pmatrix} U_{\rm Hel} \\ V_{\rm Hel} \end{pmatrix} \quad (12)$$

where K is a conversion factor between the chosen units for the velocities and the proper-motions. For example, if the velocities are in km s$^{-1}$, the distance $r$ is in pc, and the proper-motions are in arcsec yr$^{-1}$, then K is approximately equal to 4.74.

If we wish to express the velocities in an spherical system centered on the Sun, then the Heliocentric velocity can be computed from Equation (6) to (10) via a rotation matrix, in the following way:

$$\begin{pmatrix} V_{\rm rad} \\ V_l \\ V_b \end{pmatrix} = \begin{pmatrix} \cos b \cos l & \cos b \sin l & \sin b \\ -\sin l & \cos l & 0 \\ -\sin b \cos l & -\sin b \sin l & \cos b \end{pmatrix} \begin{pmatrix} U_{\rm Hel} \\ V_{\rm Hel} \\ W_{\rm Hel} \end{pmatrix} \quad (13)$$

and the proper-motions in Galactic longitude ($\mu_l$) and latitude ($\mu_b$) would be similarly given by:

$$\begin{pmatrix} \mu_l \\ \mu_b \end{pmatrix} = \frac{\rm K}{r} \begin{pmatrix} V_l \\ V_b \end{pmatrix} \quad (14)$$

Of course, the choice of computing the pair $(\mu_{\rm U}, \mu_{\rm V})$ or $(\mu_l, \mu_b)$ depends on the particular application (see, e.g., Equations 1 and 2 and comments following them).

If the peculiar velocities and the Solar peculiar motion are neglected, the equations above yield, for $r/R << 1$, the classical Oort result:

$$\begin{pmatrix} \mu_l \\ \mu_b \end{pmatrix} = \begin{pmatrix} \cos b (A \cos 2l + B) \\ -\frac{1}{2} A \sin 2l \sin 2b \end{pmatrix} \quad (15)$$

where the Oort constants are given by:

$$A = \frac{1}{2} \left( \frac{V_{\rm LSR}(R_\odot)}{R_\odot} - \frac{dV_{\rm LSR}}{dR} \Big|_{R_\odot} \right) \quad (16)$$

$$B = -\frac{1}{2} \left( \frac{V_{\rm LSR}(R_\odot)}{R_\odot} + \frac{dV_{\rm LSR}}{dR} \Big|_{R_\odot} \right) \quad (17)$$

We must emphasize that, in our kinematic model, we have *not* used the approximation given by Equation (15) to compute the proper-motions in terms of the Oort constants, rather, we have used the more general expressions described before. However, we show them here because of their usefulness for computing $V_{\rm LSR}(R_\odot)$ and $\frac{dV_{\rm LSR}}{dR}|_{R_\odot}$ from published values for A and B (see Section 5.1).

### 3.3. The Solar peculiar velocity and the motion of the Solar LSR

As it can be seen from Equation (6), the solar peculiar motion, $\vec{V}_\odot$, and the motion of the Solar LSR, $\vec{V}_{\rm LSR}(R_\odot)$, are fixed vectors for a Heliocentric observer, and are also independent of the Galactic component being considered. Therefore, it makes sense to describe the adopted values for these two vectors before discussing the velocity ellipsoid parameters for the disk, thick-disk, and halo.

There have been a number of determinations for the solar peculiar motion. The classical result, quoted in Mihalas and Binney (1981), gives $(U_\odot, V_\odot, W_\odot) = (+9.0, +12.0, +7.0)$ km s$^{-1}$ essentially based upon Delhaye's (1965) own compilation. From their analysis of the BSC, RBC89 obtained $(+11.0, +14.0, +7.5) \pm 0.4$ km s$^{-1}$. In their Figure 6, they also show that a value of $+11.5$ km s$^{-1}$ for $U_\odot$ gives a better fit to the proper-motion position angle distribution than does the value of $+9.0$ km s$^{-1}$ quoted by Delhaye. On the other hand, RO87 have adopted the value derived by Mayor (1974), namely, $(U_\odot, V_\odot, W_\odot) = (+10.3, +6.3, +5.9)$ km s$^{-1}$. As it can be seen, there is a rather big discrepancy with Delhaye's and RBC89's results in the $V_\odot$ component. More recent values for the solar peculiar motion do not seem to have converged to a single value, specially in the V-component: Ratnatunga and Upgren (1997) have found a value of $(U_\odot, V_\odot, W_\odot) = (+8, +7, +6)$ km s$^{-1}$ from Vyssotsky's sample of nearby K & M dwarfs, with uncertainties of $\pm 1$ km s$^{-1}$. Chen et al. (1997) on the other hand find $(U_\odot, V_\odot, W_\odot) = (+13.4 \pm 0.4, +11.1 \pm 0.3, +6.9 \pm 0.2)$ km s$^{-1}$ from a large sample of B, A, and F main-sequence stars. An extreme case of the discrepancies is that of Dehnen and Binney (1998) who find $(U_\odot, V_\odot, W_\odot) = (+10.00 \pm 0.36, +5.25 \pm 0.62, +7.17 \pm 0.38)$ km s$^{-1}$ from a carefully selected unbiased sample of Hipparcos stars, while Miyamoto and Zhu (1998) find $(U_\odot, V_\odot, W_\odot) = (+10.62 \pm 0.49, +16.06 \pm 1.14, +8.60 \pm 1.02)$ km s$^{-1}$ from 159 Cepheids, *also* from the Hipparcos catalogue. As suggested by RO87, these differences in the $V_\odot$ component are mainly due to difficulties in separating the asymmetric drift from the intrinsic solar motion (see Section 3.4), and also because of the peculiar motions exhibited by the very young OB stars, which are still moving under the influence of the spiral arm kinematics and/or of their parent molecular cloud. We have temporarily adopted in the model the solar peculiar motion derived by RBC89, but generally speaking it is a free vector that can be altered to determine the effect of uncertainties on the solar peculiar motion upon comparison with any kinematic survey (see Section 5.1).

The IAU adopted in 1985 a value for the motion of the Solar LSR of 220 km s$^{-1}$. Kerr and Lynden-Bell (1986) have discussed extensively the determinations of $V(R_\odot)$, as well as $R_\odot$, and the Oort constants A and B available until then. From a straight mean of different determinations they obtained $V_{\text{LSR}}(R_\odot) = 222 \pm 20$ km s$^{-1}$ (their Table 4), while from independent determinations of $R_\odot$, A, and B (their Tables 3 and 5), we find $V_{\text{LSR}}(R_\odot) = 226 \pm 44$ km s$^{-1}$. Although the uncertainties involved in $V(R_\odot)$ are larger than those of $V_\odot$, we shall see that the relative motion of disk stars is more affected by uncertainties in the Solar peculiar motion than uncertainties in the motion of the LSR, since the whole nearby disk is moving approximately with the Solar LSR.

### 3.4. Disk kinematics

Méndez and van Altena (1996) have extensively discussed the assumptions employed to describe the kinematics of the disk component in the model, and we will not repeat those here. We shall only mention that velocity dispersions are parametrized in the model as a function of spectral type and luminosity class (Table 2 on Méndez & van Altena 1996). As a result of the model comparisons presented in Méndez and van Altena (1996), a piece-wise linear increase of velocity dispersion with distance from the Galactic plane (in the same amount as predicted by theoretical models) has been found to provide a good match to the proper-motion dispersion of disk stars, and has been generally adopted here for the disk component in the amounts specified in Table 6 of Méndez and van Altena (1996). In addition, a number of dynamical and observational arguments (see Méndez and van Altena 1996 for details, also Binney and Merrifield 1998) lead to velocity dispersions having the following dependency on Galactocentric-distance:

$$\Sigma_{U_j}^2(R) = \Sigma_{U_j}^2 e^{-\frac{(R-R_o)}{H_R}} \quad (18)$$

$$\Sigma_{V_j}^2(R) = \frac{1}{2}\left(1 + \frac{d\ln V_{\text{LSR}}(R)}{d\ln R}\right)\Sigma_{U_j}^2(R) \quad (19)$$

$$\Sigma_{W_j}^2(R) = \Sigma_{W_j}^2 e^{-\frac{(R-R_o)}{H_R}} \quad (20)$$

where $H_R$ is the exponential scale-length for the population considered (index j), and $\Sigma_{U_j}(R)$, $\Sigma_{V_j}(R)$, $\Sigma_{W_j}$, and $V_{\text{LSR}}(R)$ are the velocity dispersions and circular speed (i.e., the rotation curve or, equivalently, the motion of the Local Standard of Rest) respectively at distance R from the Galactic center for that population.

Méndez and van Altena (1996) have also presented a general equation (their Equation (4), our Equation 21) that predicts the velocity lag of the disk component in a dynamically self-consistent way with the adopted velocity dispersions and the adopted rotation curve. In RBC89's model, the velocity lag was modeled as being proportional to the velocity dispersion, $\Sigma_{\rm U}$, with the proportionality factor being a free parameter that changed as a function of the kinematic group considered. On the other hand, RO87 tried a more self-consistent approach, by using a simplified version of the asymmetric drift equation to derive the lag for a particular set of stars as a function of the radial density derivative adopted for that particular component, as it follows from the collisionless Boltzmann equation (Mihalas and Binney 1981, Binney and Tremaine 1987). We have fully developed the RO87 procedure, so that the velocity lag for the disk *is not* a free parameter in itself, but it is correlated with the adopted density function for the disk and the disk rotation curve, leading to the above mentioned self-consistent expression, which we reproduce here for completeness:

$$\bar{V}(R,Z) = \sqrt{V_{\rm LSR}^2(R) - \Sigma_V^2(R) + \left(1 - \frac{2R}{H_R} + S(R,Z)\right)\Sigma_{\rm U}^2(R)} \quad (21)$$

where $V_{\rm LSR}(R)$ is the circular speed at Galactocentric distance $R$ (i.e., the motion of the LSR at that position), and $S(R,Z)$ is a function that describes the contribution to the rotational support from the cross term $\Sigma_{\rm UV}$ (usually referred to as the tilt of the velocity ellipsoid), and it is given by:

$$S(R,Z) = q\frac{(\lambda^2 - 1)R^2}{(R^2 + \lambda^2 Z^2)\lambda^2}\left(\frac{(R^2 - \lambda^2 Z^2)}{(R^2 + \lambda^2 Z^2)} - \frac{|Z|}{H_Z}\right) \quad (22)$$

where $q$ is zero if the velocity ellipsoid has cylindrical symmetry, or one if the velocity ellipsoid has spherical symmetry, $\lambda$ is the (fixed) aspect ratio of the velocity ellipsoid, defined by $\Sigma_{\rm U}/\Sigma_{\rm W}$, evaluated at $Z = 0$, and $H_Z$ is the exponential scale-height for the population considered.

It is interesting to compare the values derived by this expression with those adopted by other authors. The velocity lag is given by $V_{\rm lag} = \bar{V}(R,Z) - V_{\rm LSR}(R)$ (note that, with this definition, the lag is always negative). Representative values for $V_{\rm lag}$, computed from Equations (21) and (22), are listed in Tables 7 and 8 following the (disk) kinematic groups defined by both RBC89 and RO87. The results shown in Tables 7 and 8 have been obtained by evaluating the above equations at $Z = 0$, $R_\odot = R = 8.5$ kpc, and adopting a radial scale-length $H_R = 3.5$ kpc, a spherical velocity ellipsoid ($q = 1$), and a locally flat rotation curve with $V_{\rm LSR}(R_\odot) = 220$ km s$^{-1}$.

From Tables 7 and 8 we see that our derived values for $V_{\rm lag}$ tend to be slightly larger than those compiled by RBC89, while the agreement with RO87 is good. Since our approach is a refinement of the procedure adopted by RO87, the agreement with their results is not surprising. On the other hand, it is not unreasonable to assume that the uncertainties in Delhaye's values for $V_{\rm lag}$ could easily be of the same magnitude as the discrepancies shown in Table 7 (approximately 2 km s$^{-1}$). Figure 1 shows the effect of different assumptions in the evaluation of the mean rotational speed for the old disk ($\Sigma_{\rm U} = 30$ km s$^{-1}$) in the solar neighborhood. In general, the predicted differences are quite small and would be difficult to detect unless high accuracy radial velocities ($\sigma_{V_{\rm rad}} < 0.5$ km s$^{-1}$) and/or absolute proper-motions ($\sigma_\mu < 0.1$ mas yr$^{-1}$) are available for samples of old disk stars located at Heliocentric distances closer than 1.5 kpc. Future space astrometric missions (e.g., the US-FAME project, see http://www.usno.navy.mil/fame, or it's European counterpart, GAIA:http://astro.estec.esa.nl/GAIA/) might actually able to deliver this (or better) proper-motion accuracy for stars down to $V \sim 15$.

### 3.5. Thick-disk kinematics

#### 3.5.1. Velocity dispersion

We have adopted the velocity dispersions of $(\Sigma_{\rm U}, \Sigma_{\rm V}, \Sigma_{\rm W}) = (70, 50, 45)$ km s$^{-1}$ from Gilmore and Wyse (1987), and from the review by Gilmore et al. (1989), where a number of different results are presented and discussed. Layden's (1993, L93 hereafter) survey of field RR Lyraes gives however a value of $(\Sigma_{\rm U}, \Sigma_{\rm V}, \Sigma_{\rm W}) = (57 \pm 12, 35 \pm 9, 39 \pm 10)$ km s$^{-1}$ for the variables with $-1.0 < [Fe/H] < -0.45$, although this sample might still be contaminated by the (old) disk's RR Lyraes, so that the dispersions may be smaller than what one would expect for a "pure" thick-disk component. Indeed, a more refined analysis, employing better proper-motions from the NPM survey, yields $(\Sigma_{\rm U}, \Sigma_{\rm V}, \Sigma_{\rm W}) = (56 \pm 8, 51 \pm 8, 31 \pm 5)$ km s$^{-1}$ (Layden et al. 1996). On the other hand, RO87 have adopted values that are somewhat bigger than our adopted values, $(\Sigma_U, \Sigma_V, \Sigma_W) = (80, 60, 55)$ km s$^{-1}$, from a reinterpretation of the compilation by Delhaye (1965). In the absence of any firm observational constraints, we have assumed that the velocity dispersions for the thick-disk are isothermal, i.e., they do not depend on position in the Galaxy, this is the same approach followed by RO87.

*3.5.2. Velocity lag*

The velocity lag for the thick-disk has been a controversial issue ever since the discovery of this component by Gilmore and Reid (1983). At that time, Gilmore and Reid linked this component to the metal-rich RR Lyrae ([Fe/H] = -1.0) and the Long-Period Mira variables ($145 < P < 200$ days) that exhibit a lag of around -120 km s$^{-1}$ (Gilmore et al. 1989, Gilmore 1990). Wyse and Gilmore (1986) derived a lag of around -80 to -100 km s$^{-1}$ from the proper-motion survey by Chiu (1980), a result disputed by Norris (1987) on the basis of a zero point error in Chiu's proper-motions (see also Majewski 1992). Indeed, later surveys showed that the lag is perhaps smaller, closer to -40 km s$^{-1}$. A number of studies have produced results around this latest value (Gilmore et al. 1989), and it seems that a velocity lag as large as 100 km s$^{-1}$ has to be ruled-out (although, see Section 5.2).

In the field of global kinematic modeling, RO87 used for their thick-disk a velocity lag as computed from a simplified version of the asymmetric drift equation (see Section 3.4). Even though this approach is able to predict velocity lags as large as -37 km s$^{-1}$ (RO87, Table 1), it requires unrealistically large velocity dispersions to obtain velocity lags twice that value (the above quoted lag value was obtained considering a U-velocity dispersion for the thick-disk of 80 km s$^{-1}$, a value that is already somewhat larger than the values quoted more recently, see Section 3.5.1). Furthermore, since the variations of $\Sigma_U(R)$ and $\Sigma_V(R)$ with Galactocentric distance for this component are not known, our self-consistent approach of Section 3.4 cannot be applied. We have therefore decided to assume a constant rotation velocity for the thick-disk. This approach has been used by Armandroff (1989) in his study of the system of Galactic globular clusters, and by L93 in his study of RR Lyrae stars (this assumption was first presented by Frenk and White 1980 in the context of Globular cluster kinematics). Therefore, $V_{\text{lag}}$ has been assumed to have a fixed value for this component, whose possible range goes from around -100 km s$^{-1}$ to -40 km s$^{-1}$. It should be mentioned that it has been suggested (Majewski 1992, Majewski 1993, particularly his Fig. 6) that the thick-disk exhibits a velocity lag gradient with distance from the Galactic plane amounting to -36 km s$^{-1}$ kpc$^{-1}$ to distances from the Galactic plane of up to 7 kpc. The extent and nature of this gradient has however become more elusive from the recent surveys by Soubiran (1993) who found little or no velocity gradient from her North Galactic Pole sample, by Ojha et al. (1994) who also found no velocity gradient or even a reverse gradient (i.e. an *increase* of rotational velocity with height above the Galactic plane) from their anti-Galactic-center field, and by Guo (1995) who found little or no velocity gradient from his South Galactic Pole sample, although Guo's sampling of thick-disk stars is restricted to distances closer than 4 kpc, as opposed to 7 kpc for Majewski's sample. In Section 5.2 we actually study the effects of adopting either a solid-rigid rotation *vs.* models where we introduce a velocity shear away from the Galactic plane.

### 3.6. Halo kinematics

*3.6.1. Velocity dispersion*

We have adopted $(\Sigma_U, \Sigma_V, \Sigma_W) = (130, 95, 95)$ km s$^{-1}$, from the review by Gilmore et al. (1989). These values of the velocity dispersion are in agreement with, e.g., L93's results for the metal-poor ($[Fe/H] < -1.3$) RR Lyraes; he found $(\Sigma_U, \Sigma_V, \Sigma_W) = (160 \pm 15, 112 \pm 10, 99 \pm 10)$ km s$^{-1}$. By comparison, RO87 adopted the results by Norris et al. (1985), namely $(\Sigma_U, \Sigma_V, \Sigma_W) = (131, 106, 85)$ km s$^{-1}$.

Since the kinematical properties of the Galactic halo are only poorly constrained by present observational data, we have assumed, as with the thick-disk, that the halo velocity dispersions are also isothermal. Indeed, the locally determined velocity ellipsoid for the highest velocity subdwarfs, which spend most of their time at very large distances from the Galactic center, shows no evidence for a different anisotropy than does that for lower velocity subdwarfs (Gilmore and Wyse 1987). Furthermore, Hartwick (1983) showed that the velocity dispersion of 52 metal-weak halo giants has a similar anisotropy to that shown by the nearby RR Lyrae stars analyzed by Woolley (1978).

Ratnatunga and Freeman (1985, 1989) have found

from their own sample of field K giants, and from a number of previous determinations, that the W-velocity dispersion, $\Sigma_W$, for halo stars is constant with height up to distances of 25 kpc above the Galactic plane. They also pointed out that a constant W-velocity dispersion in spherical coordinates would be inconsistent with the observations, and therefore one of the velocity ellipsoid axes should remain parallel to the Galactic plane, and point toward the Galactic axis of rotation. Finally, they found that their data are well represented by a model where the velocity dispersions in cylindrical polar coordinates, $(\Sigma_U, \Sigma_V, \Sigma_W)$, remain constant in their three fields (SA 127 ($l = 270^o, b = +38.6^o$), SA 141 ($l = 240^o, b = -85.0^o$), SA 189 ($l = 277^o, b = -50.0^o$)) for Heliocentric distances closer than about 25 kpc, thus justifying our assumption of an isothermal halo. They argue that this model is also an excellent fit to Pier's (1983) kinematical data for BHB stars in the inner halo. Incidentally, they found that their observed velocity distribution does not differ significantly from a Gaussian, thus lending support to the assumed shape of our velocity distribution function.

### 3.6.2. Velocity lag

The same assumption of a constant rotation velocity used for the thick-disk has been applied to the halo component. The halo has been claimed to be counter-rotating at -60 km s$^{-1}$ ($V_{\text{lag}} = -280$ km s$^{-1}$, Majewski 1992) and rotating at +20 km s$-1$ ($V_{\text{lag}} = -200$ km s$^{-1}$, L93). As with the thick-disk, the halo velocity lag has also been assumed to be fixed for this component, with a typical value of -220 km s$^{-1}$ (i.e., no net rotation for the halo), close to the value -229 km s$^{-1}$ adopted by RO87, but whose possible range goes from around -320 km s$^{-1}$ to -100 km s$^{-1}$. Also, our approach of a constant velocity rotation for the halo is similar to that of RO87's model.

## 4. Model comparisons to magnitude and color counts

### 4.1. Characterizing the stellar populations content of the catalogue

Before proceeding to the kinematic comparisons between the SPM data and the model predictions, it is relevant to characterize the stellar populations that are represented in the catalogue as a function of apparent brightness. This is also a critical step for an assessment as to which kinematic parameters can actually be constrained as a function of apparent magnitude.

Figure 2 shows the expected counts as a function of $B_J$ magnitude for the SPM-SGP region. Because of the large solid angle covered by the survey, some experimentation was needed to decide upon the best angular resolution to employ in the integration scheme in the starcounts model in order to properly account for projection effects on both the starcounts and the kinematic model predictions. It was found that, with a resolution coarser than 4 deg in both RA and DEC, the predicted counts and kinematics from the model varied by less than 0.5%. Therefore, we adopted this angular resolution. It must be emphasized that a finer angular resolution could easily be performed with the model, but then the required CPU time becomes increasingly large, making these computations impractical (e.g., already, at the adopted resolution, the model has to be evaluated at 36 positions within the SPM-SGP region). Also, the solid angle being integrated had the very same borders in RA and DEC as the observed area, such that projection effects are similar in both the observations and the model.

The model predictions in this section have been performed with the standard Galactic and stellar population parameters for the model as described in Section 3.1 and by Méndez and van Altena (1996), with the modifications indicated by comparisons to faint magnitude and color counts to the Hubble Deep Field (HDF, Méndez et al. 1996) and to the HDF Flanking Fields (Méndez and Guzmán 1998). We should also point out that the evaluations performed here are used merely as a guide to the type of stellar population mapped at different magnitude intervals, and that *no* attempt has been made to fit the observed *magnitude* counts because of the incomplete nature of the sample, as described in Section 2.1 (although the color counts in one $B_J$ magnitude intervals are fully accounted for sample incompleteness through a free scale factor, see Section 4.2). It has been demonstrated that the model run in this "blind mode" is able to accurately reproduce (with less than 1 % uncertainty) the observed magnitude and color counts over more than 10 magnitudes at the SGP (Prandoni et al. 1998) as well as to other lines-of-sight as derived from comparisons to multi-color data from the ESO Imaging Survey covering several tens of square-degrees at different Galactic positions (da Costa et al. 1998, Nonino et al. 1998, Zaggia et al. 1998).

From Figure 2 we can clearly see that the disk is the dominant source of the counts at *all* magnitudes. However, for $B_J > 14$ the thick-disk starts to be important, with a contribution to the counts larger than 10% of that of the disk, *per magnitude interval*. Similarly, the halo becomes important, in the same sense, for $B_J > 17$.

In the range where the disk is the dominant source of the counts (i.e, for $B_J < 14$), we have a mixture of main-sequence, giant, and subgiant stars. Figure 3 indicates that in this magnitude range, giants are slightly dominant over main-sequence stars for $B - V > 1.0$, while for bluer color we basically should observe only main-sequence stars. Therefore, any attempt to look at the kinematics, in a differential way, between disk main-sequence and giant stars should approximately follow these two distinct color intervals. Indeed, for colors bluer than $B - V = 1.0$ the model predicts a *ratio* of $(9.5 \pm 0.2) \times 10^{-2}$ disk giants and subgiants per main-sequence star, while for redder color this ratio becomes $1.58 \pm 0.05$ (the uncertainties come from Poisson statistics on the expected counts for a complete sample). This, and all subsequent color plots, have been convolved with the color uncertainties in the respective magnitude intervals as derived from the catalog itself to obtain a better idea of how well we can or can not separate distinct populations of stars.

In the range where the thick-disk becomes important, while the contribution from the halo is still minimal, i.e., $14 \leq B_J < 17$, most disk stars are on the main-sequence, while we have a mixture of giants and main-sequence stars from the thick-disk (see Figure 4). For $B - V > 1.0$ we expect very little contribution from the thick-disk at all (these would then be mostly disk main-sequence stars), while for bluer colors we have a mixture of disk main-sequence and thick-disk main-sequence and giants (for $B - V < 1.0$ the ratio between thick-disk and disk stars is 0.336, while the ratio decreases to 0.048 for redder colors). Unfortunately, as can be seen from Figure 4, the color range encompassed by thick-disk main-sequence stars and giants mostly overlap. However, since the thick-disk is parametrized in the model by a single kinematic population, the distinction between main-sequence and giants from the thick-disk is not as critical as for the disk, where the kinematic parameters *are* assumed to be dependent upon the spectral type and luminosity class. This fact allows us to treat all thick-disk stars in a common fashion, irrespective of their luminosities.

Finally, in the faintest range $(17 \leq B_J < 19)$, where all three components do contribute to the stellar counts, the disk still dominates for $B - V > 1.0$, while a mixture of thick-disk and halo stars appears at bluer colors, producing a characteristic double-peaked color distribution (Figure 5). As can be seen from Figure 5, it is not possible to separate thick-disk and halo stars from colors alone and, in principle, a simultaneous fit to the kinematic parameters for both populations has to be performed at these magnitudes. We can also see that giants from both populations fall at approximately the same color interval and at about the same rate. Similar colors are also expected for halo and thick-disk main-sequence stars, although the model predicts more thick-disk main-sequence stars than halo main-sequence stars; The predicted overall ratio (for all colors) in this magnitude interval is $n_{\mathrm{halo_{m-s}}}/n_{\mathrm{thick-disk_{m-s}}} = 0.447 \pm 0.003$. These considerations are important because they indicate that the derived kinematics for the halo would necessarily be based on relatively few stars from this population falling in our sample.

The mean distance for the different populations sampled as a function of magnitude are shown in Table 9. We see that disk stars are sampled to less than 1 kpc from the plane, even at the faintest magnitude bins. The thick-disk is sampled on a range of about 2 kpc, and up to almost 4 kpc, while the halo sample is based on distant stars located at typical distances of 5 kpc from the plane.

### 4.2. The color counts

As expressed before, the magnitude counts suffer from a selective incompleteness as a function of magnitude. However, at a given magnitude bin, the color distribution (just as the proper-motion distributions) will differ from the complete-sample distribution, only by a scale factor. In this section we thus compare the observed color histograms to the model predictions in the pre-selected $B_J$ magnitude bins of the survey, using a free scale factor to go from the model-predicted to the observed color counts.

Figures 6 and 7 shows a comparison of the standard model being used here, and the SPM color counts. As shown in Section 4.1, for $B_J < 14$ we have mostly disk stars. Figure 6 shows the error-convolved color distributions in the bright-magnitude portion of the survey, while Figure 7 shows the color distributions for the faint portion. In all cases a scale factor has been applied to the model counts to bring them onto the observed counts. At $B_J < 16$ the scale factor was computed by forcing the model to have the same number of stars as observed in the color range $0.0 \leq B - V \leq 1.5$, where the majority of the stars are found, while for fainter magnitudes the scale factor was computed from the extended range $0 \leq B - V \leq 2.0$.

In general, we notice a very good fit to the observed color counts at all magnitudes. However, there are several features worth mentioning. At faint magnitudes ($B_J > 14$) the fit is extremely good, apparently without any further refinements needed to the model, at least within the uncertainties of the counts and the photometric errors. At fainter magnitudes ($B_J > 17$), the appearance of the characteristic double-peaked color distribution due to blue halo and thick-disk turn-off stars and red M-dwarfs from the disk becomes less distinct due to the photometric errors. Nevertheless, the important point here is that the fit to the overall counts is very good, with the slight indication of a small systematic effect in our photometry in the range $0.2 \leq B - V \leq 0.8$ on the amount of -0.05 mag in $B - V$. We notice that, while the effect of changing the scale-height of these (bright) main-sequence stars has a minimal impact on the predicted color counts in the range $14 < B_J < 17$, the effect becomes more important in the last two magnitude bins (where photometric errors are quite large), indicating that the currently adopted value of 325 pc for bright M-dwarfs provides a better fit to the color counts than does the smaller value of 250 pc suggested from studies of local *fainter* M-dwarfs in the disk (Méndez and Guzmán 1998).

At bright magnitudes, the situation is more confusing: On one hand, it is apparent that the model is predicting slightly more giants than observed (the red peak in the distributions for $B_J < 12$). On the other hand, it seems that our model predictions are bluer than observed for $B - V \leq +0.7$ in *certain* magnitude bins, while in others the fit is quite good (see Figure 6). To explore the origin of these discrepancies, first, it is interesting to note that in the magnitude range $13 < B_J < 14$ we expect to see almost no disk giants, and, *at the same time*, the contribution from thick-disk stars is negligible. This magnitude range is therefore ideal to explore the origin of the discrepancy, where the model is actually bluer than the observed counts, even for $B - V \leq +1.0$. We have run several models to see whether the observed discrepancy can be accounted for in a reasonable way by tuning-up some of the model parameters. Since the major contributor to the counts in this magnitude range comes from disk main-sequence stars, we have concentrated on what determines the *shape* of their expected counts. There is only one *overall* relevant model parameter, the scale-height, determining the contribution of disk main-sequence stars in this magnitude range. We first notice that the model predictions indicate that the absolute magnitude range sampled by this color distribution encompasses the range $+3.2 \leq M_V \leq +4.2$, i.e., these are still quite bright main-sequence stars, at a point where their scale-height is known to be increasing quickly with (fainter) absolute magnitude. In our model, we have adopted a variable scale-height for main-sequence stars to account for the known fact that older stars have diffused to larger distances from the Galactic plane than younger stars (Wielen and Fuchs 1983). The functional form giving the scale-height as a function of absolute visual magnitude described by Miller and Scalo (1979), and Bahcall and Soneira (1980), which seems to be a good *representation* of the available observational data (see also Gilmore and Reid 1983), has been included in our model in the way indicated by Bahcall (1986). Figure 8 shows the results of these runs with extreme parameters for the scale-height of main-sequence stars ($H_Z(MS)$). The compilations by Miller and Scalo (1979) and Bahcall and Soneira (1980), as well as the results from Gilmore and Reid (1983), seem to indicate that $H_Z(MS)$ is approximately constant for $M_v \leq +2$ ($\sim 90\ pc$) and for $M_v > +5$ ($\sim 325\ pc$). Following Bahcall (1986), we have used a linear interpolation between $M_v = +2$ and $M_v = +5$. We have considered two extreme cases, taken from the range allowed by observational data (Gilmore and Reid 1983), by assuming a "lower" envelope and an "upper" envelope for $H_Z(MS)$. The lower envelope is described by a scale-height of 50 pc for early type stars and 300 pc for later type stars, the upper envelope is described by a scale-height of 120 pc for early type stars and 400 pc for later type stars. The shape of the upper and lower envelopes are self-similar, in that the slope of the linear interpolation for $H_Z(MS)$ between the early and late type stars was kept constant at the same value adopted by Bahcall (1986), namely, $\sim 84\ pc/M_v$. It is apparent from Fig. 8 that we cannot effectively distinguish between a lower $H_Z(MS)$, an increase in reddening of +0.05 mag in E(B-V), or a -0.05 mag systematic effect on the colors. However, we can rule out the first two alternatives on the grounds of previous studies. For example, Méndez and van Altena (1996) were able to set the overall level for $H_Z(MS)$ in the range $+2 \leq M_v \leq +4$ from comparisons to magnitude and color counts in two intermediate-latitude fields. They found a scale-height very similar to the one adopted in the standard model used here - thus ruling out the small scale-height solution. Our model runs have adopted a reddening of E(B-V) = 0.03 at the SGP, therefore an increase of 0.05 mag in this quantity would imply a *mean* reddening of $E(B-V) = 0.08$. However, this value is too high, certainly beyond any of the values found by different investigators of the reddening distribution in and around the SGP. Indeed, the maximum value reported in the literature, comes from an estimate based on HI and *IRAS* 100-micron flux maps, and gives $E(B-V) = 0.06$ (Nichol and Collins 1993), while smaller maximum values are reported in the more recent maps based on the combined *COBE*/DIRBE and *IRAS*/ISSA data (Schlegel, Finkbeiner, and Davis 1998). There is, however, an even stronger argument in favor of the idea that our colors suffer from a small systematic effect, and that is that there is no way for our model to fit *simultaneously* the apparently bluer colors at $11 \leq B_J < 14$, while preserving the already existing good fit at all other (brighter and fainter) magnitudes. For this reason, we believe that our colors *do* have a small, but noticeable, systematic effect at some *specific* magnitudes, most noticeable at $11 \leq B_J < 14$. Indeed, it is not far fetched to assume that our photographic photometry could have a systematic error of such amount, especially when considering that a number of internal calibration procedures had to be applied in order to make use of all the grating images on the plates, and that the faint photometric zero-points were established only from one or two CCD frames placed arbitrarily in the field, and covering a tiny fraction of a full plate (for details see Platais et al. 1998).

The other point concerns the fit to the red peak at bright magnitudes. It does seem as though the model is over-predicting the contribution of disk giants. Indeed, as shown in Figure 9, the predicted number of red giants is extremely sensitive to the adopted value for its scale-height ($H_Z(G)$). We have performed a $\chi^2$ fit to the observed colors in the magnitude range $9 \leq B_J < 12$ where the contribution from Giants is most important. The model predictions included a "minimal model" with a scale-height of 150 pc, and a "maximal' model with a scale-height of 250 pc. The minimum $\chi^2$ was computed in the color range $+0.7 \leq B - V \leq +1.5$ by interpolating between the two extreme model predictions, and by always scaling to the total observed color counts in the same color range. Separate $\chi^2$ fits were performed in the ranges $9 \leq B_J < 10$, $10 \leq B_J < 11$, and $11 \leq B_J < 12$, and the mean (weighted by the number of stars used in the fit) and its standard deviation value for the Giant's scale-height turned out to be $H_Z(G) = 172 \pm 7$ pc. Figure 10 shows the resultant models adopting this $H_Z(G)$ for disk giants, and where some attempt has been made to correct for slight systematic shifts in the SPM photometry. It is important to notice that the fits were performed *only* in the color range $+0.7 \leq B-V \leq +1.5$, and therefore the improved fit outside this range provides an indication of the properness for the parameters adopted to describe the main-sequence disk color counts. As expected, runs with this new scale-height render a much better fit to the overall color counts in the whole range $9 \leq B_J < 14$, and we adopt this value throughout.

## 5. Model comparisons to the absolute proper-motions

In this section we present comparisons between the observed absolute proper-motions derived from the SPM-SGP data and our model predictions. The basic standard kinematic parameters employed in the model have been described in Méndez and van Altena (1996), and in Section 3 above. Here we present only a summary of the basic assumptions.

For the peculiar solar motion we have adopted the value derived by RBC89, namely, $(U_\odot, V_\odot, W_\odot) = (+11.0, +14.0, +7.5)$ km s$^{-1}$, and a flat rotation curve with $V_{\rm LSR}(R_\odot) = +220$ km s$^{-1}$. It must be emphasized that, since we are analyzing data near the SGP, our model predictions do not span a large range in Galactocentric distance, and therefore the model is *not* overly sensitive to the value adopted for the *slope* of the rotation curve near $R_\odot$ (although see Section 5.1). As for the disk kinematics, we have adopted the velocity dispersions indicated in Section 3.4 with the scale-heights adopted in Section 4.2, a scale-length of 3.5 kpc (again, the model is not sensitive to this last parameter, as it enters as a function of the Galactocentric distance, which for the SPM-SGP data is quasi-constant, see Equation 21), and a value of $q = 0$ for the velocity ellipsoid (see Equation 22, $q = 0$ implies a velocity ellipsoid parallel to the Galactic plane at all heights from the Galactic plane, while $q = 1$ is for a velocity ellipsoid that points towards the Galactic center).

For the thick-disk we assume an isothermal velocity dispersion equal to $(\Sigma_U, \Sigma_V, \Sigma_W) = (70, 50, 45)$ km s$^{-1}$, as a compromise value between different determinations (see Section 3.5.1), and a constant velocity lag of 40 km s$^{-1}$ with respect to the motion of the LSR. For the halo instead, we adopt $(\Sigma_U, \Sigma_V, \Sigma_W) = (130, 130, 95)$ km s$^{-1}$ from Gilmore et al. (1989), and a zero net rotation velocity about the Galactic center.

The three-dimensional integration of the velocity ellipsoid at all distance shells required by the model (see Equation 4) is very expensive in terms of CPU cycles. Therefore, some trial runs were needed to select the proper integration resolution. It was found that a Gaussian-normalized resolution of 0.2 between -3.0 and +3.0 in $U'/\Sigma_U$, $V'/\Sigma_V$, and $W'/\Sigma_W$ leads to differences in the derived proper-motion median and dispersions smaller than 0.1 mas yr$^{-1}$, except at the brightest bins, where the differences were in any case smaller than 0.3 mas yr$^{-1}$.

We should note that, because of the geometry for the SGP data, our model comparisons are most sensitive to the motions in the direction of Galactic rotation, and along the Galactic center-anticenter direction, and *not* to the motion perpendicular to the Galactic plane (see Equations 1 and 2, and the discussion following them).

Finally, all model predictions have been convolved with the proper-motion errors as found in Section 2.1 (Tables 5 and 6) before computing any kinematic parameter to be compared with the observed distributions.

We have investigated the effects of changing the mean reddening at the SGP from $E(B - V) = 0$ to $E(B - V) = 0.06$ on the predicted kinematic parameters, and found that the maximum changes occurred at the brighter magnitudes, but were in all cases smaller than 0.2 mas yr$^{-1}$ in the median $\mu_U$ and 0.4 mas yr$^{-1}$ in the median $\mu_V$. The proper-motion dispersions changed by, at most, 0.4 mas yr$^{-1}$ in $\Sigma_{\mu_U}$, and by 0.3 mas yr$^{-1}$ in $\Sigma_{\mu_V}$. As it can bee seen from Tables 2, 3, and 4, these changes are smaller than the $1\sigma$ observed uncertainties, and therefore do not play an important role in this discussion.

## 5.1. The Solar motion and the LSR speed

In this subsection we present our model comparisons as a function of apparent magnitude, and the sensitivity of those predictions with respect to the assumed values for the Solar peculiar motion and the speed of the LSR.

Figures (11) and (12) show the observed and the model predictions in the median proper-motions as well as the proper-motion dispersions respectively, while Table 10 indicates the values for the different kinematic parameters derived from the standard model (called Run 1). In Table 10 the first two lines for each magnitude entry indicate the values for the median and dispersion on the U-component of the proper-motion, while the last two rows indicate the same parameters for the V-component. Table 10 lists only model predictions for the most representative runs, as otherwise the table would be too cluttered without adding much information for the reader. Table 11, instead, gives a summary description of all simulations presented in this paper. For the standard run (Run 1), the model parameters are those described before. Figure (11) clearly shows that, while the standard model produces a very good fit to the U-component of the proper-motion (along the Galactocentric direction), the V-component (along Galactic rotation) is grossly underestimated, specially for $B_J < 14$, where the disk component dominates the overall kinematics. A comparison with a model having a scale-height for disk Giants of 250 pc (Run 2) does not resolve the problem. Indeed, this solution also produces a bad fit to the predicted motion in U at the brightest magnitudes, and therefore it reinforces the value of 172 pc found before from the color counts alone.

We have also tried a run with $q = 1$ (see Equation 22, Run 3). In this case, the proper-motion in U is not affected (as expected from the projection effects), while the motion in V is slightly shifted downwards (i.e., more negative lags) in an almost systematic way, in the sense of *increasing the discrepancy* with the observed proper-motion. A more radical change in the model predictions occurs when we change the peculiar Solar motion from the standard value of $V_\odot = +14$ km s$^{-1}$ to $V_\odot = +5$ km s$^{-1}$ (Run 4). The largest effect occurs at the brightest bins, i.e., for nearby disk stars where the Solar peculiar motion dominates the reflex motion, while the change becomes less important (although still notice-able) at fainter magnitudes where one is sampling objects from the other Galactic components located at larger distances, where the dominant effect is that of the overall rotation of the disk, and the relative state of rotation between the different Galactic components. This is clearly shown (Fig. 11) by a run where we keep the old Solar peculiar velocity, but change the overall rotation speed for the LSR from the IAU adopted value of $+220$ km s$^{-1}$ to $+270$ km s$^{-1}$ (Run 5), as suggested by recent Hipparcos results (Miyamoto and Zhu 1998). At the brightest bins, the effect of changing $V_{\mathrm{LSR}}(R_\odot)$ becomes also noticeable because of the larger fraction of bright giants, which can be seen to large distances, and where the differential rotation effects become amplified. A run where we simultaneously change the Solar peculiar motion *and* the LSR rotational speed to $+5$ km s$^{-1}$ and $+270$ km s$^{-1}$ respectively is given by Run 6. We note that, in this run, the velocity lags for the thick-disk and halo are increased in proportion to the increase of the disk's rotational speed, as the net rotation of those two components is kept constant at 180 km s$^{-1}$ and 0 km s$^{-1}$ respectively (see Sections 3.5.2 and 3.6.2). We conclude that Run 6 clearly provides a much better fit to the median motion in V than does Run 1.

The median proper-motion in the U-component shows a good fit to the observed values, and the changes in the V-component described above do not affect this parameter in a major way because of the orthogonality of the projection effects toward the Galactic poles. These results do show us, though, that the adopted value for the Solar peculiar motion in this direction is the correct one. To explore the sensitivity of the model predictions to this parameter, Runs 8 and 9 (Figure 14) show the effect of changing the standard $U_\odot = +11$ km s$^{-1}$ by $\pm 3$ km s$^{-1}$. An eye-ball fit from Figure 14 suggests for the U component of the solar motion a value of $U_\odot = +11.0 \pm 1.5$ km s$^{-1}$. Also, there is no indication of a local expansion or contraction of the Galactic disk, as also found from the kinematics of local molecular clouds (Belfort and Crovisier 1984).

The predicted proper-motion dispersions (Figure (12)) are less affected than the median proper-motions by the changes described above. In particular, we see from Figure (12) that the biggest change comes from a change of the Giant-star scale-height, and even in this case the differences are minimal, with the standard scale-height of 250 pc producing a slightly better fit than the adopted 172 pc value. This is a natural consequence of Giants having a slightly larger velocity dispersion than main-sequence stars, but being sampled to much larger distances, and having more representation in the total counts for the larger $H_Z(G)$ value (see Fig. 9), thus decreasing the predicted dispersions. If we insist on the the small $H_Z(G)$ value, these results would mean that our adopted value of $(\Sigma_\mathrm{U}, \Sigma_\mathrm{V}, \Sigma_\mathrm{W}) = (30, 20, 20)$ km s$^{-1}$ for Giants and Subgiants are perhaps a bit overestimated. We also see from Figure (12) that, while the U-component of the proper-motion dispersion is well matched by the model predictions, there seems to be an overall *overestimation* of the velocity dispersions in the V component, a point to which we will return later. In the lower panel on Figure (12) we can also see that a change from $V_\odot = +14$ km s$^{-1}$, $V_{\mathrm{LSR}}(R_\odot) = +220$ km s$^{-1}$ (Run 1) to $V_\odot = +5$ km s$^{-1}$, $V_{\mathrm{LSR}}(R_\odot) = +270$ km s$^{-1}$ (Run 6) has a very small impact on the predicted proper-motion dispersion in the V-component.

When comparing the model predictions to the observed motions it is relevant to specify the degree of accuracy of our motions in terms of any remaining systematic effects. Different internal and external tests on the catalogue indicate that the uncertainty in the correction to absolute zero point is around 1 mas yr$^{-1}$ per field (Platais et al. 1998). However, this random uncertainty gets averaged over the total of 30 fields in the SGP area, leading to an rms uncertainty of less than 0.2 mas yr$^{-1}$. After publishing version 1.0 of the SPM catalogue (Platais et al. 1998), an extensive investigation was carried out on the estimation of the residual magnitude equation for stars and galaxies on the SPM plates. This resulted in a slight shift (only for the purpose of magnitude equation) between the magnitude system of these two types of objects in the catalogue, leading to version 1.1 of the catalogue. Also, a detailed comparison between the SPM proper-motions and the Hipparcos motions at the bright end of our sample revealed a small offset between the Hipparcos and the SPM 1.1 proper motions, mostly in declination, in the sense $\mu_{\mathrm{Hipp}} - \mu_{\mathrm{SPM}} = -0.73 \pm 0.06$ (m.e.). This discrepancy is actually quite small if we consider that the SPM motions were tied to the extragalactic system at faint magnitudes where galaxies are measurable in large numbers, while the Hipparcos motions rely on the calibrated observations of bright stars. We have thus to bridge a range of about 10 magnitudes to compare the SPM and the Hipparcos proper-motion zero-points, and it should not be surprising to find a small offset between the two systems. In order to assess the effect of the small offset observed between the SPM and Hipparcos catalogues, we have compared the derived SPM motions from SPM 1.0, SPM 1.1, and SPM 1.1 reduced to the Hipparcos system. This is shown in Fig. (13) for the median proper-motions. The effect of these uncertainties on the U-component is quite small, but it is slightly larger for the V-component. Still, the use of *any* of the catalogues does not invalidate our previous conclusions, nor the discussion that follows. The proper-motion dispersions on the other hand show a variation of less than 0.02 mas yr$^{-1}$, and are therefore considered negligible in comparison with the modeling effects discussed here. For definiteness, we adopt version 1.1 of the catalogue, *without* the correction for the declination offset to the Hipparcos system. The fact that our overall motions are not affected by the systematic effects discussed previously is important in the context of the velocity lags for the thick-disk and halo; see e.g., Section 3.5.2.

We have pointed out that a change of the orientation of the velocity ellipsoids from a cylindrical to a spherical projection tends to produce a slightly larger value for the velocity lag, especially at fainter magnitudes (compare Runs 1 and 3 in Fig. (11)). The upper solid line on Fig. (11) shows that while the fit of the model to the observed data for $V_\odot = +5$ km s$^{-1}$, $V_{\rm LSR}(R_\odot) = +270$ km s$^{-1}$ is good at bright magnitudes ($B_{\rm J} < 14$), the model underestimates the lag at fainter magnitudes. However, as indicated above, by changing the orientation of the velocity ellipsoid we can actually increase the lag. This is shown in Figure (14) (lower panel) where a model with $V_\odot = +5$ km s$^{-1}$, $V_{\rm LSR}(R_\odot) = +270$ km s$^{-1}$ and $q = 1$ (Run 7) gives a better overall fit to the observed median motion in the V component than a model with $q = 0$. The model predictions from Run 7 are also indicated in Table 10.

We have also further explored the origin of the slight overestimate on the predicted proper-motion dispersion in the V-component seen in Figure 12 (lower panel). Even though the overestimate is small, it is clearly systematic and it could, therefore, be due to a (set of) wrong assumptions in the kinematic model. The velocity dispersion in the V-component is *not* an entirely free parameter in the model. It is actually derived from the U-velocity dispersion and the slope of the rotation curve, as shown in Eq. (19), and therefore can be used to explore either of these parameters. In addition to the Galactocentric dependency made explicit by Eqs. (18) to (20), there is also a dependency on distance from the Galactic plane, which has been described in Méndez and van Altena (1996), in the sense that the velocity dispersion for disk stars *increases* as a function of distance to the plane ($|Z|$). This increase as a function of $|Z|$ has been found to be necessary in the model, through comparisons to intermediate latitude proper-motions, mostly in the U and W-components (Méndez & van Altena 1996), but it *has not* been tested so far for motions in the V-component. Therefore, a natural test was to turn off the above mentioned expected increase in $\Sigma_{\rm V}$ *vs.* $|Z|$, to see what would be the effect upon the derived dispersion in the respective component of the proper-motions: This is shown in Fig. 15 (upper panel) by the dashed line (Run 10). As expected, the predicted dispersions are smaller, and agree quite well with the observed ones. However, by decreasing the velocity dispersions in V, the circular speed for the disk *increases* (see Eq. (21)), thus producing a *smaller lag*, and as a result, the fit to the median motion as a function of $B_{\rm J}$ gets significantly worse (Fig. (15), lower panel, dashed line). Theoretically, it would be hard to explain why the U and W-components of the motion do show this increase, while the V component does not (Fuchs and Wielen 1987). Therefore, we have explored whether changes in the slope of the rotation curve, or a straight reduction in the local values of $\Sigma_{\rm U}$ (from which $\Sigma_{\rm V}$ is derived, see Eq. (19)) could be held responsible for the poor fit. Figure (15) shows the results when leaving the $|Z|$-gradient untouched, but for the case of a very large slope in the local rotation curve, $dV_{\rm LSR}(R)/dR = -11.7$ km s$^{-1}$ (corresponding to the lower $3\sigma$ value derived from the Oort constants, namely $A = 14.4 \pm 1.2$ km s$^{-1}$, $B = -12.0 \pm 2.8$ km s$^{-1}$, Kerr and Lynden-Bell 1986, Run 11), and for a 10% decrease in the local values for the U-velocity dispersion (Run 13), respectively. The large LSR slope solution (Run 11) is somewhere in between Run 7 and Run 10, producing a worse fit in $\Sigma_{\mu_{\rm V}}$ than Run 10, especially in the magnitude range $13 \leq B_{\rm J} < 16$. A 10% reduction in $\Sigma_{\rm U}$ (Run 13) produces an effect on the proper-motion dispersion that is similar to a large gradient in the LSR speed (Run 11, upper panel on Fig. (15)), but it *also* produces an undesired large change in the computed mean motion for disk stars, in the sense of *increasing* the net rotation, and therefore decreasing the lag predicted by the asymmetric drift-equation, as it is indeed expected from Eq. (21). The overall effect of a 10% reduction in $\Sigma_{\rm U}$ is that Run 13 does not seem to provide a good representation of the SPM-SGP data. The change in the predicted dispersion in the U-component from Run 13 is actually quite small, and we obtain a fit that is as good as the one from Run 1 in Fig. (12) (upper panel). Finally, we notice that a small change in the slope of the rotation curve, from zero-slope in Run 7 (continuous line on Fig. (15)) to $dV_{\rm LSR}(R)/dR = -2.4$ km s$^{-1}$ (as predicted by Oort's constants, dotted line on Fig. (15), Run 12) does have a correspondingly small change in, both, the median and the dispersions, and we adopt this last run as our best-matching model to the observed motions, from changing the kinematic parameters for the disk component alone (see also Table 10). However, from this discussion it is clear that we do not seem to be able to fit simultaneously, and without a small magnitude-dependent bias ($\sim 0.5$ mas yr$^{-1}$), the proper-motion dispersion and the median motion in the V-component with the current assumptions in the model.

A possible origin for the discrepancy between the observed and predicted dispersions in the V-component could be due to an overestimation of our proper-motion measuring errors. Indeed, as explained before, all of our model predictions are convolved with the uncertainties in the catalogue. This convolution, of course, increases the expected width of the proper-motion distributions, in accordance with the scatter introduced by the measuring errors. We have tested whether this could actually have an impact on our derived (model) dispersions. For this, we have decreased all of our measuring errors as given in Tables 5 and 6 by an amount equal to 0.5 mas yr$^{-1}$. We then convolved our model predictions with these new errors, and computed the median and dispersion in exactly the same manner as done before. The result of this test is that our derived values were minimally affected by the decrease in the measuring errors. In particular, the proper-motion dispersions changed (decreased) depending on the magnitude interval, but in the magnitude range $8 \le B_{\rm J} < 14$ the change was smaller than 0.05 mas yr$^{-1}$, and in the range $14 \le B_{\rm J} < 17$ smaller than 0.1 mas yr$^{-1}$. So, while some of the discrepancies could be due to this effect, it is quite possible that a fraction of the effect seen is due to inadequacies in the model. An interesting outcome of this experiment was that the computed median motion is largely insensitive to the error convolution, with variations of less than 0.03 mas yr$^{-1}$ for a decrease of 0.5 mas yr$^{-1}$ in the uncertainties of the observed motions. Therefore, given the already good fit to the observed median proper-motions in U and V as a function of $B_{\rm J}$, it seems quite likely that, whatever the inadequacy in the model is, it is mostly affecting the predicted proper-motion dispersions, and not the predicted mean motions.

### 5.2. The kinematics of the thick-disk and halo

We have run several models under different assumptions regarding the thick-disk and halo velocity dispersions and velocity lags. The results are shown in Figs. 16 and 17, where we have also included Run 12 (the best-fit model from previous Section). In general, we do not have a sensitivity to the assumed velocity dispersions in the model for neither the thick-disk nor the halo, although we clearly see changes in the model predictions for different assumptions involving the net motion of both components. Changes in the (constant) mean motion of the thick-disk and halo are magnitude-dependent because of the changing mixture of these populations as a function of apparent brightness. Figure 16 (upper panel) shows that changes in the net rotation of the thick-disk act mostly as a zero-point shift in the proper-motions at faint magnitudes, without altering the shape of the predicted motions. The best fit on Fig. 16 (dashed line) is provided by a model with a thick-disk velocity of $+160$ km s$^{-1}$ (Run 14), while we can clearly see that, for an assumed LSR speed of 270 km s$^{-1}$, a run with a velocity lag of -40 km s$^{-1}$ (dotted line) does not provide a good fit to the the median motion in V. A net rotation for the thick-disk of $+160$ km s$^{-1}$ would imply then, for our preferred LSR speed, a large velocity lag of -110 km s$^{-1}$ for this component, this would be in agreement with earlier results by Wyse and Gilmore (1986) from an analysis of Chiu's absolute proper-motions. One way of decreasing this velocity lag is to compensate it by applying a lag to the thick-disk. Indeed, the thick-disk has been ascribed to have a velocity gradient away from the Galactic plane (see Section 3.5.2). We have tested this possibility by including a rather steep gradient of 36 km s$^{-1}$ kpc$^{-1}$ in the mean motion for this component as per Majewski (1993, his Figure 6). In this case, the mean motion for the thick-disk ($\bar{V}(R, Z)_{\rm thick-disk}$) is computed from:

$$\bar{V}(R,Z)_{\rm thick-disk} = V_{\rm LSR}(R) - 10 - 36 \times \frac{|Z|}{1000} \quad (23)$$

where $|Z|$ is in pc. The results from this model (Run 15, see also Table 10) are shown in Fig, 16 (lower panel, dashed line). This model does indeed produces a better fit to the median motion in the V component than do Runs 12 and 13, by predicting a flatter secular proper-motion at bright magnitudes, while increasing the lag at the faintest bins. Because Eq. 23 gives the lag with respect to the currently adopted value for $V_{\rm LSR}$, and Majewski had adopted $220\,{\rm km\,s^{-1}}$, we have also tried a model where we have normalized the mean motion to a similar value by adjusting the zero point above from $10\,{\rm km\,s^{-1}}$ to $60\,{\rm km\,s^{-1}}$. This run (shown also on Fig. 16, dot-dashed line), produces a lag that is too large, incompatible with the data at the intermediate and fainter bins. Similarly, a model where we keep Eq. 23 but adopt $V_{\rm LSR}=220\,{\rm km\,s^{-1}}$, produces the opposite effect (dotted line); In this case we get a bad fit to the secular proper-motions at bright magnitudes. We thus conclude that the SPM data is better fit by a model (Run 15, Table 10) with a thick-disk having a velocity lag similar to that proposed by Majewski (1993).

As for the kinematics of the halo, our photometric errors prevent us from using colors to cleanly separate halo stars from disk stars, and therefore our sensitivity to model changes on parameters describing the dispersions and velocity lag for the halo is unfortunately small. This is demonstrated by Fig. 17, where we show the model predictions for a halo rotating at $+40\,{\rm km\,s^{-1}}$, and a counter-rotating halo at $-40\,{\rm km\,s^{-1}}$. It is clear that we can not effectively distinguish between these three models, and that fainter proper-motions, or better colors to minimize the disk contamination, would help to discriminated between these different model runs.

### 5.3. The final proper-motion histograms

In this section we compare the detailed observed proper-motion histograms with those derived from Run 15. This comparison is mostly intended to show the extent of the discrepancies between the observed and predicted motions, and to advance some possible explanations for these discrepancies. Figures 18, 19, and 20 show the predicted *vs.* observed motions in one magnitude intervals in $B_{\rm J}$. Both observations and model predictions have been convolved with the observational errors, and the resultant numbers binned in $2\,{\rm mas\,y^{-1}}$ bins. Furthermore, model predictions have been scaled to the total one-magnitude bin counts in the range $-80\,{\rm mas\,y^{-1}} < \mu \leq +80\,{\rm mas\,y^{-1}}$, independently in the U and V components. In general, we see a very good agreement between the model & predicted motions along the U and V components. In the range $10 < B_{\rm J} < 13$, we seem to have more stars at small proper-motions than those predicted by the model. The effect of this is that the observed median motion is shifted to smaller (absolute) values in the sense of decreasing the observed lag. This effect can be clearly seen in Fig. 14, where the model median proper-motion in V in the range $11 < B_{\rm J} < 13$ is underestimated. In this magnitude range we can also see some structure in the U-component of the proper-motion (left panels on Fig. 18). Both of these effects could be due to the presence of moving groups. The extent and characterization of these sub-structures would require a clustering analysis which is beyond the scope of this paper. We note however that, whatever the cause of the discrepancies between the observed and predicted model median proper motion, it goes away at fainter magnitudes. In the range $13 < B_{\rm J} < 17$ the fit to the observed histograms is remarkably good (see Fig. 19). In the last two magnitude bins however the model seems to grossly underestimate the observed dispersions (Fig. 20). The cause for this could be an underestimate in the observational proper-motion errors, or a much larger velocity dispersion for early G to late K dwarfs in the disk than that adopted in the model ($\Sigma_U, \Sigma_V, \Sigma_W = (30, 20, 15)\,{\rm km\,s^{-1}}$). We notice that changing the velocity dispersions for either the thick-disk or halo components by 10% did not change the predicted dispersions. Indeed, the proper-motion dispersion at these magnitudes is still dominated by disk stars with $+6 < M_V < +8$ (which also dominate the magnitude counts, see Section 4.1). On the other hand, we would need to increase the observational errors by 10% in the range $17 < B_{\rm J} < 18$, and by 70% in the range $18 < B_{\rm J} < 19$, if the model is required to match the observed dispersions with the currently adopted velocity dispersions (see Table 12 and Figure 19). While an increase of 10% in our estimated uncertainties can not be ruled out, a 70% increase is highly questionable - and in this case an increased velocity dispersion for disk stars might have to be advocated. In any event, a detailed analysis of the SPM fainter data is beyond the scope of this pa-per, and we will also defer this discussion for a future paper.

## 6. Discussion and limitations of our modelling

Figure 11 (lower panel) clearly demonstrates that there are two regimes for discussion. One is for magnitudes brighter than about $B_J \leq 14$, where the V proper-motion is very far from the expectation, the other is the region $15 \leq B_J \leq 18$, where the V motion is approximately independent of magnitude. We have clearly demonstrated that trying to explain both these phenomena with the change of a single parameter is impossible within the context of our kinematic model. Thus, one is forced to adopt both a small solar peculiar velocity and an extremely high disk rotation, to fit the bright stars, and a very steep velocity shear, to fit the very faint stars on our sample. While it is possible this is the correct situation, other studies are not in complete agreement and therefore it is important to emphasise that these two results do not necessarily have a single solution.

Bright stars *do seem* to pose a particular problem in that, e.g., the scaled models of Fig. 6 suggest some anomaly in the bright stars, whereas the fainter counts fit very well (Fig. (7)). The same effect is seen in the proper motion distribution functions (Fig. (18) and (18)), and of course in the median V motion. Perhaps these consistent anomalies *are* correlated, and suggest a real astrophysical effect, limited to the brighter disk stars? We are exploring whether the SPM data are good enough to have independently seen the strange phase-space structure evident in the Dehnen (1998) analysis of Hipparcos data. If that is the case, then fitting a single model to both these stars and the better-behaved fainter stars should be taken with caution. Fundamentally, the external constraints from other studies on both the gradient of the rotation curve (found to be zero from Cepheids by Metzger et al. 1998), and its amplitude (270 km s$^{-1}$ is hard to fit with other constraints, although see Méndez et al. 1999), must induce reservations about the fit to the brighter stars.

At fainter magnitudes, however, the parameters which are required to fit the data are quite reasonable. Fitting to the gradient, the data actually require a mean thick disk rotation of about 180 km s$^{-1}$ at $B_J = 16$, and 150 km s$^{-1}$ at $B_J = 18$ (see Eq. (23) and Table 9), which is is certainly plausible. In fact, such gradients *are* consistent with the (limited) data on vertical shear in observed edge-on spirals. We note that Wyse and Gilmore (1990) also see a rather broad distribution of asymmetric drift in the thick disk, from radial velocities. This (marginal) result seems to be holding up in a more recent radial velocity study (Wyse and Gilmore 1999).

## 7. Conclusions

The color-counts and secular proper-motions for a randomly selected sample of stars derived from the SPM catalogue has been fit to the predictions from a star-counts Galactic structure and kinematic model. In general, a good representation of the data is obtained. Tuning of the model parameters requires a scale-height for Giant stars of 170 pc in order to reproduce the observed color counts at bright magnitudes, while all other parameters are unchanged from the original model presented by Méndez & van Altena (1996). It is somewhat puzzling that Méndez & van Altena (1996) had found from star and color-counts towards two intermediate latitude fields that the scale-height for sub-Giant stars is closer to 250 pc. One would then expect that the scale-height for a slightly more evolved population would be correspondingly larger, and yet we find that for Giants toward the SGP, the SPM color-counts indicate a value for their scale-height of 170 pc. This could imply that either our parameterization for the density laws of these populations are not the most appropriate, or that their value is indeed a function of, e.g., Galactic latitude. Alternatively, these discrepancies could imply that the assumed density of stars in the model in the Giant and sub-Giant section of the Hess-diagram is incorrect. This issue clearly deserves further attention, and could certainly be tackled by model comparisons to unbiased counts derived from current near-infrared surveys (2Mass, DENIS) at low Galactic latitudes, where the bright magnitude counts are dominated by these types of stars. Small systematic shifts are observed in the SPM photographic colors as a function of apparent magnitude when compared with the model predictions, and the catalogue is corrected for this effect.

The absolute proper-motions in the U-component indicates a solar peculiar motion of $11.0 \pm 1.5$ km s$^{-1}$, with no need for a local expansion or contraction term. In the V-component, the absolute proper-motions can only be reproduced by the model if we adopt a solar peculiar motion of $+5$ km s$^{-1}$, a large LSR speed of 270 km s$^{-1}$, and a (disk) velocity ellipsoid that always points towards the Galactic center. The fainter secular motions show an indication that the thick-disk must exhibit a rather steep velocity gradient of about -36 km s$^{-1}$ kpc$^{-1}$. We are not able to set constraints on the overall rotation of the halo, nor on the thick-disk or halo velocity dispersions. At bright magnitudes, the model shows a slightly larger proper-motion dispersion than observed, while the opposite is true at fainter magnitudes. We show that, at bright magnitudes, these discrepancies in the dispersions are not due to wrong assumptions regarding the SPM proper-motion errors, but that at fainter magnitudes our errors could be underestimated by 10% in the range $17 < B_{\rm J} < 18$ and by 70% in the range $18 < B_{\rm J} < 19$, although this issue deserves further attention. Some substructure in the U & V proper-motions could be present in the brighter bins $10 < B_{\rm J} < 13$, and it might be indicative of (disk) moving groups. Their existence could be related to the already known moving-groups in the Galactic disk (Dehnen 1998), or could be due to Galactic bar stars presently passing by the solar neighborhood, as found by Raboud et al. (1998) from a sample of NLTT stars.

Our derived value for the LSR speed would imply a mass for the Galaxy within the Solar circle larger, by about a factor of 1.5, than previous values. The implications of this finding, as well as supporting new evidence (coming mainly from a new measurement of the proper motion of the LMC (Anguita et al. 1999) and from the binary motion of the M31 - Milky Way and Leo I - Milky Way pairs (Zaritsky 1999)), are further discussed in Méndez et al. (1999). We also note here the interesting prospects of space astrometry where, e.g., space interferometric space missions like SIM (http://sim.jpl.nasa.gov/), will be able to directly measure the disk rotational speed throughout the entire Galaxy.

The SPM is a joint project of the Universidad Nacional de San Juan, Argentina and the Yale Southern Observatory. Financial support for the SPM has been provided by the US NSF and the UNSJ through its Observatorio Astronómico "Félix Aguilar". We would like to also acknowledge the invaluable assistance of Lic. Carlos E. López, who participated in, and supervised, all of the SPM second-epoch observations. We are indebted to an anonymous referee who provided many useful comments on the limitations and potentialities of the Galactic model employed here. These, and other comments by the referee, have greatly helped to clarify many points of the original manuscript. R.A.M acknowledges continuous support from a Cátedra Presidencial en Ciencias to Dr. M.T. Ruiz, and to the many CTIO Blanco-4m telescope users who, unaware to them, provided access to the use of that telescope's main data-reduction CPU to run (properly niced) some of the models presented here. Finally, R.A.M. and W.F.vA. acknowledge an interesting discussion and numerical simulations provided by Dr. Stan Peale on the issue of microlensing event rates towards the Galactic center.

van Altena, W.F., Platais, I., Girard, T.M., & López, C.E. 1994, in Workshop Galactic and Solar System Optical Astrometry, eds. L.V. Morrison, & G.F. Gilmore (Cambridge University), 26

Wesselink, A.J. 1974, in IAU Symp. 61, New Problems in Astrometry, eds. W. Gliese, C.A. Murray, & R.H. Tucker (Dordrecht:Reidel), 201

Wielen, R., and Fuchs, B., 1983, in Kinematics, Dynamics, and Structure of the Milky Way, ed. W. L. H. Shuter, Dordrecht: Reidel, 81

Woolley, R., 1978, MNRAS, 184, 311

Wyse, R.F.G, and Gilmore, G., 1986, AJ, 91, 855

Wyse, R.F.G., and Gilmore, G., 1990, in Chemical and Dynamical Evolution of Galaxies, eds F. Ferrini, J. Franco and F. Matteucci (ETS Editrice: Pisa), 19

Wyse, R.F.G., and Gilmore, G., 1999, in preparation

Yentis, D.J., Cruddace, R.G., Gursky, H., Stuart, B.V., Wallin, J.F., MacGillivray, H.T., & Collins, C.A., 1992, in Workshop Digitised Optical Sky Surveys, ed. H.T. MacGillivray & Thomson, E.B. (Dordrecht:Kluwer), 67

Zaggia, S., Hook, I., Méndez, R., da Costa, L., Olsen, L.F., Nonino, M., Wicenec, A., Benoist, C., Bertin, E., Deul, E., Erben, T., Guarnieri, M.D., Hook, R., Prandoni, I., Scodeggio, M., Slijkhuis, R., Wichman, R., 1998, A&A, submitted (astro-ph/9807152)

Zaritsky, D. 1999, in The Third Stromlo Symposium: The Galactic halo, ASP Conf. Ser., Vol. 165, eds. B. K. Gibson, T. S. Axelrod and M.E. Putnam (Astron. Soc. of the Pacific, California), page 34

TABLE 1
UNCERTAINTY IN $B-V$ COLORS AS A FUNCTION OF $B_J$ MAGNITUDE.

| $B_J$ range mag | Mean $\sigma_{B-V}$ mag | Median $\sigma_{B-V}$ mag | St. Dev. of $\sigma_{B-V}$ mag | Number of stars used | Number of stars total |
|---|---|---|---|---|---|
| 9 - 10  | 0.056 | 0.054 | 0.017 | 190  | 197  |
| 10 - 11 | 0.040 | 0.036 | 0.013 | 486  | 528  |
| 11 - 12 | 0.043 | 0.042 | 0.012 | 972  | 1021 |
| 12 - 13 | 0.054 | 0.054 | 0.014 | 4319 | 4415 |
| 13 - 14 | 0.068 | 0.064 | 0.017 | 3426 | 3524 |
| 14 - 15 | 0.084 | 0.081 | 0.022 | 2897 | 3007 |
| 15 - 16 | 0.095 | 0.092 | 0.029 | 4132 | 4334 |
| 16 - 17 | 0.125 | 0.120 | 0.038 | 5951 | 6170 |
| 17 - 18 | 0.200 | 0.192 | 0.060 | 5878 | 6176 |
| 18 - 19 | 0.290 | 0.283 | 0.079 | 1589 | 1651 |

TABLE 2
MEAN AND MEDIAN PROPER-MOTION IN U AS A FUNCTION OF $B_J$ MAGNITUDE.

| $B_J$ range mag | Mean $\mu_U$ mas yr$^{-1}$ | Median $\mu_U$ mas yr$^{-1}$ | St. Dev. of the Mean mas yr$^{-1}$ | Number of stars used | Number of stars total |
|---|---|---|---|---|---|
| 9 - 10 | -12.44 | -12.50 | 2.68 | 193 | 197 |
| 10 - 11 | -8.71 | -7.95 | 1.36 | 508 | 528 |
| 11 - 12 | -6.54 | -6.90 | 0.77 | 971 | 1021 |
| 12 - 13 | -5.89 | -5.30 | 0.30 | 4216 | 4415 |
| 13 - 14 | -4.73 | -4.20 | 0.29 | 3397 | 3524 |
| 14 - 15 | -4.68 | -4.00 | 0.26 | 2890 | 3007 |
| 15 - 16 | -3.96 | -3.50 | 0.19 | 4152 | 4334 |
| 16 - 17 | -2.85 | -2.60 | 0.14 | 5843 | 6170 |
| 17 - 18 | -2.27 | -1.90 | 0.16 | 5909 | 6176 |
| 18 - 19 | -2.74 | -2.20 | 0.44 | 1581 | 1651 |

Table 3
Mean and Median proper-motion in V as a function of $B_J$ magnitude.

| $B_J$ range mag | Mean $\mu_V$ mas yr$^{-1}$ | Median $\mu_V$ mas yr$^{-1}$ | St. Dev. of the Mean mas yr$^{-1}$ | Number of stars used | Number of stars total |
|---|---|---|---|---|---|
| 9 - 10 | -15.12 | -12.80 | 1.35 | 177 | 197 |
| 10 - 11 | -14.25 | -12.10 | 0.82 | 492 | 528 |
| 11 - 12 | -11.16 | -9.65 | 0.48 | 944 | 1021 |
| 12 - 13 | -10.84 | -8.85 | 0.20 | 4162 | 4415 |
| 13 - 14 | -10.07 | -8.70 | 0.20 | 3344 | 3524 |
| 14 - 15 | -9.49 | -8.60 | 0.18 | 2819 | 3007 |
| 15 - 16 | -8.75 | -8.10 | 0.14 | 4099 | 4334 |
| 16 - 17 | -8.35 | -7.60 | 0.11 | 5819 | 6170 |
| 17 - 18 | -9.34 | -8.60 | 0.14 | 5911 | 6176 |
| 18 - 19 | -9.61 | -8.60 | 0.39 | 1574 | 1651 |

TABLE 4
PROPER-MOTION DISPERSION IN U AND V AS A FUNCTION OF $B_J$ MAGNITUDE.

| $B_J$ range mag | $\Sigma_{\mu_U}$ mas yr$^{-1}$ | St. Dev. in $\Sigma_{\mu_U}$ mas yr$^{-1}$ | $\Sigma_{\mu_V}$ mas yr$^{-1}$ | St. Dev. in $\Sigma_{\mu_V}$ mas yr$^{-1}$ |
|---|---|---|---|---|
| 9 - 10  | 37.18 | 12.24 | 17.98 | 4.01 |
| 10 - 11 | 30.63 | 6.17  | 18.27 | 2.61 |
| 11 - 12 | 23.95 | 2.98  | 14.65 | 1.38 |
| 12 - 13 | 19.33 | 1.02  | 12.92 | 0.53 |
| 13 - 14 | 17.13 | 0.95  | 11.39 | 0.50 |
| 14 - 15 | 13.92 | 0.75  | 9.61  | 0.42 |
| 15 - 16 | 11.91 | 0.49  | 8.68  | 0.30 |
| 16 - 17 | 10.35 | 0.33  | 8.27  | 0.23 |
| 17 - 18 | 12.12 | 0.43  | 10.41 | 0.33 |
| 18 - 19 | 17.68 | 1.46  | 15.48 | 1.17 |

TABLE 5
MEAN AND MEDIAN PROPER-MOTION ERRORS IN U AS A FUNCTION OF $B_J$ MAGNITUDE.

| $B_J$ range mag | Mean $\sigma_{\mu_U}$ mas yr$^{-1}$ | Median $\sigma_{\mu_U}$ mas yr$^{-1}$ | St. Dev. mas yr$^{-1}$ | Number of stars used | Number of stars total |
|---|---|---|---|---|---|
| 9 - 10  | 2.32 | 2.20 | 0.57 | 188  | 197  |
| 10 - 11 | 2.02 | 1.90 | 0.45 | 498  | 528  |
| 11 - 12 | 1.85 | 1.80 | 0.38 | 965  | 1021 |
| 12 - 13 | 2.10 | 2.00 | 0.48 | 4247 | 4415 |
| 13 - 14 | 2.62 | 2.50 | 0.65 | 3353 | 3524 |
| 14 - 15 | 3.50 | 3.10 | 1.27 | 2822 | 3007 |
| 15 - 16 | 2.58 | 2.50 | 0.50 | 3816 | 4334 |
| 16 - 17 | 3.40 | 3.20 | 0.94 | 5845 | 6170 |
| 17 - 18 | 5.88 | 5.70 | 1.82 | 5848 | 6176 |
| 18 - 19 | 7.98 | 7.90 | 1.68 | 1522 | 1651 |

TABLE 6
MEAN AND MEDIAN PROPER-MOTION ERRORS IN V AS A FUNCTION OF $B_J$ MAGNITUDE.

| $B_J$ range mag | Mean $\sigma_{\mu_V}$ mas yr$^{-1}$ | Median $\sigma_{\mu_V}$ mas yr$^{-1}$ | St. Dev. mas yr$^{-1}$ | Number of stars used | Number of stars total |
|---|---|---|---|---|---|
| 9 - 10 | 2.30 | 2.20 | 0.54 | 188 | 197 |
| 10 - 11 | 2.00 | 1.90 | 0.44 | 494 | 528 |
| 11 - 12 | 1.85 | 1.80 | 0.38 | 963 | 1021 |
| 12 - 13 | 2.10 | 2.00 | 0.47 | 4230 | 4415 |
| 13 - 14 | 2.61 | 2.50 | 0.64 | 3343 | 3524 |
| 14 - 15 | 3.52 | 3.10 | 1.31 | 2837 | 3007 |
| 15 - 16 | 2.59 | 2.50 | 0.50 | 3811 | 4334 |
| 16 - 17 | 3.41 | 3.20 | 0.94 | 5820 | 6170 |
| 17 - 18 | 5.91 | 5.70 | 1.82 | 5829 | 6176 |
| 18 - 19 | 7.95 | 7.90 | 1.60 | 1506 | 1651 |

Table 7

Velocity lag as a function of spectral type compiled by RBC89 *vs.* those computed from our model (Eq. (21)).

| Spectral Type | Luminosity Class | $\Sigma_{\rm U}$ km s$^{-1}$ | $\lambda$ | $V_{\rm lag}$ (RBC89) km s$^{-1}$ | $V_{\rm lag}$ (This paper) km s$^{-1}$ |
|:---:|:---:|:---:|:---:|:---:|:---:|
| sg | I-II | 12 | 1.5 | 0 | -1.3 |
| OB | V | 10 | 1.7 | 0 | -0.9 |
| A | V | 15 | 1.7 | 0 | -1.9 |
| F | V | 25 | 1.9 | 0 | -5.0 |
| G | IV | 25 | 2.1 | -1 | -5.0 |
| K | III | 31 | 1.8 | -5 | -8.0 |
| M | III | 31 | 1.9 | -6 | -8.0 |

Table 8
Velocity lag as a function of age predicted by RO87 vs. those computed from our model (Eq. (21)).

| Age range Gyr | $\Sigma_{\rm U}$ km s$^{-1}$ | $\lambda$ | $V_{\rm lag}$ (RO87) km s$^{-1}$ | $V_{\rm lag}$ (This paper) km s$^{-1}$ |
|---|---|---|---|---|
| 0.00 - 0.15 | 16.7 | 2.8 | -1.6 | -2.2 |
| 0.15 - 1.00 | 19.8 | 2.0 | -3.6 | -3.2 |
| 1 - 2 | 27.2 | 2.1 | -6.7 | -6.1 |
| 2 - 3 | 30.2 | 1.6 | -8.5 | -7.9 |
| 3 - 5 | 36.7 | 1.6 | -12.6 | -11.8 |
| 4 - 10 | 43.1 | 1.7 | -17.2 | -16.2 |

TABLE 9

Predicted typical distances from the plane for different Galactic components as a function of $B_{\rm J}$ magnitude.

| $B_{\rm J}$ range | Disk M-S stars pc | Disk Giant stars pc | Thick-disk stars pc | Halo stars |
|---|---|---|---|---|
| 9 - 10 | 160 | 380 | ⋯ | ⋯ |
| 10 - 11 | 190 | 500 | ⋯ | ⋯ |
| 11 - 12 | 250 | 620 | ⋯ | ⋯ |
| 12 - 13 | 330 | 790 | ⋯ | ⋯ |
| 13 - 14 | 440 | 1070 | ⋯ | ⋯ |
| 14 - 15 | 570 | ⋯ | 2070 | ⋯ |
| 15 - 16 | 710 | ⋯ | 2060 | ⋯ |
| 16 - 17 | 810 | ⋯ | 2310 | ⋯ |
| 17 - 18 | 890 | ⋯ | 2920 | 5290 |
| 18 - 19 | 840 | ⋯ | 3600 | 5600 |

TABLE 10

OBSERVED AND PREDICTED MEDIAN PROPER-MOTION AND PROPER-MOTION DISPERSION FOR THE MOST REPRESENTATIVE KINEMATIC MODEL RUNS.

| $B_J$ | Observed | Run 1 | Run 7 | Run 12 | Run 15 |
|---|---|---|---|---|---|
| 8-9 | ⋯ | -11.85 | -11.95 | -11.95 | -11.95 |
|  | ⋯ | 38.94 | 38.95 | 38.95 | 38.95 |
|  | ⋯ | -21.23 | -12.96 | -12.40 | -12.12 |
|  | ⋯ | 22.29 | 23.08 | 22.08 | 21.99 |
| 9-10 | -12.50 ± 2.68 | -10.16 | -10.26 | -10.25 | -10.26 |
|  | 37.18 ± 12.24 | 36.60 | 36.61 | 36.61 | 36.61 |
|  | -12.80 ± 1.35 | -19.75 | -12.73 | -12.26 | -11.72 |
|  | 17.98 ± 4.01 | 21.74 | 22.20 | 21.28 | 21.13 |
| 10-11 | -7.95 ± 1.36 | -8.75 | -8.83 | -8.82 | 8.83 |
|  | 30.63 ± 6.17 | 31.26 | 31.27 | 31.27 | 31.27 |
|  | -12.10 ± 0.82 | -18.57 | -12.28 | -11.99 | -11.26 |
|  | 18.27 ± 2.61 | 20.51 | 20.79 | 20.19 | 20.06 |
| 11-12 | -6.90 ± 0.77 | -7.21 | -7.28 | -7.28 | -7.28 |
|  | 23.95 ± 2.98 | 25.58 | 25.60 | 25.60 | 25.59 |
|  | -9.65 ± 0.48 | -15.72 | -10.85 | -10.73 | -10.14 |
|  | 14.65 ± 1.38 | 16.56 | 16.93 | 16.85 | 16.79 |
| 12-13 | -5.30 ± 0.30 | -5.62 | -5.68 | -5.68 | -5.69 |
|  | 19.33 ± 1.02 | 21.35 | 21.36 | 21.36 | 21.36 |
|  | -8.85 ± 0.20 | -13.62 | -9.68 | -9.48 | -9.20 |
|  | 12.92 ± 0.53 | 14.76 | 14.96 | 14.66 | 14.59 |
| 13-14 | -4.20 ± 0.29 | -4.40 | -4.44 | -4.41 | -4.40 |
|  | 17.13 ± 0.95 | 18.40 | 18.42 | 18.53 | 18.53 |
|  | -8.70 ± 0.20 | -11.67 | -8.70 | -8.51 | -8.42 |
|  | 11.39 ± 0.50 | 12.83 | 12.99 | 12.72 | 12.57 |
| 14-15 | -4.00 ± 0.26 | -3.48 | -3.52 | -3.53 | -3.52 |
|  | 13.92 ± 0.75 | 15.42 | 15.44 | 15.43 | 15.43 |
|  | -8.60 ± 0.18 | -10.09 | -8.14 | -8.08 | -7.90 |
|  | 9.61 ± 0.42 | 11.07 | 11.23 | 11.10 | 10.86 |
| 15-16 | -3.50 ± 0.19 | -2.77 | -2.81 | -2.82 | -2.81 |
|  | 11.91 ± 0.49 | 12.98 | 13.00 | 13.00 | 13.00 |
|  | -8.10 ± 0.14 | -8.63 | -7.74 | -7.69 | -7.43 |
|  | 8.68 ± 0.30 | 9.53 | 9.68 | 9.58 | 9.36 |
| 16-17 | -2.60 ± 0.14 | -2.20 | -2.23 | -2.23 | -2.23 |
|  | 10.35 ± 0.33 | 11.17 | 11.18 | 11.19 | 11.19 |
|  | -7.60 ± 0.11 | -7.59 | -7.59 | -7.43 | -7.36 |
|  | 8.27 ± 0.23 | 9.11 | 9.11 | 8.89 | 8.73 |
| 17-18 | -1.90 ± 0.16 | -1.91 | -1.94 | -1.92 | -1.91 |
|  | 12.12 ± 0.43 | 12.27 | 12.27 | 11.97 | 11.97 |
|  | -8.60 ± 0.14 | -7.47 | -7.75 | -7.66 | -7.98 |
|  | 10.41 ± 0.33 | 10.95 | 10.89 | 10.49 | 10.28 |
| 18-19 | -2.20 ± 0.44 | -1.86 | -1.88 | -1.89 | -1.86 |
|  | 17.68 ± 1.46 | 14.50 | 14.50 | 14.51 | 14.52 |
|  | -8.60 ± 0.39 | -7.85 | -8.02 | -7.99 | -8.53 |
|  | 15.48 ± 1.17 | 13.15 | 13.01 | 12.92 | 12.63 |
| 19-20 | ⋯ | -1.57 | -1.58 | -1.72 | -1.69 |
|  | ⋯ | 13.31 | 13.30 | 14.23 | 14.23 |
|  | ⋯ | -7.96 | -8.03 | -7.90 | -8.40 |
|  | ⋯ | 13.90 | 13.85 | 12.67 | 12.36 |

NOTE.—The first two lines for each magnitude entry indicate the values for the median and dispersion on the U-component of the proper-motion, while the last two rows indicate the same parameters for the V-component. Only the model predictions for the most representative runs are indicated as, otherwise, the table will be too cluttered without adding much information to the reader

TABLE 11
BASIC FEATURES OF ALL MODEL SIMULATIONS.

| Run number | Model parameters | General comments |
|---|---|---|
| 1 | Standard but $H_Z(G) \sim 170$ pc | Poor fit to V-component |
| 2 | Standard model[a] | Poor fit to both U- and V-components |
| 3 | As Run 1 but $q = 1$ | Decreases predicted V-motion, poorer fit |
| 4 | As Run 1 but $V_\odot = +5$ km s$^{-1}$[b] | Better fit to brighter motions in V |
| 5 | As Run 1 but $V_{\rm LSR}(R_\odot) = +270$ km s$^{-1}$[c] | Better fit to fainter motions in V |
| 6 | As Run 1 but $V_\odot = +5$ km s$^{-1}$ and $V_{\rm LSR}(R_\odot) = +270$ km s$^{-1}$ | Better fit to overall motions in V |
| 7 | As Run 6 but $q = 1$ | Improves fit to fainter V-motion |
| 8-9 | As Run 5, but changing $U_\odot$ by $\pm 3$ km s$^{-1}$[d] | No expansion/contraction of disk |
| 10 | No-$|Z|$ gradient in (disk) velocity dispersion in V | Inconsistent with U and W observed gradients |
| 11 | Large LSR slope ($dV_{\rm LSR}(R)/dR = -11.7$ km s$^{-1}$) | Bad fit to $\Sigma_{\mu_V}$ |
| 12 | As Run 7,[e] but $dV_{\rm LSR}(R)/dR = -2.4$ km s$^{-1}$[f] | Best-fit model from disk |
| 13 | 10% decrease in model $\Sigma_U$ | Bad fit to $\Sigma_{\mu_V}$ and predicted lag too small |
| 14 | Thick-disk rotational velocity of $+160$ km s$^{-1}$[g] | Improves fit to fainter motions in V |
| 15 | Thick-disk velocity gradient of -36 km s$^{-1}$ kpc$^{-1}$[h] | Best-fit to fainter motions in V |

[a] As in Méndez & van Altena (1996), with $H_Z(G) \sim 250$ pc.

[b] Standard model adopted $V_\odot = +12$ km s$^{-1}$.

[c] Standard model adopted $V_{\rm LSR}(R_\odot) = 220$ km s$^{-1}$.

[d] Standard value is $U_\odot = +11$ km s$^{-1}$.

[e] Run 7 had a zero slope for $V_{\rm LSR}(R_\odot)$.

[f] This slope is the value derived from the Hipparcos and ground-based values for Oort's constants A and B.

[g] For an assumed disk speed of $+270$ km s$^{-1}$ this implies a velocity lag of -110 km s$^{-1}$ for this galactic component.

[h] See Eq. 23, from Majewski 1993.

TABLE 12
PROPER-MOTION DISPERSION IN U AND V IN THE RANGE $17 < B_J \leq 19$ FOR DIFFERENT ASSUMPTIONS ABOUT THE PROPER-MOTION ERRORS.

| $B_J$ range mag | $\Sigma_{\mu_U}$ mas yr$^{-1}$ | $\Sigma_{\mu_V}$ mas yr$^{-1}$ | Comments |
|---|---|---|---|
| 17 - 18 | $12.12 \pm 0.43$ | $10.41 \pm 0.33$ | Observed values |
|  | 11.97 | 10.28 | Errors as in Table 6 |
|  | 12.19 | 10.54 | Errors increased by 10% |
|  | 13.69 | 12.22 | Errors increased by 70% |
| 18 - 19 | $17.68 \pm 1.46$ | $15.48 \pm 1.17$ | Observed values |
|  | 14.52 | 12.63 | Errors as in Table 6 |
|  | 14.91 | 13.10 | Errors increased by 10% |
|  | 18.97 | 16.15 | Errors increased by 70% |

Fig. 1.— Mean rotational speed as a function of Galactocentric distance for the old disk population predicted from Equations 21 and 22 for different parameters of the velocity ellipsoid. The solid line is for $H_R = 3.5$ kpc and $\lambda = 1.5$, the dashed line is for $H_R = 3.5$ kpc and $\lambda = 2.0$, the dot-dashed line is for $H_R = 4.0$ kpc and $\lambda = 1.5$, and the dotted line is for $H_R = 4.0$ kpc and $\lambda = 2.0$. Only very accurate radial velocities and/or proper-motions for disk stars within a few kpc from the Sun would we able to differentiate between these different models.

Fig. 2.— Predicted starcounts as a function of $B_J$ magnitude for the SPM-SGP region. The expected contribution from the different stellar populations is indicated. The disk dominates the counts for $B_J < 14$, while the thick-disk becomes an important contributor at fainter magnitudes. By $B_J \sim 17$ there is an appreciable contribution from all three major Galactic components, the disk, the thick-disk, and the halo, albeit the disk dominates the total counts at all magnitudes.

Fig. 3.— Error-convolved predicted color counts in the magnitude range where the disk dominates the counts ($B_J < 14$). The solid lines indicate the overall counts, the dashed line the contribution from disk Giants, and the dot-dashed line that from main-sequence disk stars. The dotted line shows the small contribution from the thick-disk. Because of the selectively incomplete nature of the SPM catalogue as a function of magnitude, data points can not be placed on this figure, unless we consider one magnitude bins in $B_J$. This is also true for Figs. 4 and 5 (but see, e.g., Fig. 6).

Fig. 4.— Same as Figure 3, except for the magnitude range $14 \leq B_J < 17$. In this case the disk contributes only with main-sequence stars, while the thick-disk contributes with a mixture of Giants and main-sequence that mostly overlap in their color distribution for our photometric uncertainties. The solid line indicates the overall counts, while the dashed line shows the expected contribution from disk main-sequence stars. The dot-dashed line indicates the overall contribution from thick-disk stars, which is in turn divided into main-sequence (double-dotted-dashed line) and giants (dotted line). Finally, the very small contribution from halo (giant) stars is shown as the continuous line at the lowest expected counts.

Fig. 5.— Same as Figure 3, except for the magnitude range $17 \leq B_J < 19$. In this case we have contributions from all three Galactic components: The disk, thick-disk and halo. The upper panel shows the contribution from disk main-sequence stars as the dashed line, the overall thick-disk as the dot-dashed line, and the overall halo as the dotted line. The lower panel shows the contribution from thick-disk main-sequence stars as the dotted line, and the small contribution from thick-disk giants & subgiants as the dashed line. halo main-sequence stars are indicated by the double-dotted-dashed line, and the smaller contribution by halo giants as the dot-dashed line. For all populations, the major contributor to the counts in this magnitude range comes from main-sequence stars, and it is difficult to separate thick-disk and halo stars from colors alone, given our photometric uncertainties.

Fig. 6.— Error-convolved observed and predicted color counts for the bright portion of the SPM-SGP survey. The model predictions have been scaled to match the total observed counts at each magnitude bin. The solid line indicates the overall predictions from our starcounts model, while the dashed line indicates the contribution from disk giants. It is apparent that the model is predicting slightly more giants than observed.

Fig. 7.— Same as Figure 6, but for the fainter portion of the survey. The solid line indicates our model predictions for a scale-height of 325 pc for M-dwarfs, while the dashed line indicates the model predictions for a smaller scale-height of 250 pc. The fit of the model predictions to the observed counts does not suggest any modifications to the standard parameters in the model.

Fig. 8.— Color counts in the range $13 \leq B_J < 14$ compared to model predictions for extreme ranges of the scale-height of main-sequence disk stars. In this magnitude range, the contribution from disk giants and thick-disk stars is minimal. The solid line indicates our standard model (see text), the dashed line indicates a lower scale-height, and the dot-dashed line a higher scale-height. The dotted line shows the predictions from the standard model for an extreme reddening of $E(B-V) = 0.08$ in the SGP-SPM region. We cannot effectively distinguish between a model with a lower scale-height for main-sequence stars, an increase in reddening of +0.05 mag in E(B-V), or a -0.05 mag systematic error in the colors.

Fig. 9.— Color counts in the range $10 \leq B_J < 11$ compared to model predictions for extreme ranges of the scale-height of disk Giants ($H_Z(G)$). The solid line shows our standard model (with $H_Z(G) = 250$ pc), the dashed line a model with a lower scale-height of $H_Z(G) = 150$ pc, and the dot-dashed line a model with a large scale-height of $H_Z(G) = 350$ pc. The red-peak predicted counts are very sensitive to the adopted value for $H_Z(G)$ and, clearly, a model with a scale-height between 150 and 250 pc should be preferred.

Fig. 10.— Same as Figure 6, except that our best $\chi^2$ scale-height of 172 pc for disk giants has been used. The solid line is the run with the standard model, while the dashed line indicates the run with the modified scale-height. In the the magnitude ranges $11 \leq B_J < 12$, $12 \leq B_J < 13$, and $13 \leq B_J < 14$, systematic shifts of -0.05, -0.08 and -0.05 magnitudes in $B - V$, respectively, have been applied to the data.

Fig. 11.— Median observed (filled dots with error bars) and predicted (lines) absolute proper-motions along U (Galactocentric direction, upper panel), and V (Galactic rotation, lower panel). In the upper panel, the solid line indicates the run from the standard model (with a scale-height of 172 pc for Giants, Run 1 in Table 9), while the dashed line indicates a run with a scale-height of 250 pc for Giants (Run 2). In the lower panel, the lower solid line is from Run 1, while the dashed line is from Run 2. The dot-dashed line is for a model with $q = 1$ (Run 3, see text), while the dotted line is for a model with a disk having a rotational speed of $+270$ km s$^{-1}$ (Run 4). The triple-dotted dashed line indicates the predictions for a model with a Solar peculiar motion in the V-component of $+5$ km s$^{-1}$ (Run 5) instead of the classical value $+14$ km s$^{-1}$ adopted in the standard model (Run 1), while the upper solid line shows the run with, both, $V_\odot = +5$ km s$^{-1}$, and $V_{LSR}(R_\odot) = +270$ km s$^{-1}$ (Run 6).

Fig. 12.— Same as Figure 11, except that for the proper-motion dispersions. In the upper panel (U-component of the dispersion), the solid line is for Run 1 (Table 9), and the dashed line is for Run 2. In the lower panel (V-component of the dispersion), the meaning is the same as for the upper panel, except that the dot-dashed line is for Run 6 with, both, $V_\odot = +5$ km s$^{-1}$, and $V_{LSR}(R_\odot) = +270$ km s$^{-1}$.

Fig. 13.— Median observed absolute proper-motions along U (Galactocentric direction, upper panel), and V (Galactic rotation, lower panel). The solid dots are for version 1.1 of the catalogue, the open triangles are for version 1.0, and the open squares for version 1.1 with a declination offset as found from a comparison to the Hipparcos motions at the bright end of our sample. An offset of $\pm 0.15$ mags has been applied to the points in order to avoid crowding. Although the changes are noticeable, the possible systematics effects still present in the SPM-SGP motions are negligible in comparison with the discrepancies to the model seen in Figures 11 and 12. The proper-motion dispersions are the same for all three catalogues, and are therefore not shown.

Fig. 14.— Symbols as in Figure 11. In the upper panel the solid line is for Run 1, the dashed line is for a model with a Solar peculiar motion in the U-component of $+8$ km s$^{-1}$ (Run 8), while the dot-dashed line is for a model with $U_\odot = +14$ km s$^{-1}$ (Run 9). Run 1, with $U_\odot = +11$ km s$^{-1}$ provides the best fit to the U-component of the observed motion. The agreement to the model also indicates the absence of any significant expansion/contraction in the local disk (see text). In the lower panel, the solid line is for Run 6 (see Figure 11), while the dashed line is for a model with $q = 1$ (Run 7, Table 9), indicating that a better fit to the observed median motion in the V-component is achieved by a velocity ellipsoid whose major axis points to the Galactic center as we move away from the Galactic plane.

Fig. 15.— Proper-motion dispersion (upper panel), and median proper-motion (lower panel) in the V-component as a function of apparent $B_J$ magnitude. Symbols are as in Figures 12 (upper panel) and 11 (lower panel). For both panels, solid lines are for Run 7, dashed lines are for a model where no-gradient in the velocity dispersion away from the Galactic plane is considered (Run 10), the dot-dashed line is for a model where a large gradient in the (local) rotation curve is considered ($dV_{LSR}(R)/dR = -11.7$ km s$^{-1}$ kpc$^{-1}$, Run 11), the dotted line is for a model where a rotation curve with a gradient as predicted by the mean values for the Oort constants A and B is considered ($dV_{LSR}(R)/dR = -2.4$ km s$^{-1}$ kpc$^{-1}$, Run 12) and, finally, the triple-dot dashed line shows the predictions for a model (Run 13) where the same rotation curve as for Run 12 is considered, but the velocity dispersion in U has been decreased by 10% (see text).

Fig. 16.— Median proper-motion in the V-component as a function of apparent $B_J$ magnitude. In both panels, the solid line refers to the standard Run 12, with a thick-disk rotation velocity of $+180$ km s$^{-1}$. In the upper panel, the dashed line is for a model (Run 14) with a thick-disk rotation velocity of $+160$ km s$^{-1}$, the dot-dashed line for a model with a rotation velocity of $+200$ km s$^{-1}$, and the dotted line is for a model with a velocity lag of 40 km s$^{-1}$ with respect of the current LSR speed of 220 km s$^{-1}$. It is apparent that the best model is the one with the smaller rotational velocity for the thick disk. In the lower panel, the dashed line is for a model with a gradient on the thick-disk velocity lag of 36 km s$^{-1}$ kpc$^{-1}$ (Run 15), the dot-dashed line for a model with a similar gradient, but with a smaller net rotation by 50 km s$^{-1}$ (see text), and the dotted line is for a model with the same velocity lag gradient and a LSR speed of 220 km s$^{-1}$. The lower panel seems to indicate that, alternatively, the best matching model is Run 15 with a steep gradient for the motion of the thick-disk away from the Galactic plane.

Fig. 17.— Median proper-motion in the V-component as a function of apparent $B_J$ magnitude. The solid line is for Run 15 (see Fig. 16), the dashed line is for a halo moving at $+40$ km s$^{-1}$, and the dot-dashed line is for a counter-rotating halo at -40 km s$^{-1}$. The SPM data can not unambiguously distinguish between these different models.

Fig. 18.— Proper-motion histograms in U (left panel) and V (right panel) for the magnitude range 9 top panel $< B_J \le$ 13 bottom panel in one magnitude-intervals. Solid dots (with Poisson error bars) represent the observed convolved counts in 2 mas y$^{-1}$ bins (4 mas y$^{-1}$ bins for the brighter bin where fewer stars were measured). The solid lines indicate the (convolved) model predictions using Run 15 and the median error values shown in Table 6. The apparent discrepancies in the magnitude range $11 < B_J \le 13$ are discussed in the text, and could be due, e.g., to moving groups in the Galactic disk.

Fig. 19.— Same as Fig. 18, except that for the range 13 top panel $< B_J \le$ 17 bottom panel. Here we see an excellent agreement with the model predictions.

Fig. 20.— Same as Fig. 18, except that for the range 17 top panel $< B_J \le$ 19 bottom panel. Run 15 predicts too narrow distributions when we adopt the standard proper-motion error values given in Table 6. The effect of an increase of 70% in the adopted proper-motion errors is indicated by the dashed line. Indeed, only a 10 % increase is needed to reproduce the observed dispersions in the range $17 < B_J \le 18$ (top panel), while an increase of 70% is needed to reproduce the observations on the faintest SPM bin (lower panel), see Table 12.

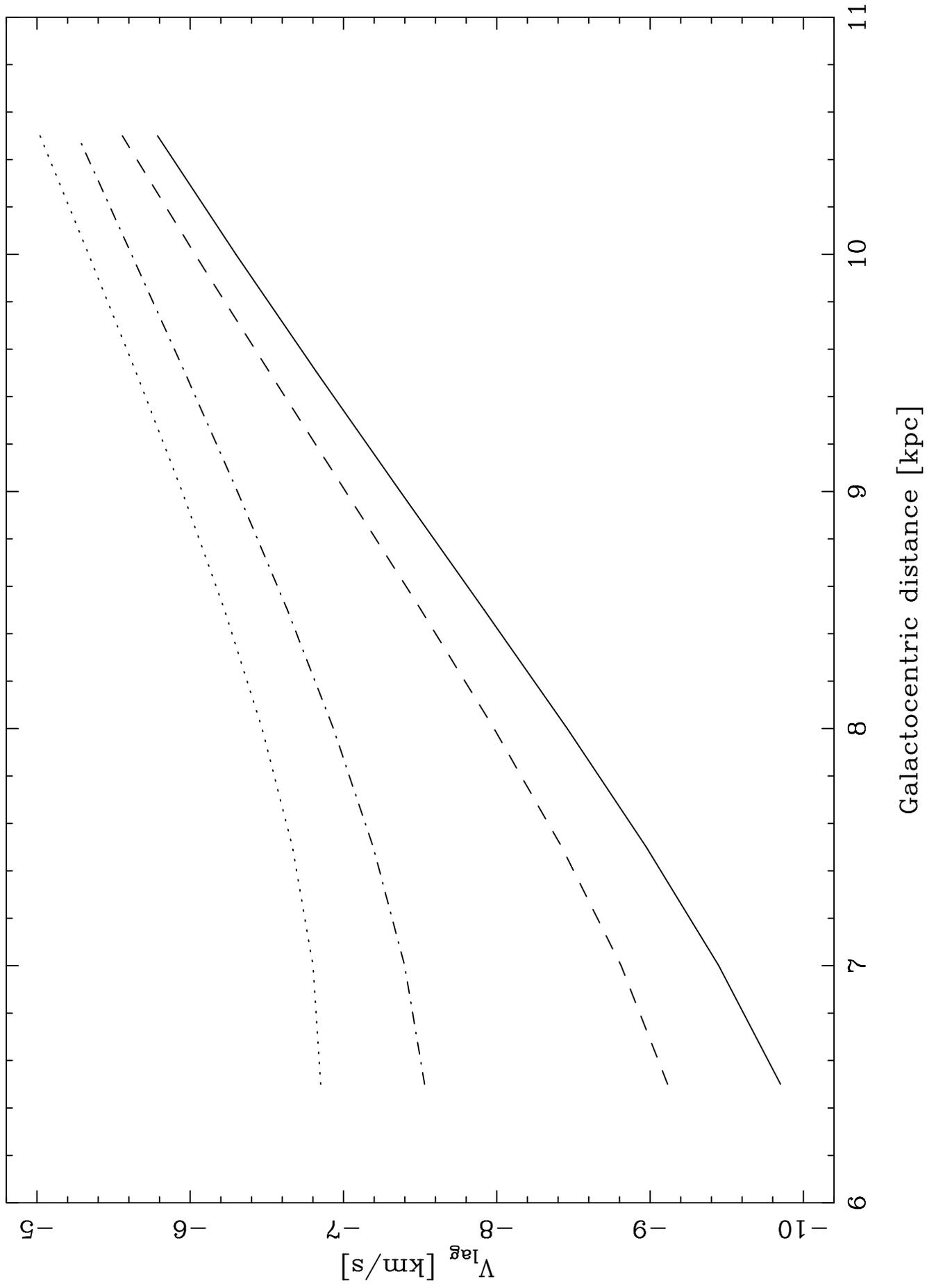

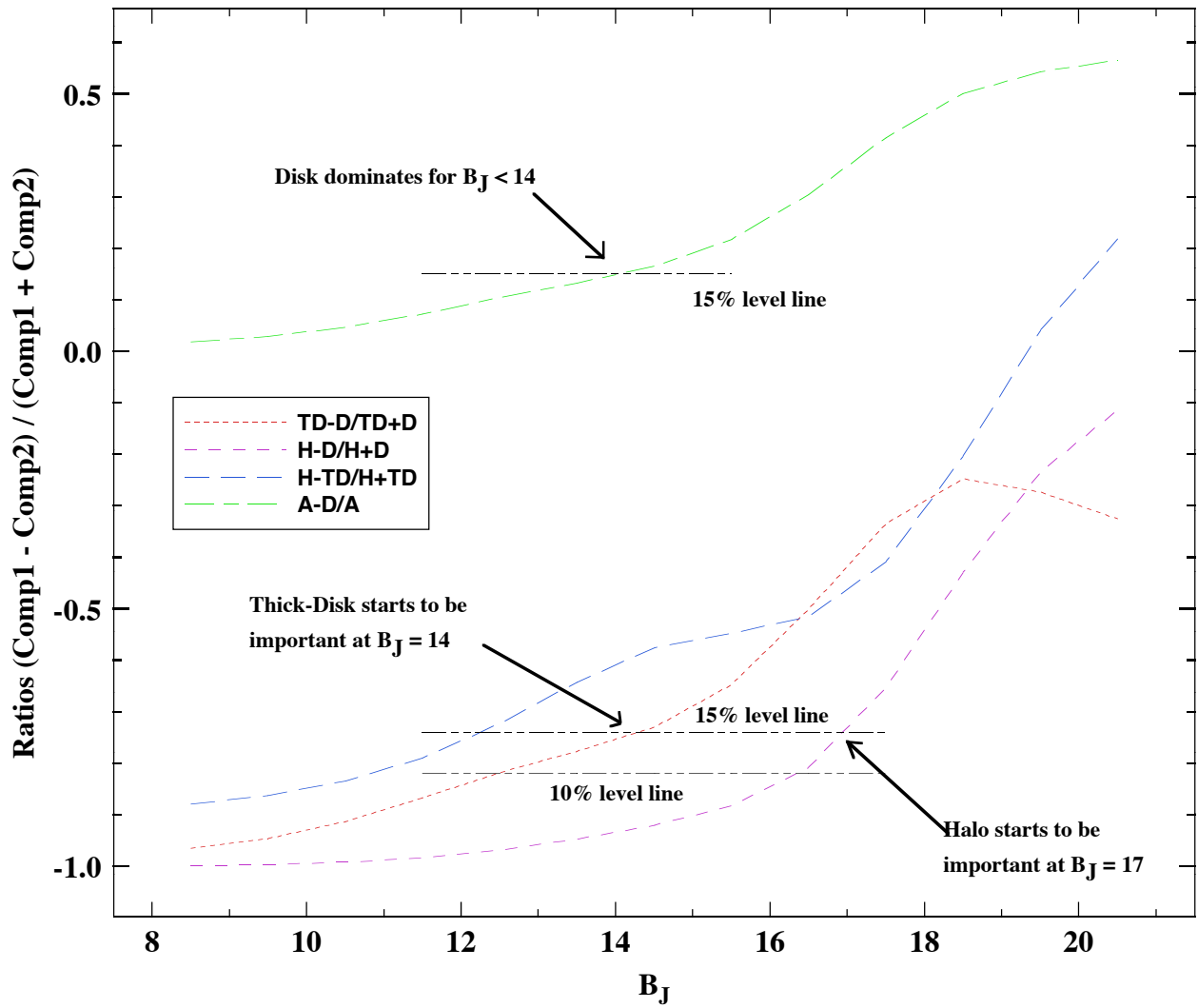

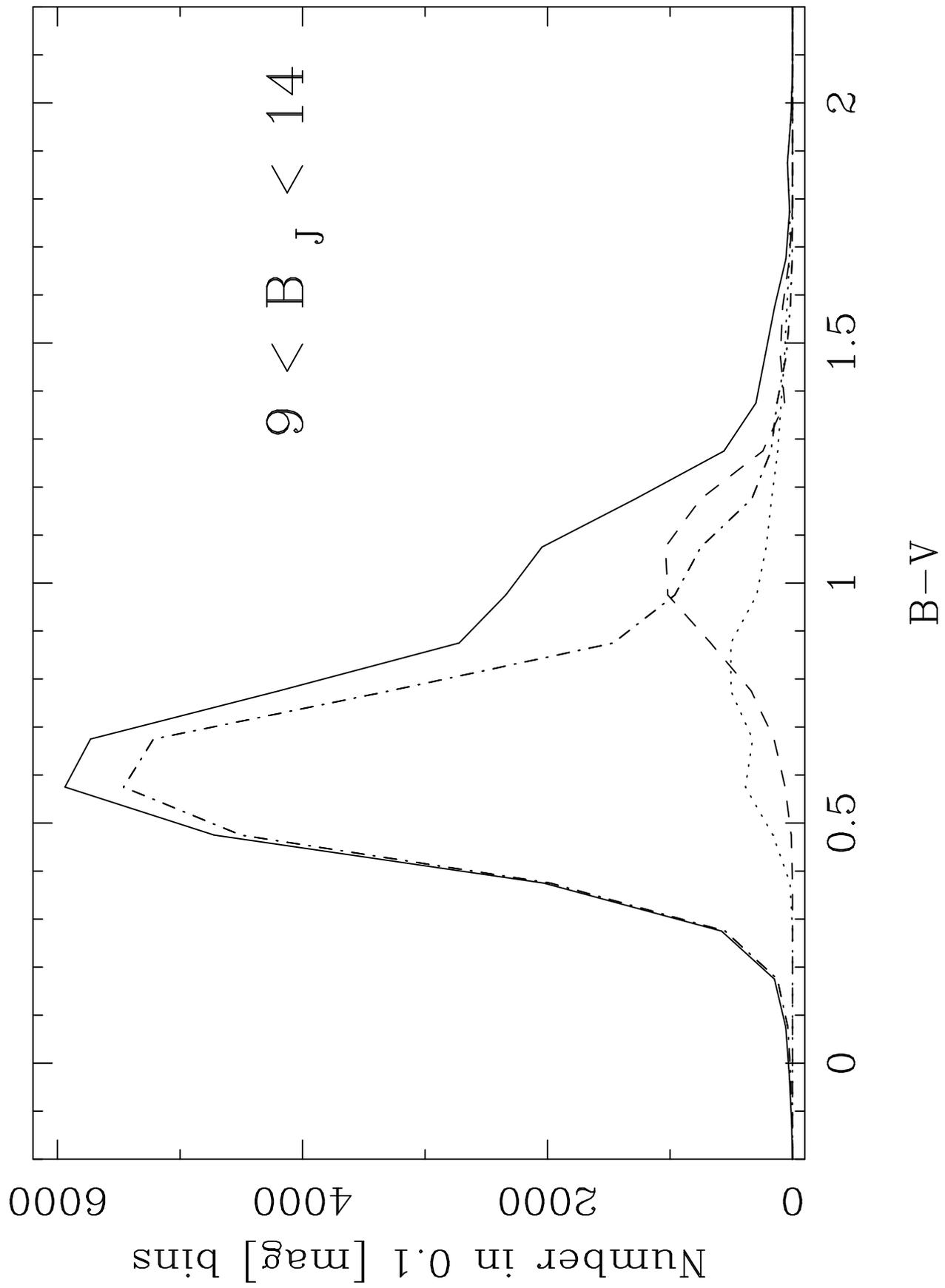

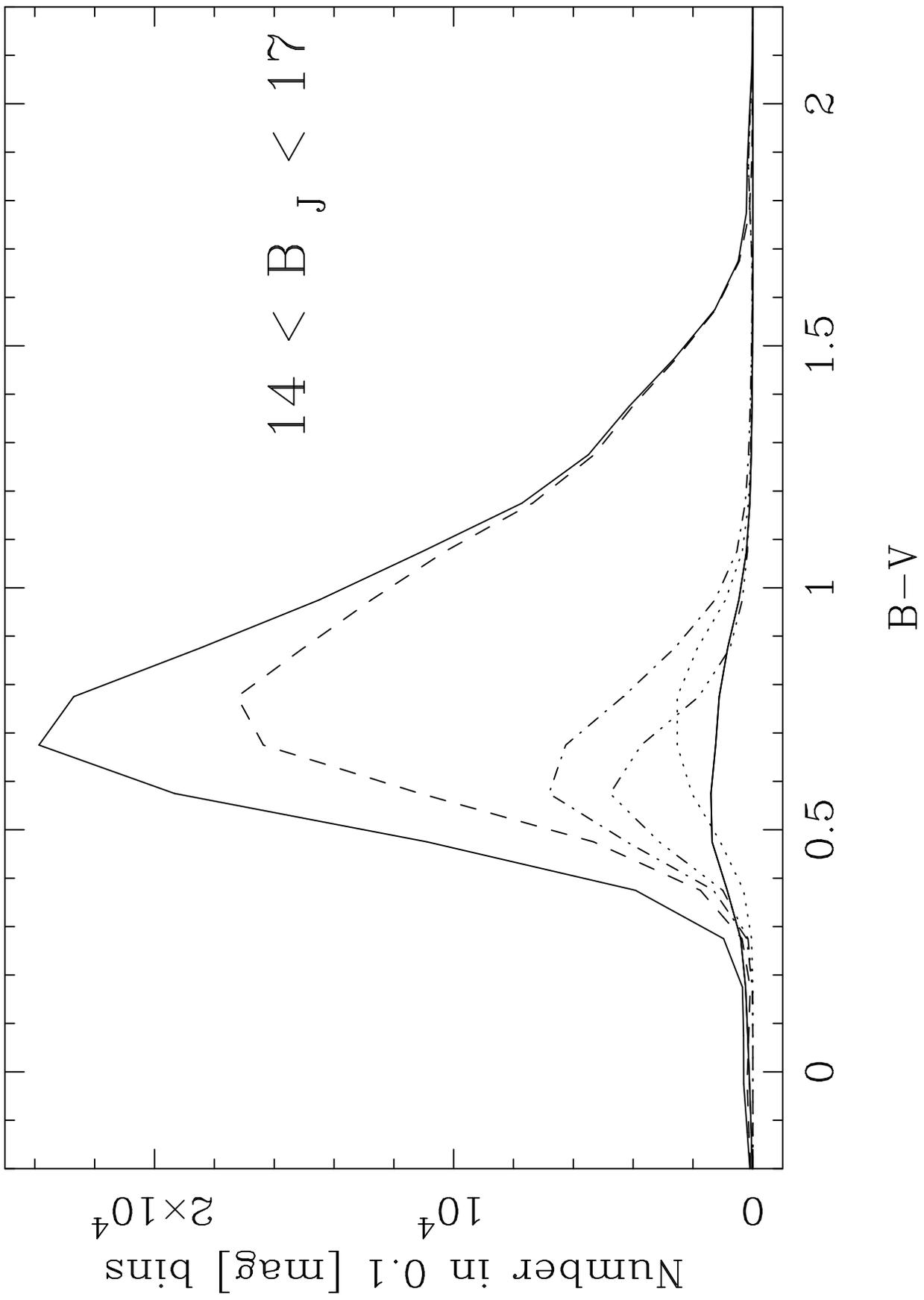

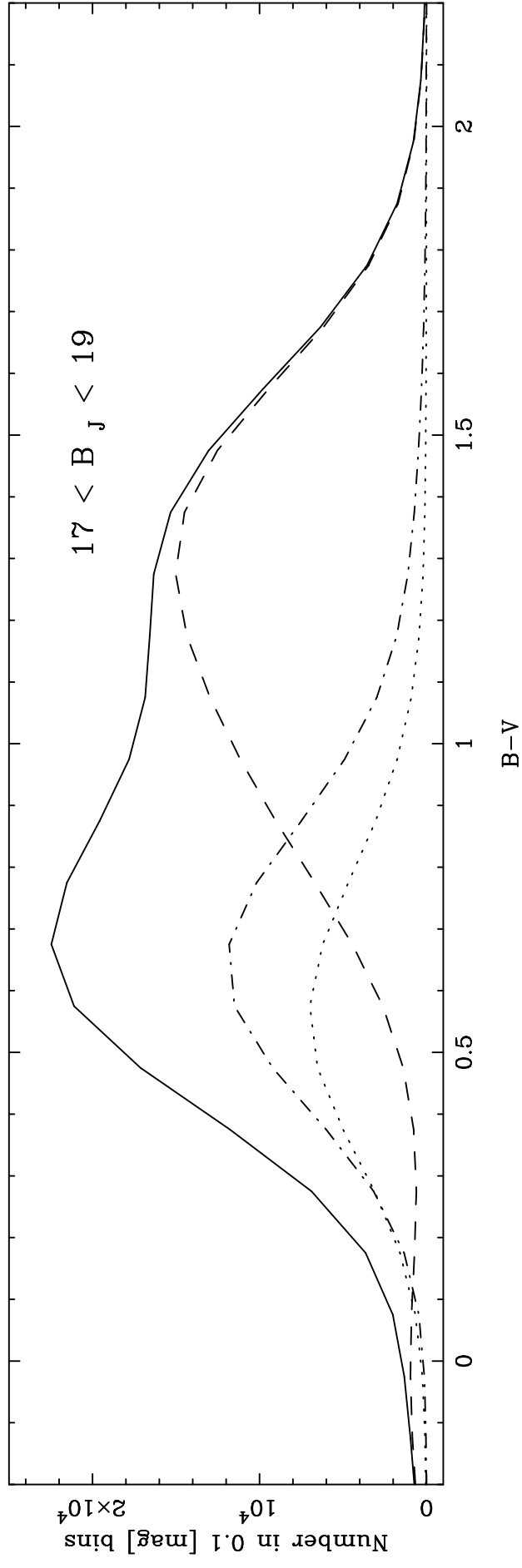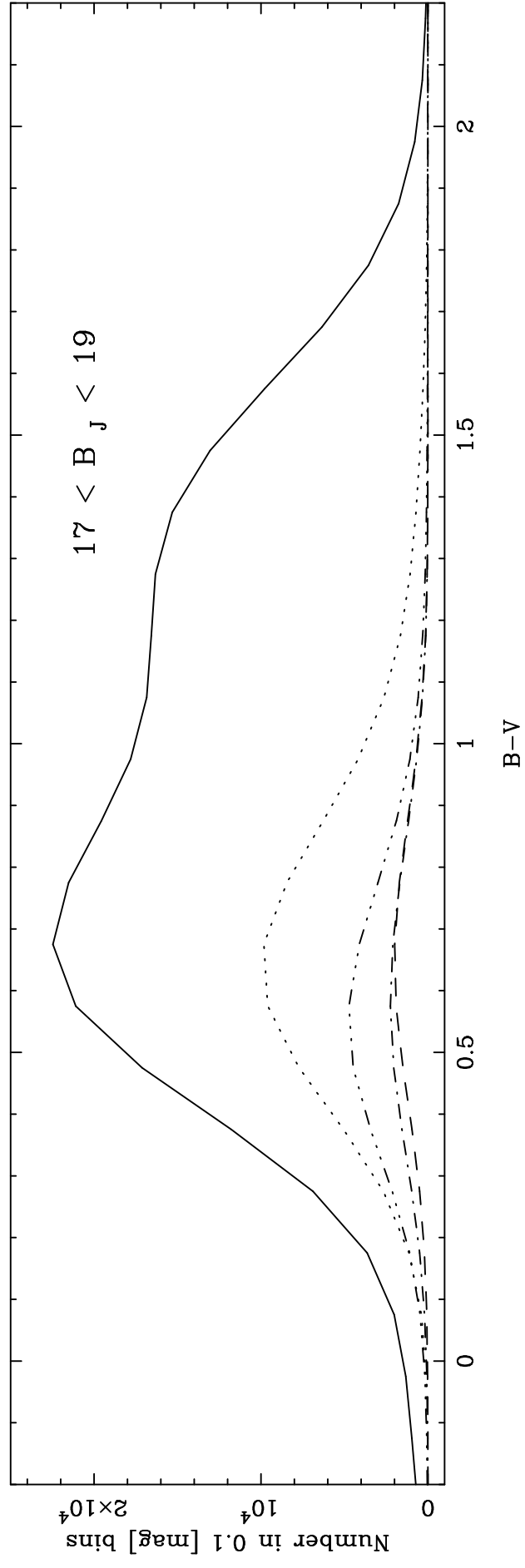

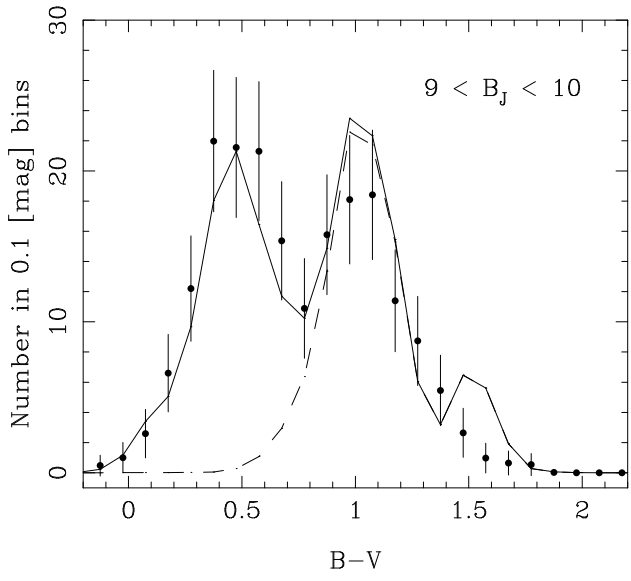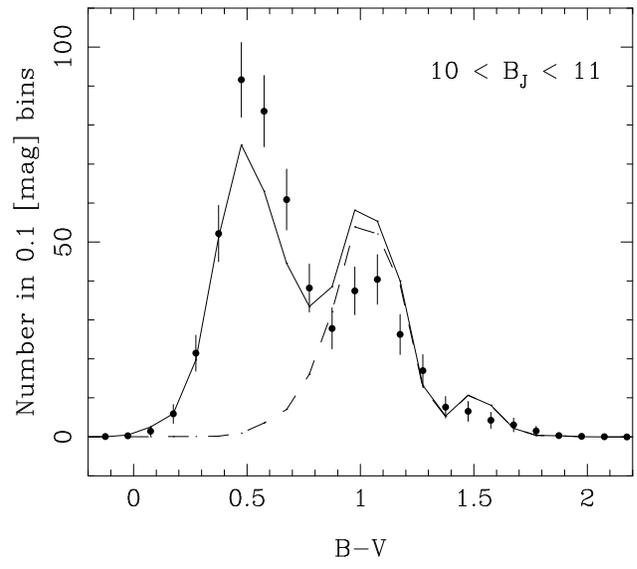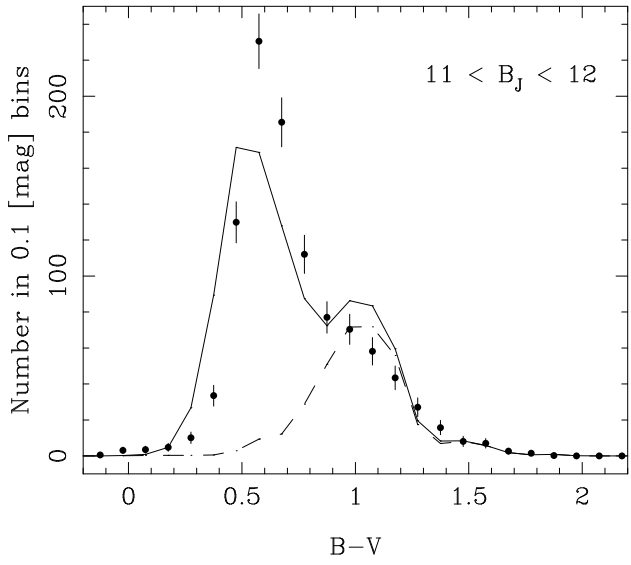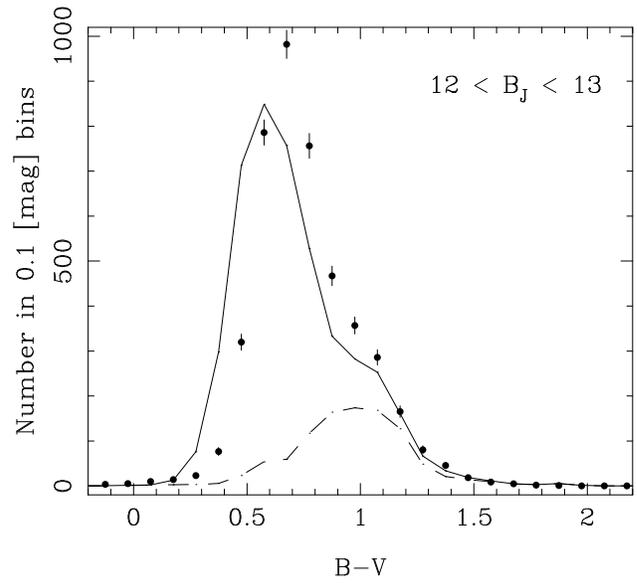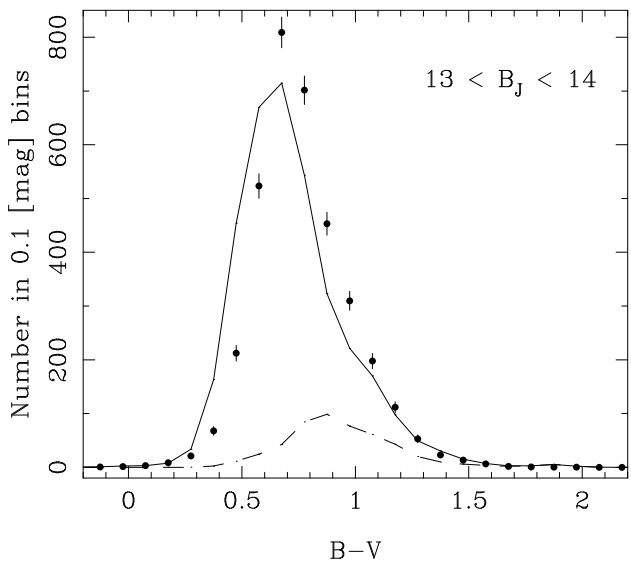

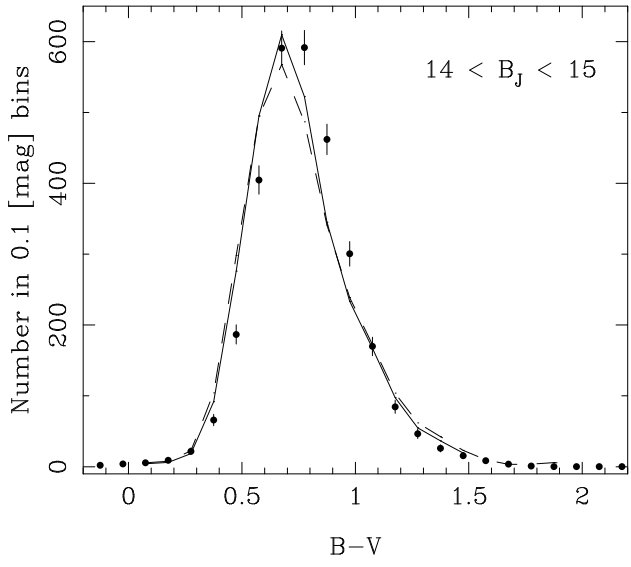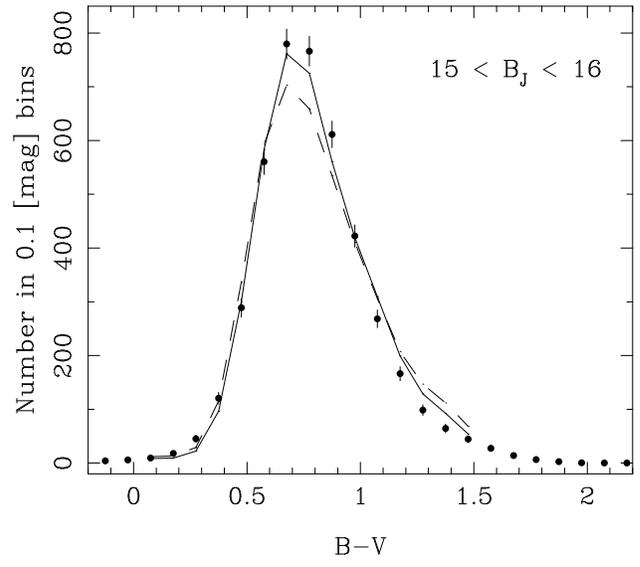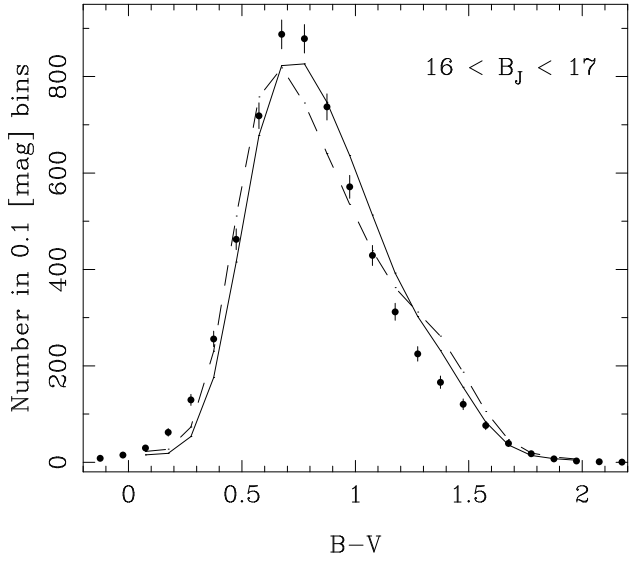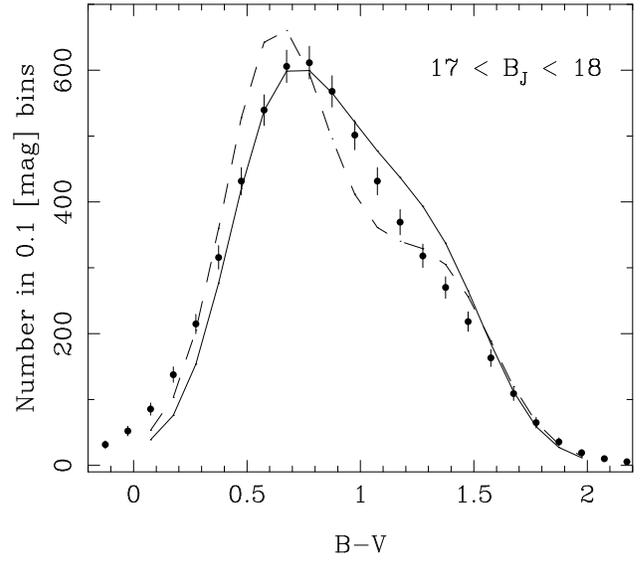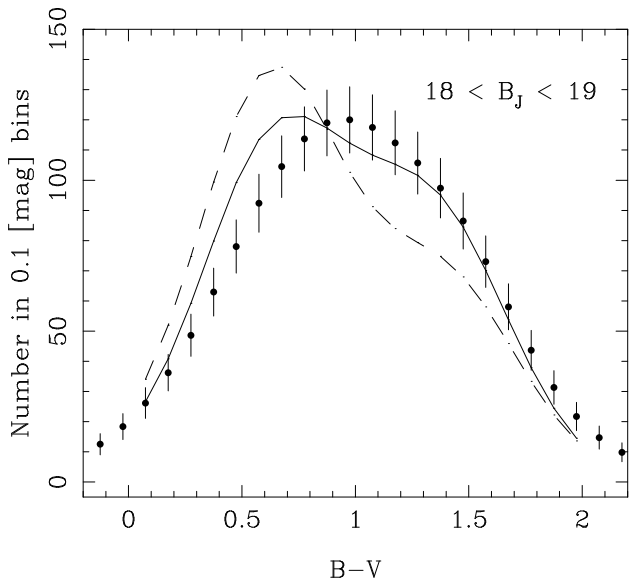

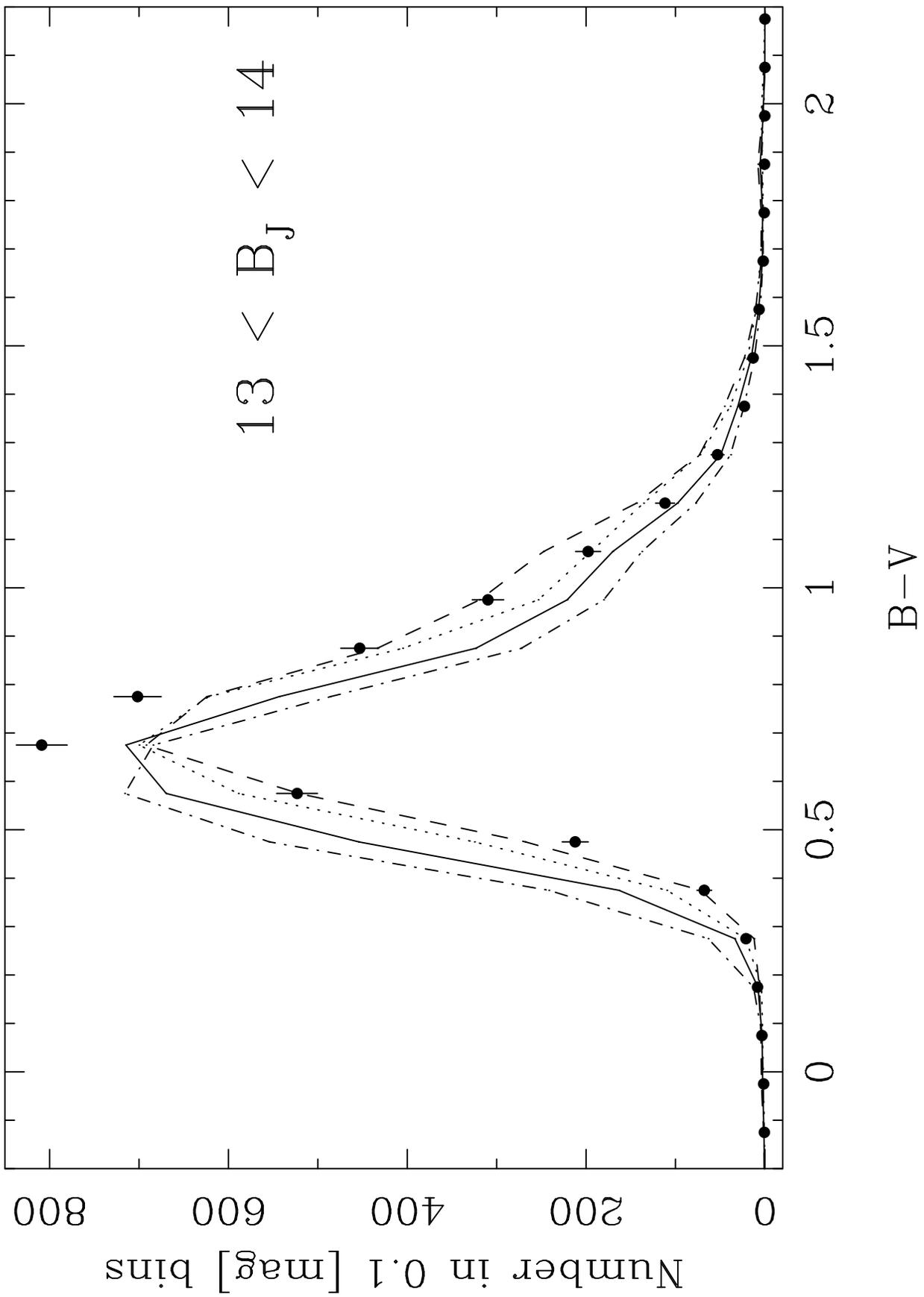

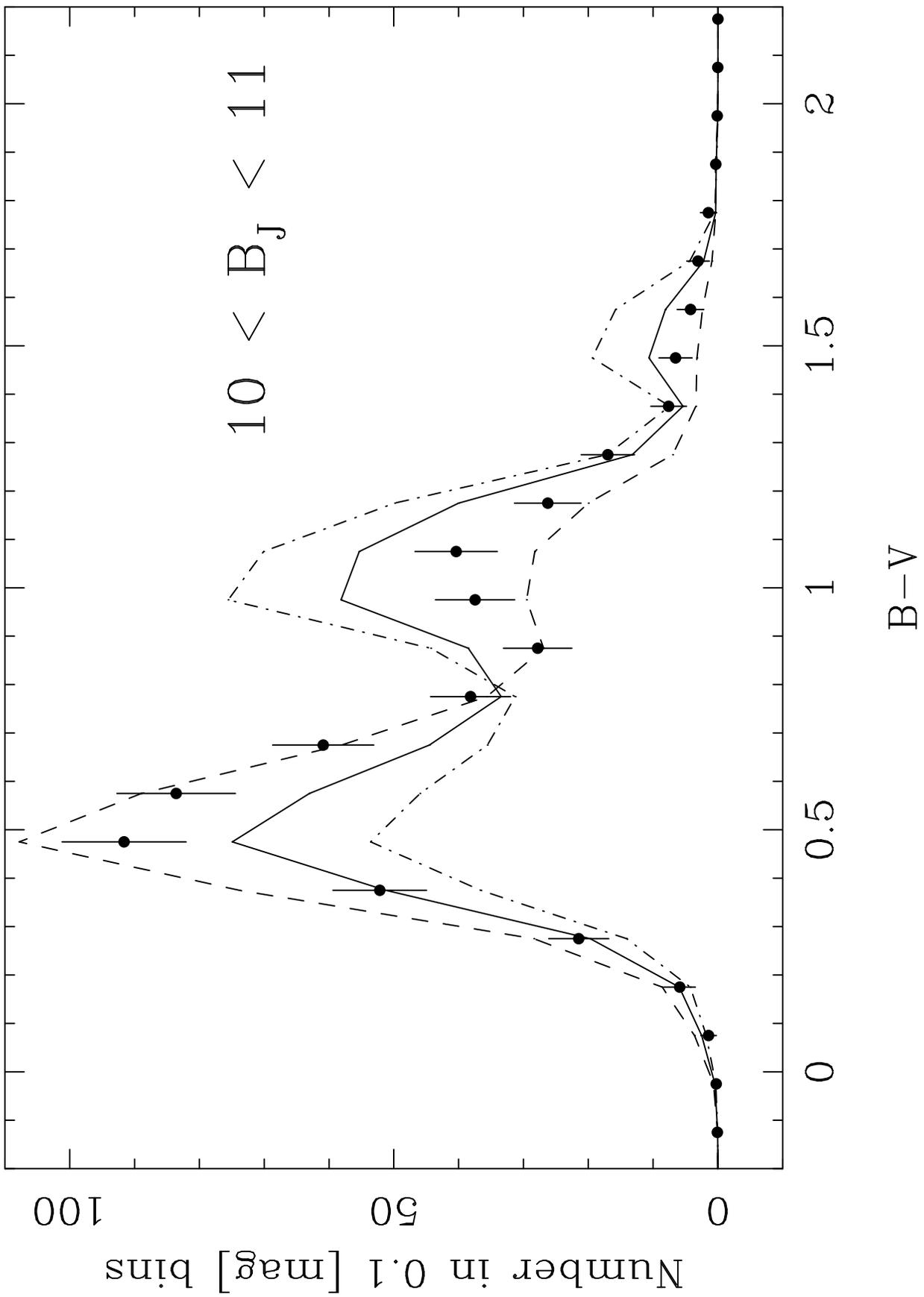

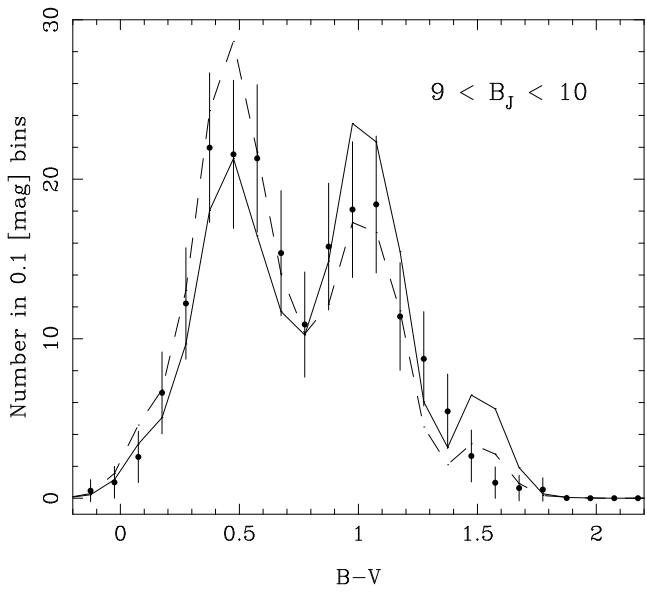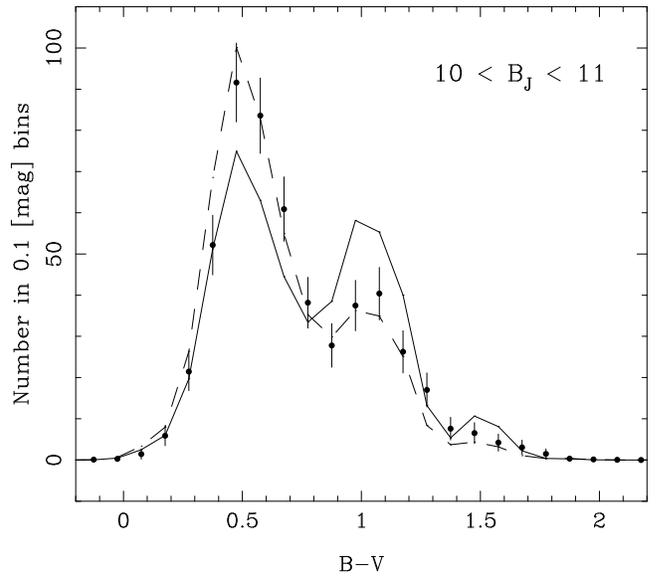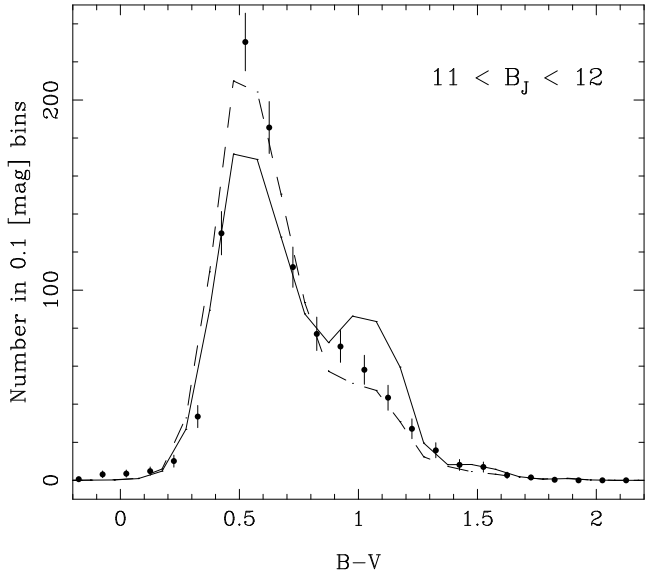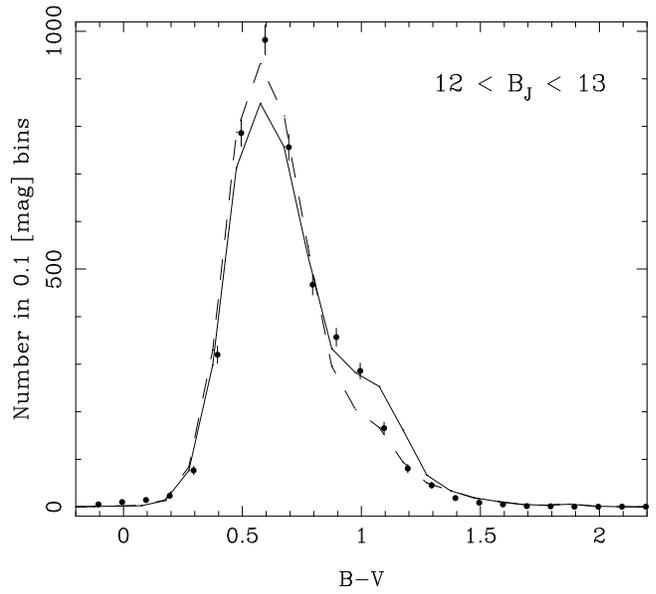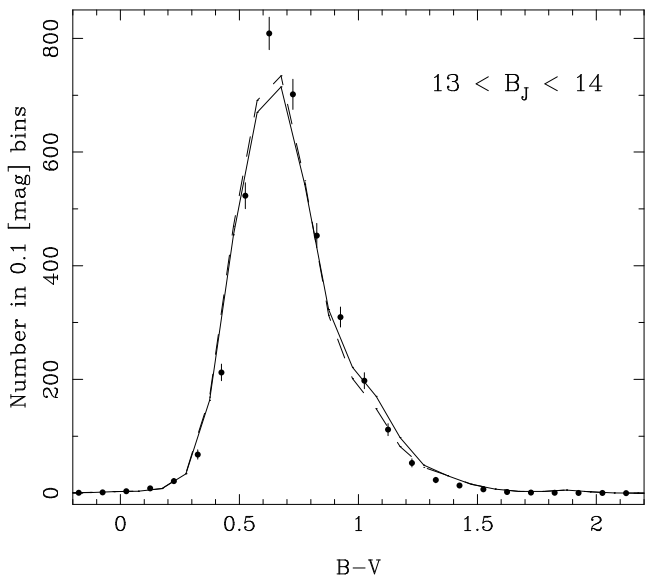

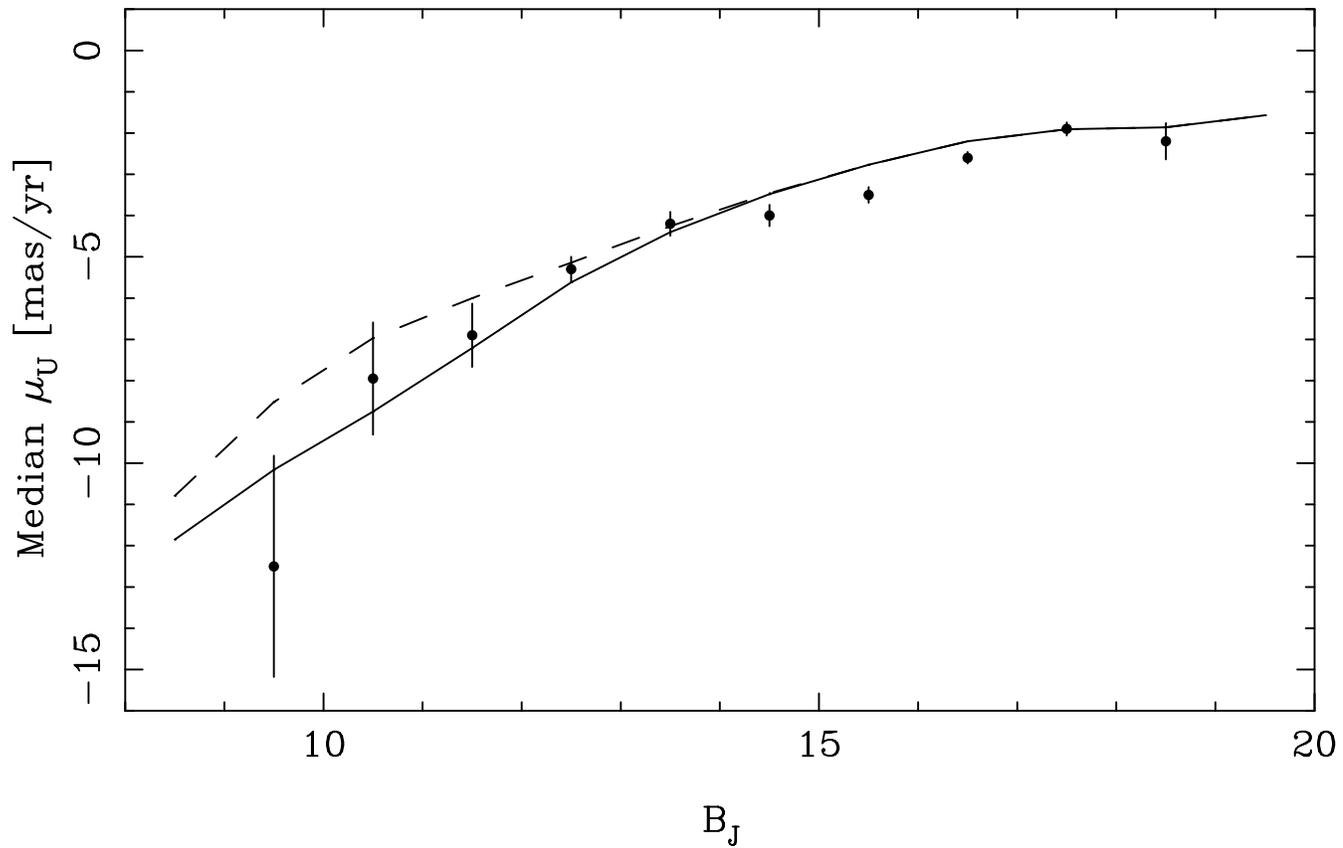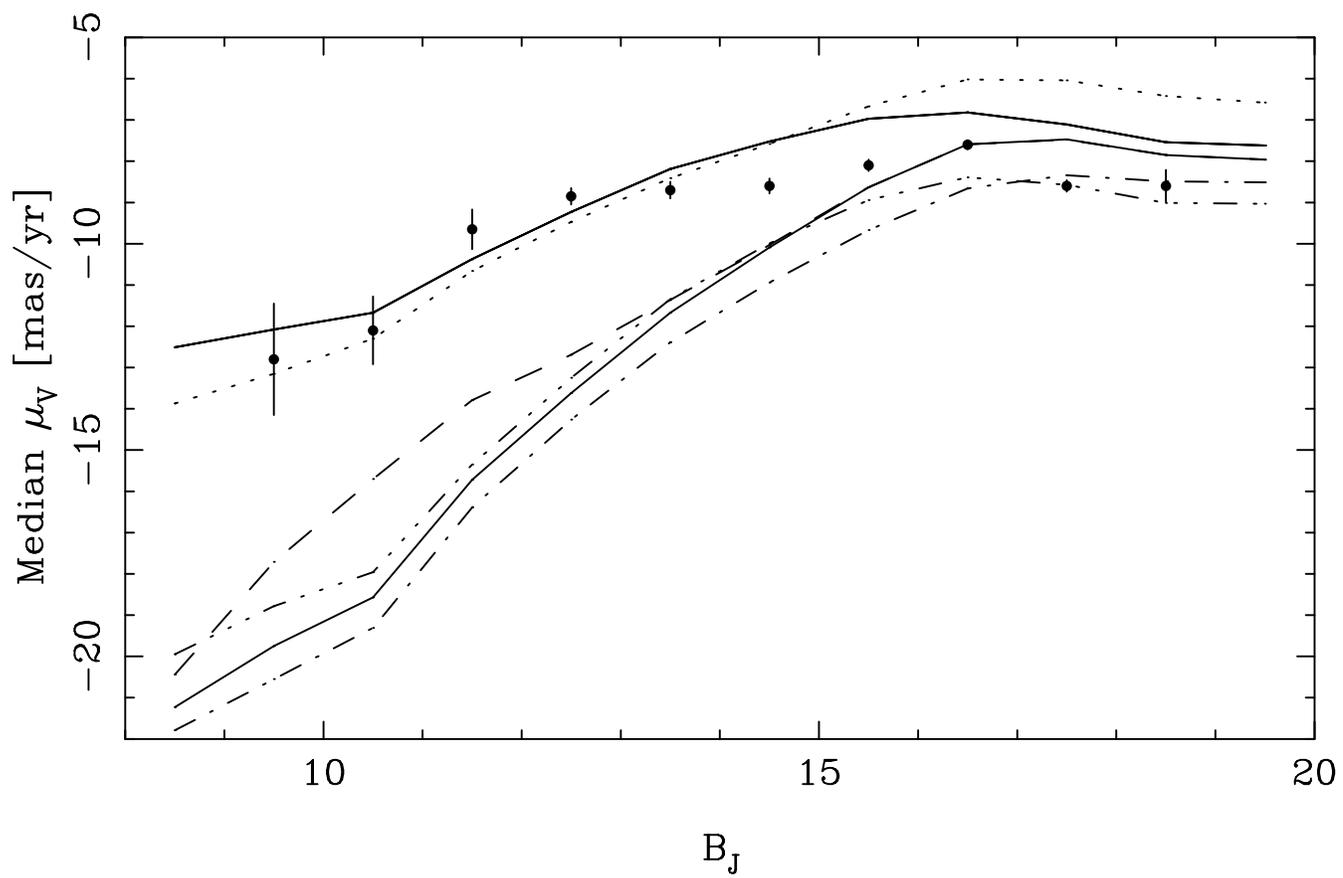

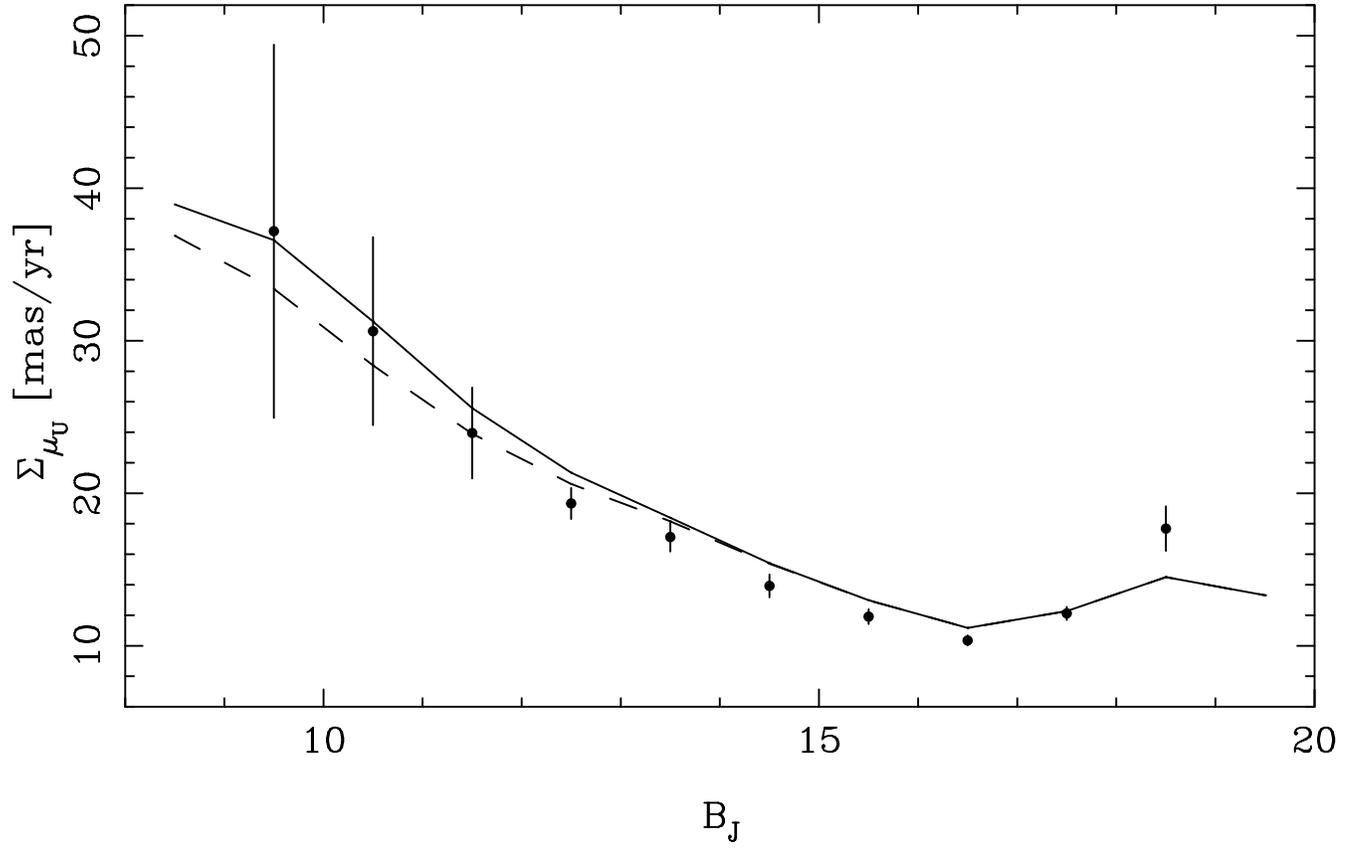
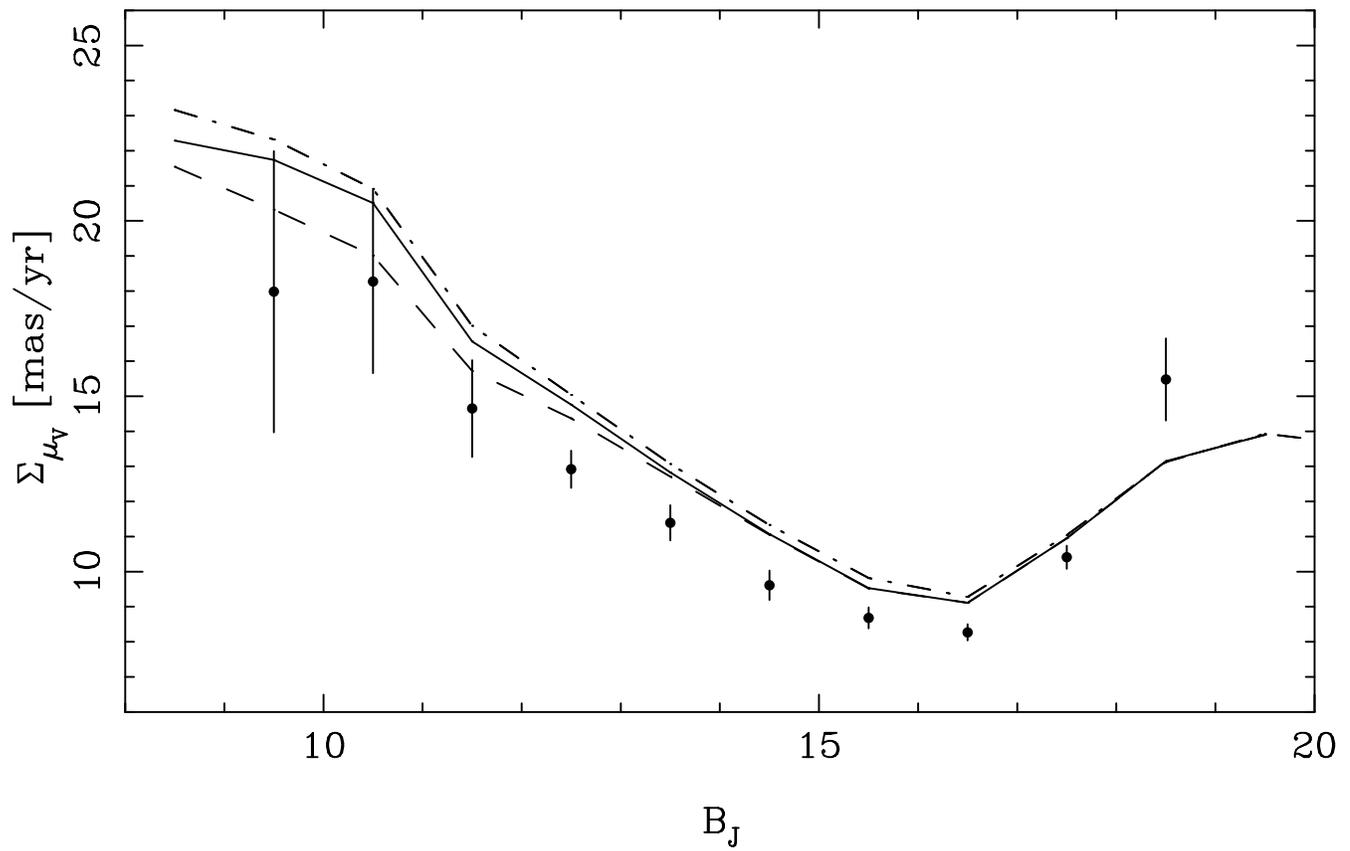

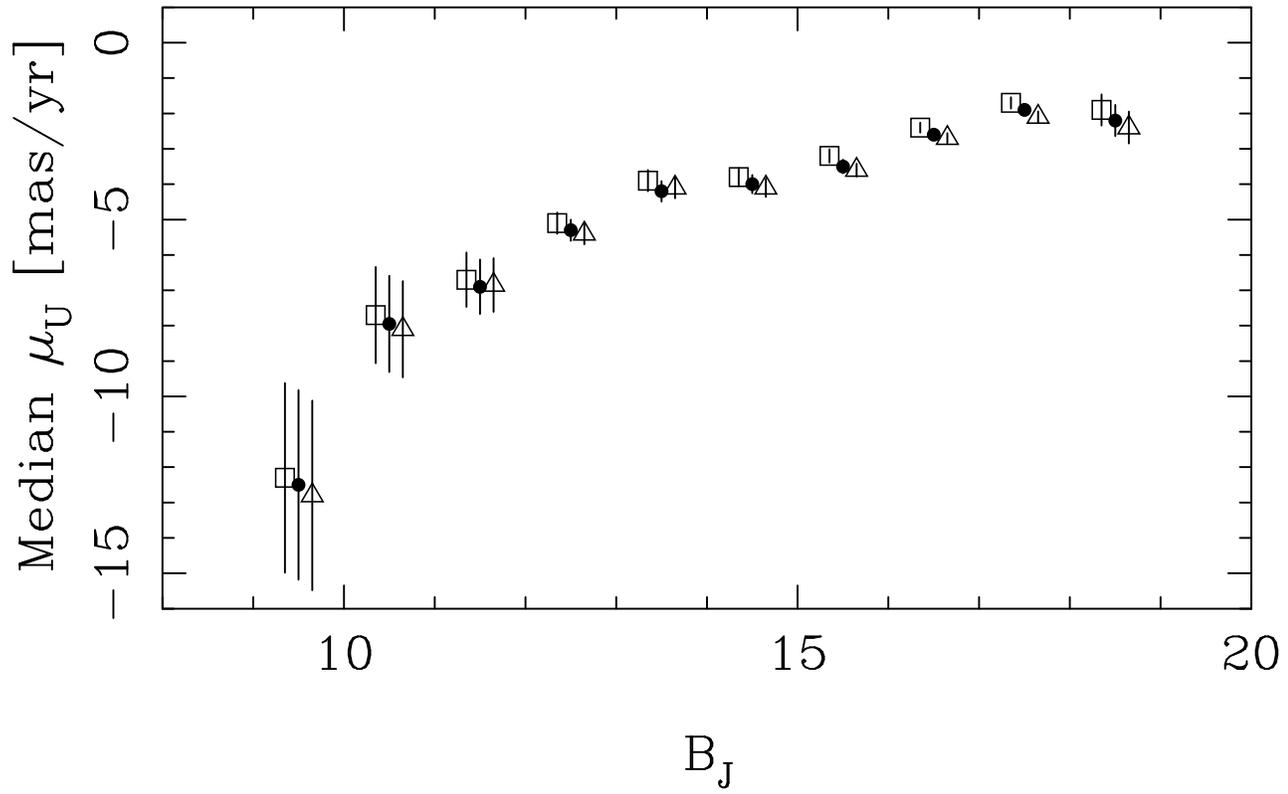
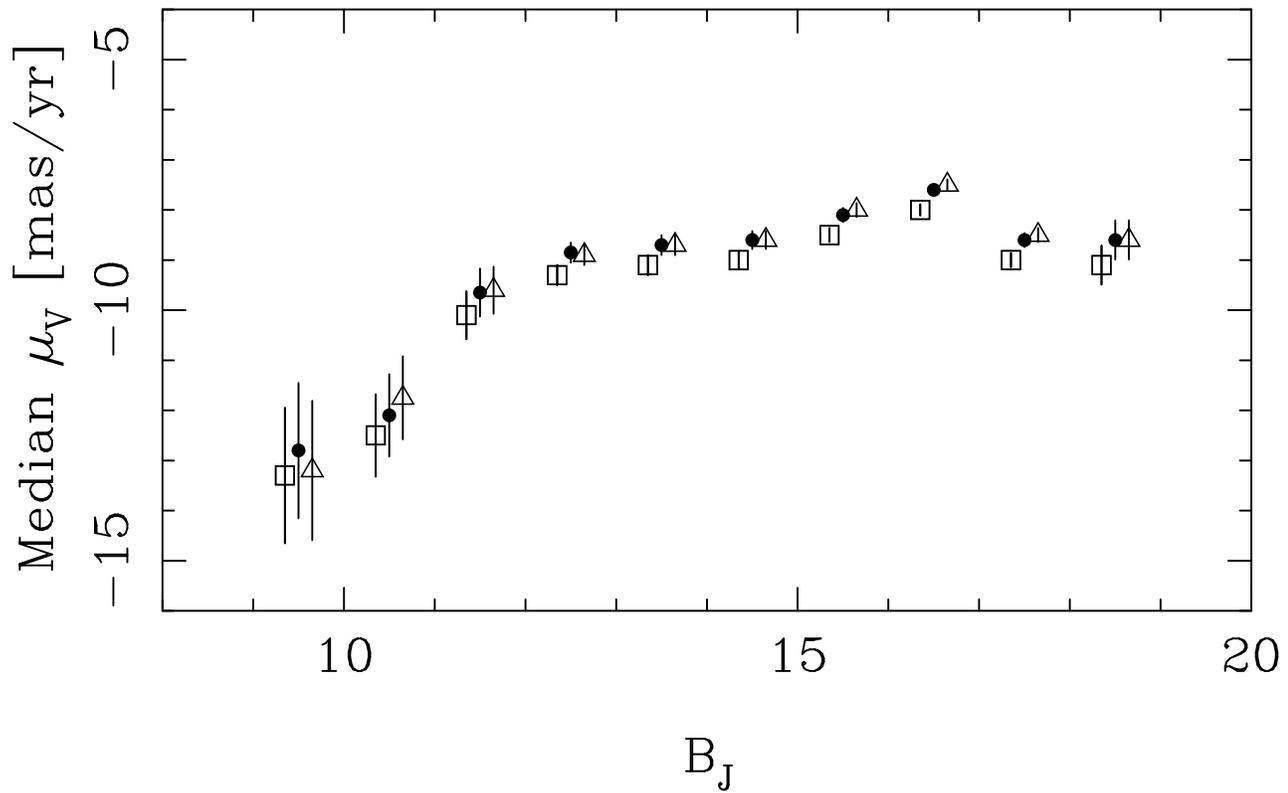

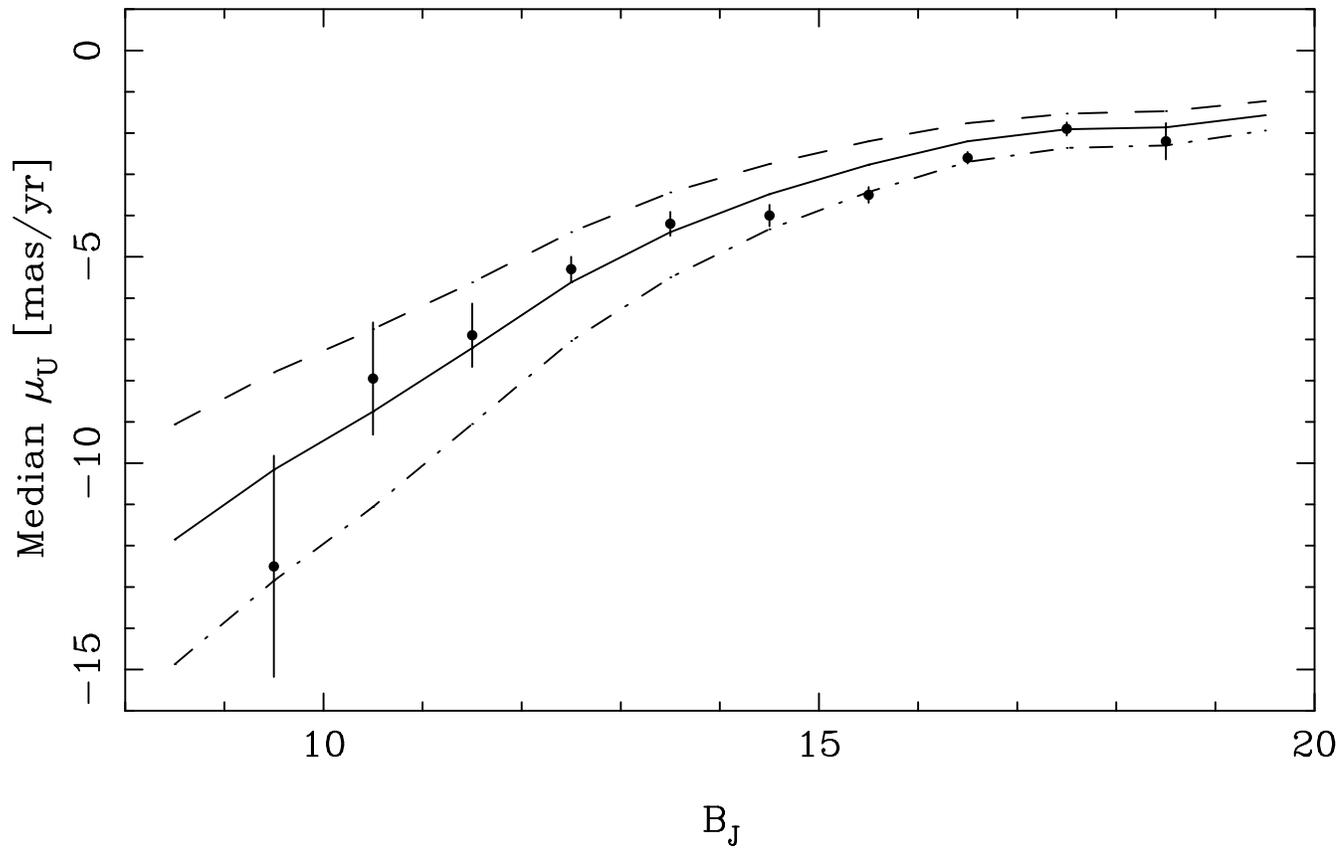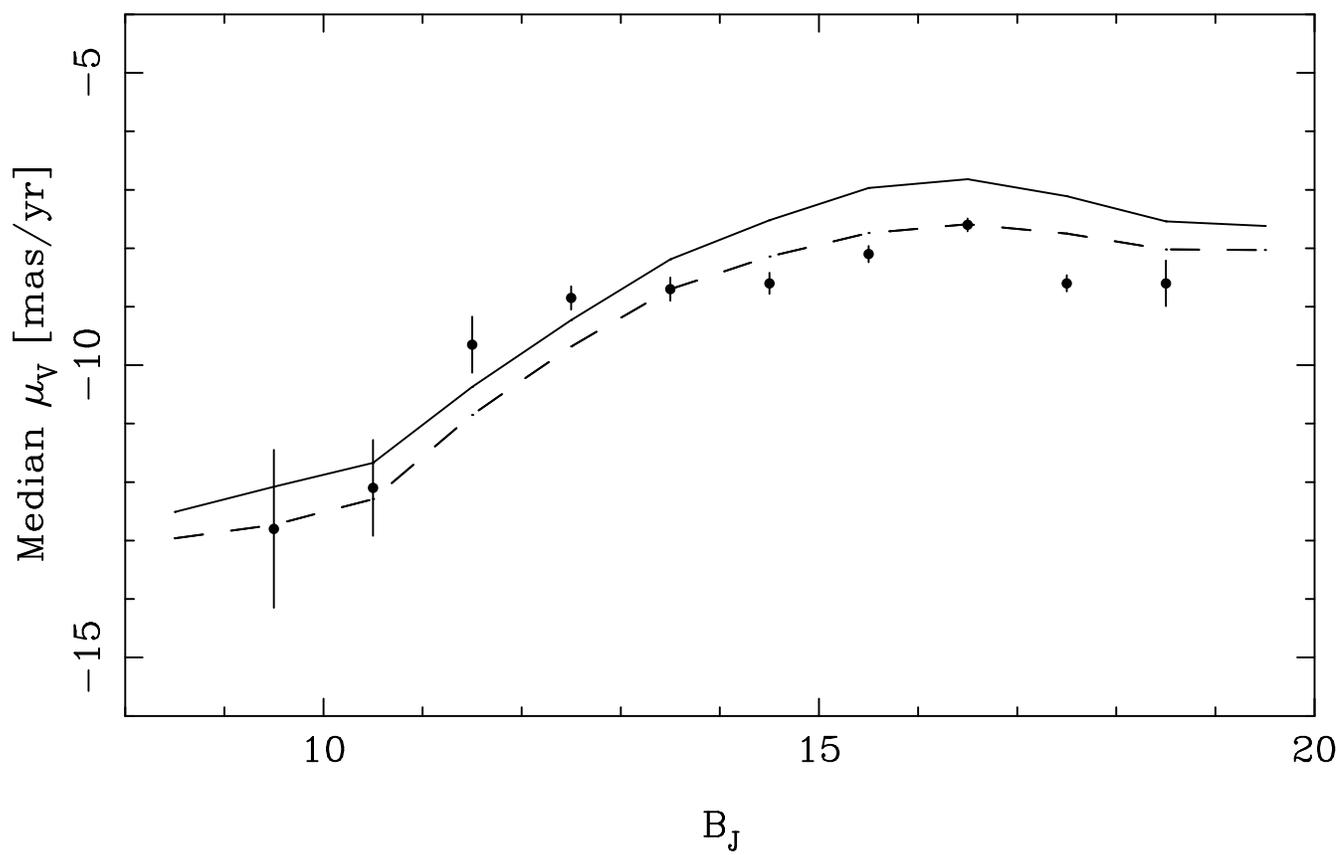

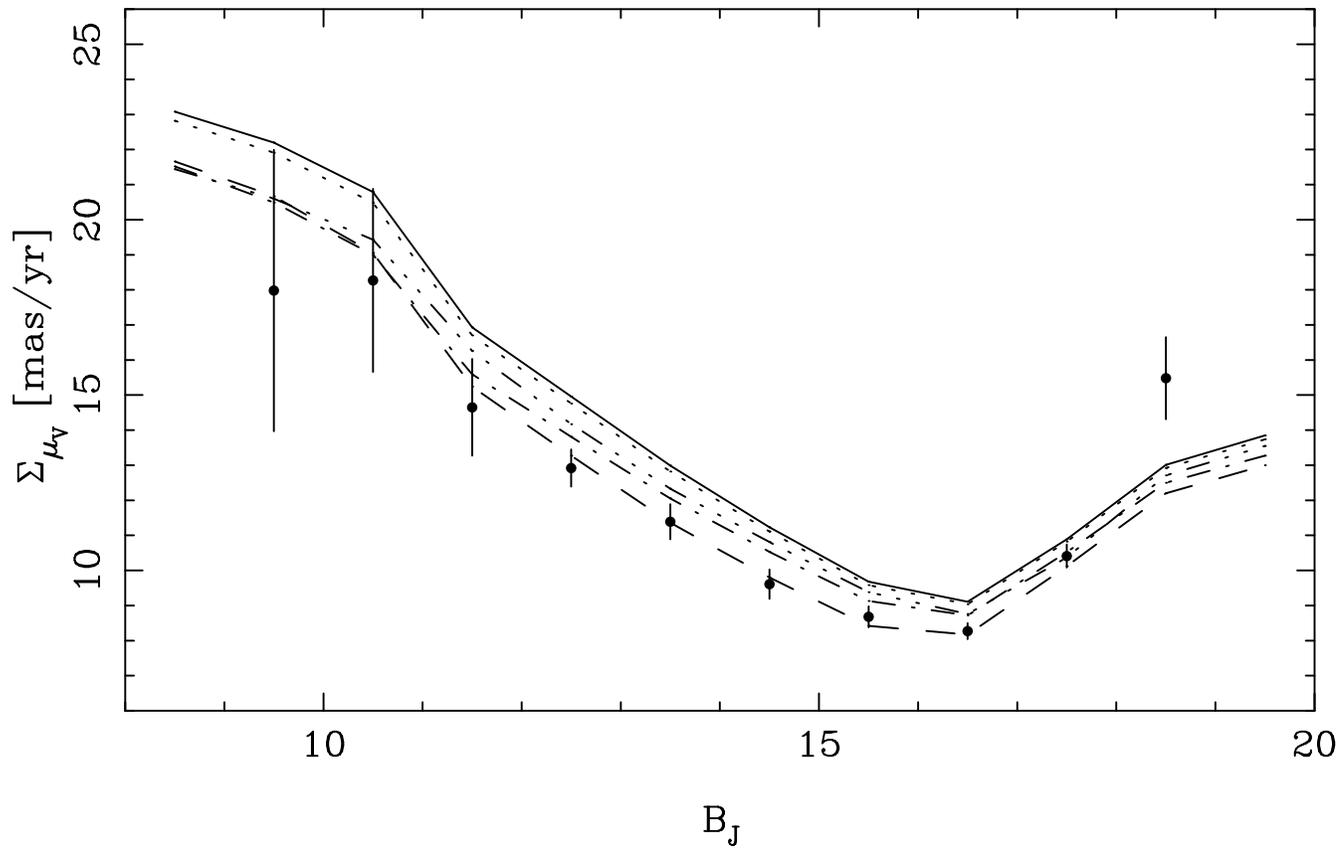

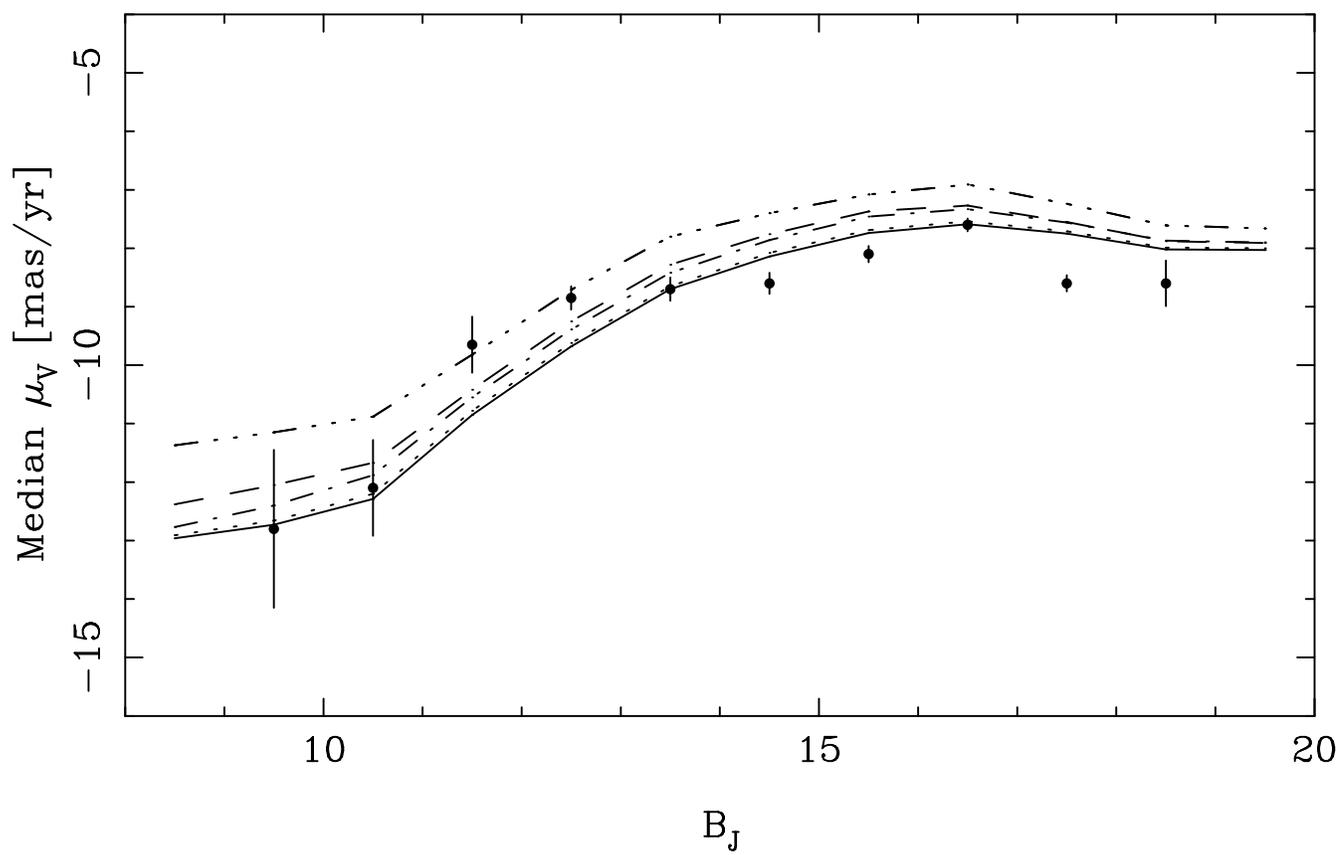

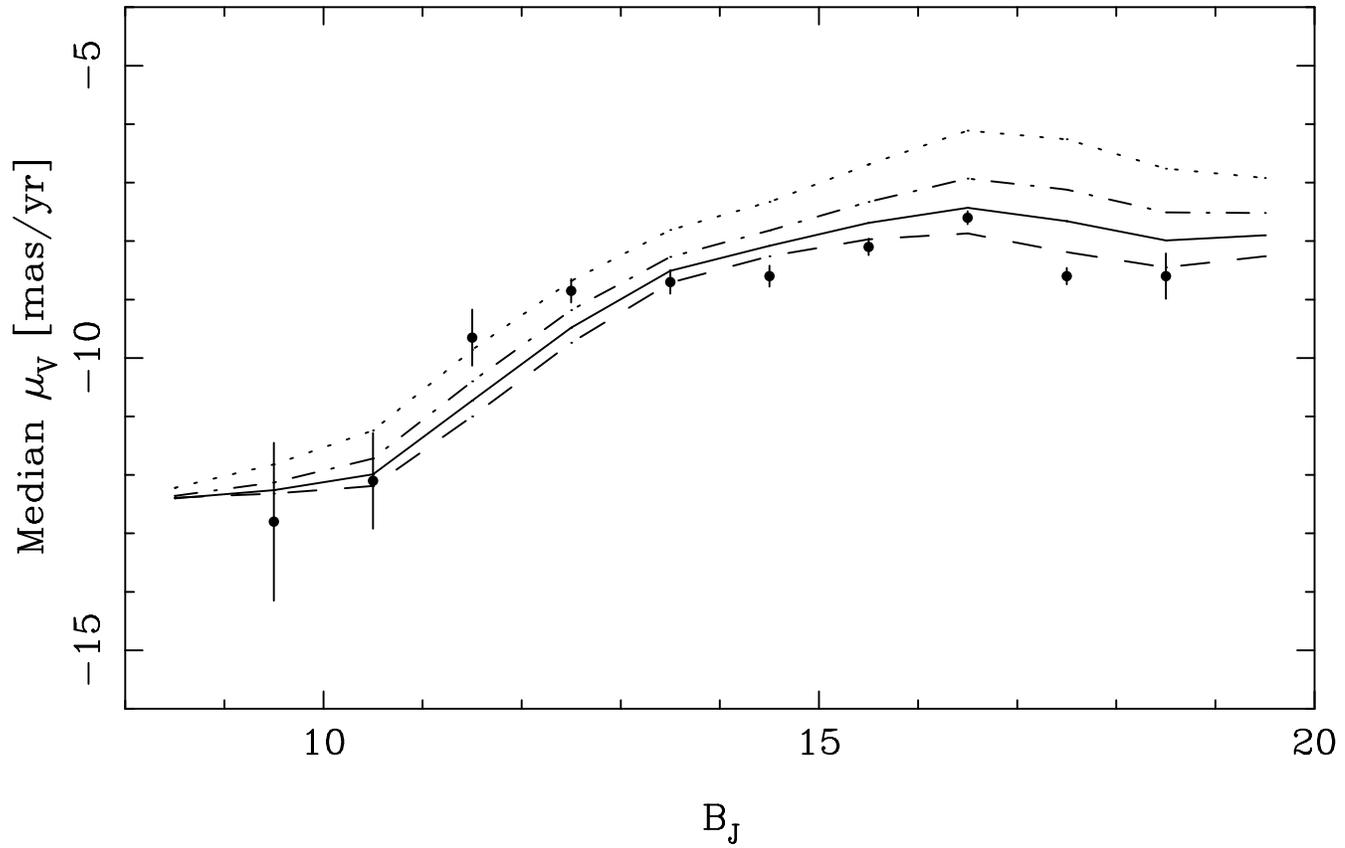
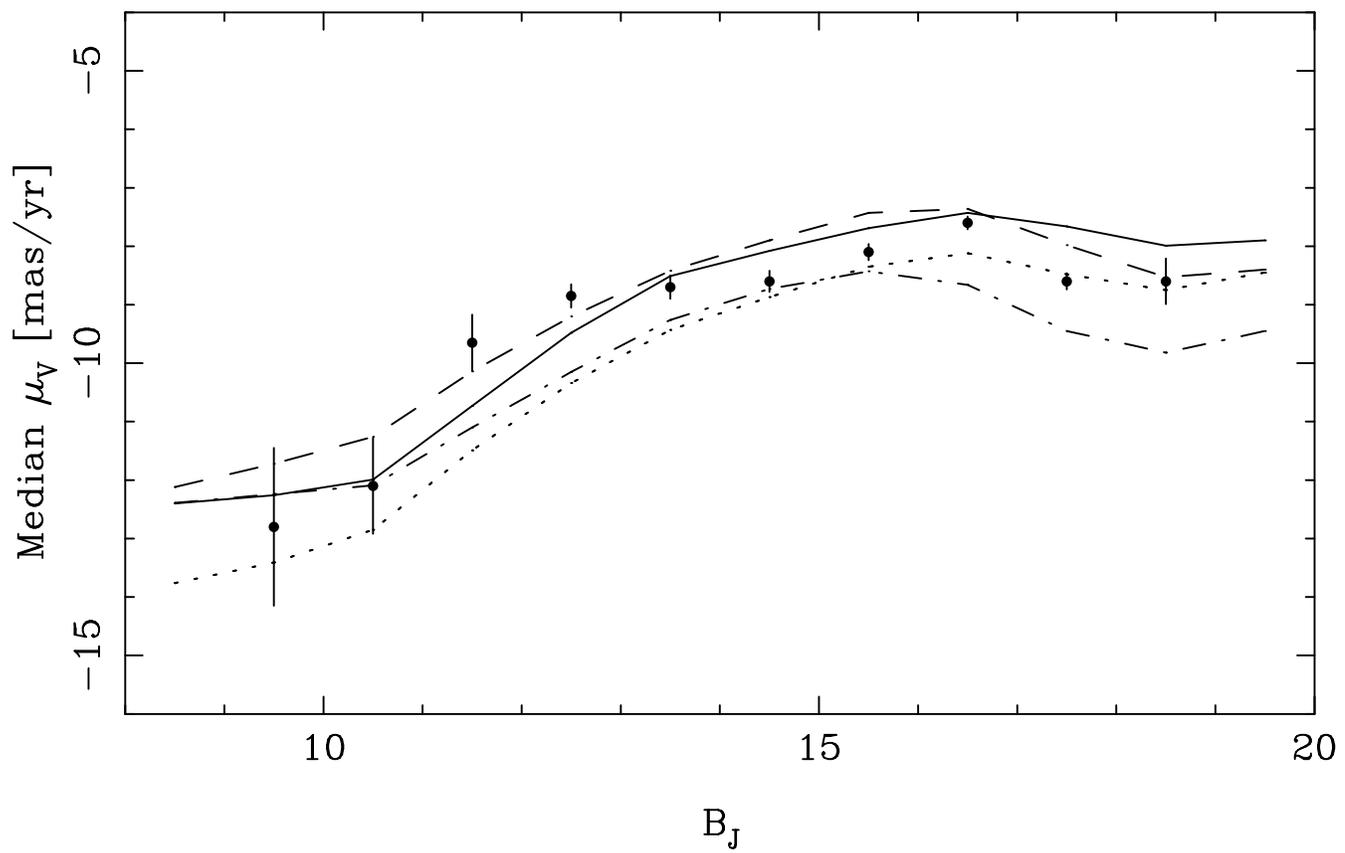

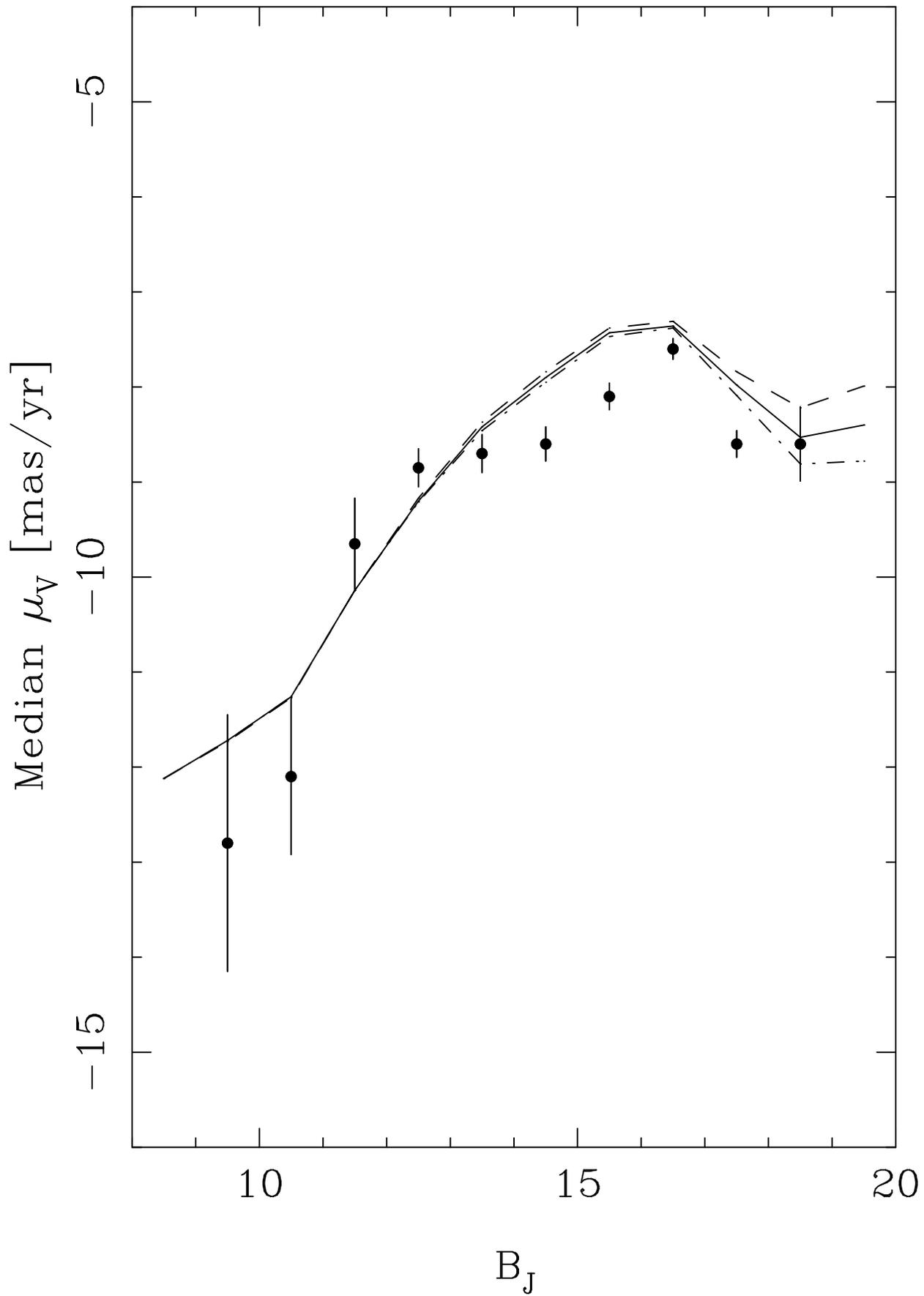

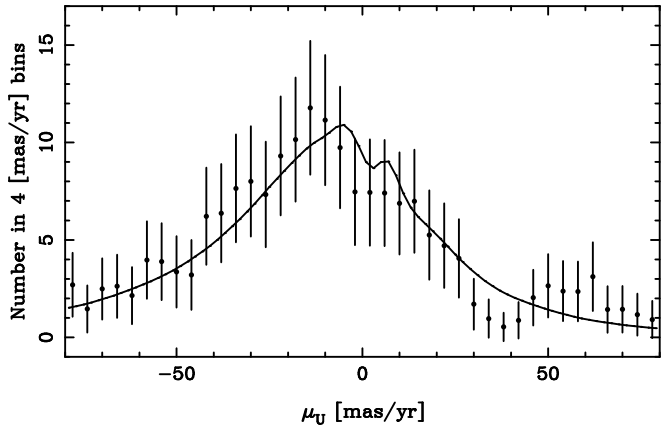
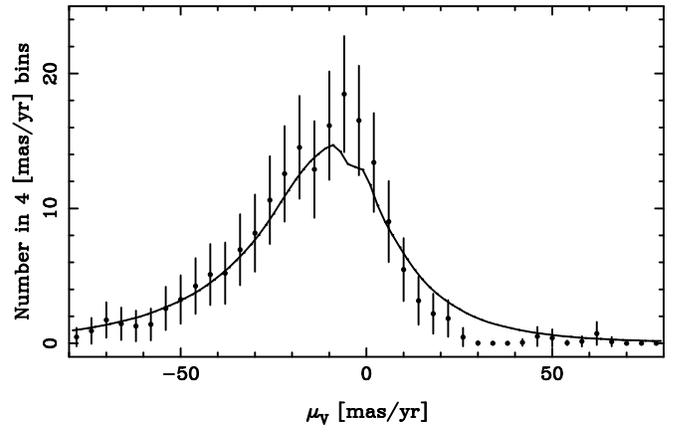
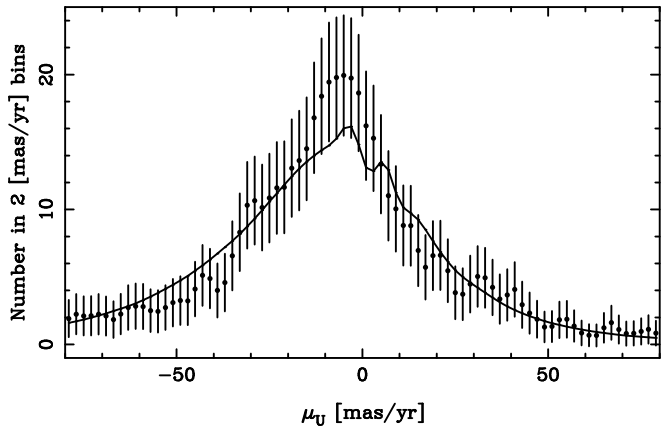
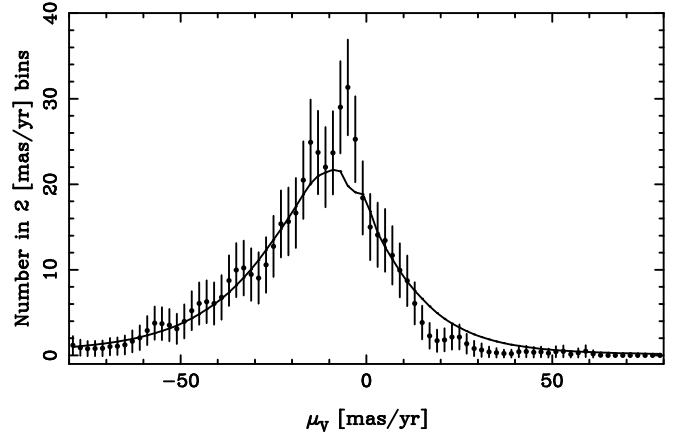
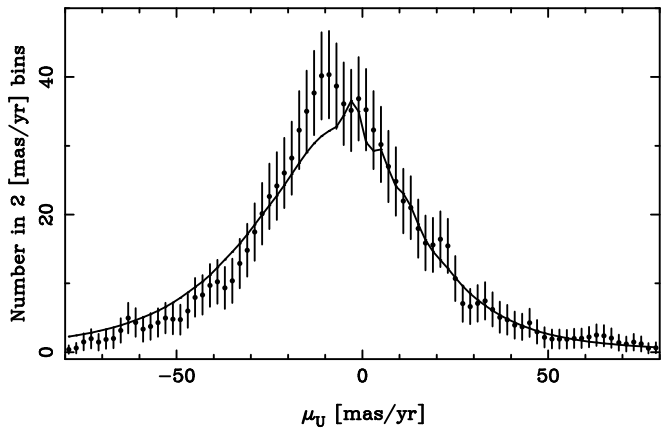
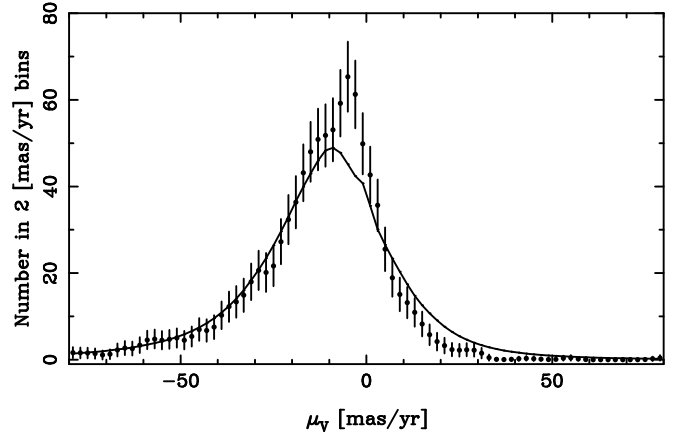
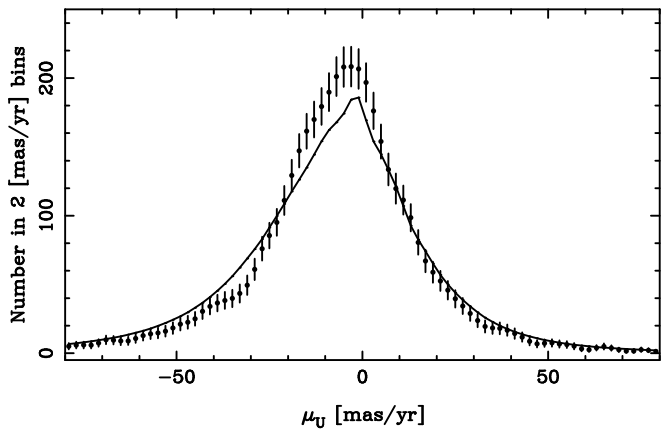
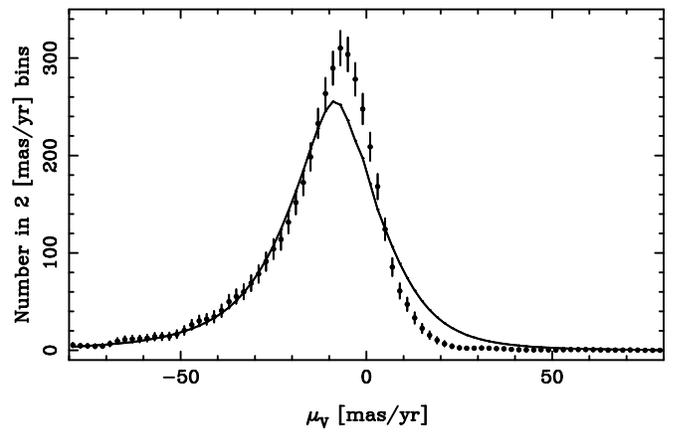

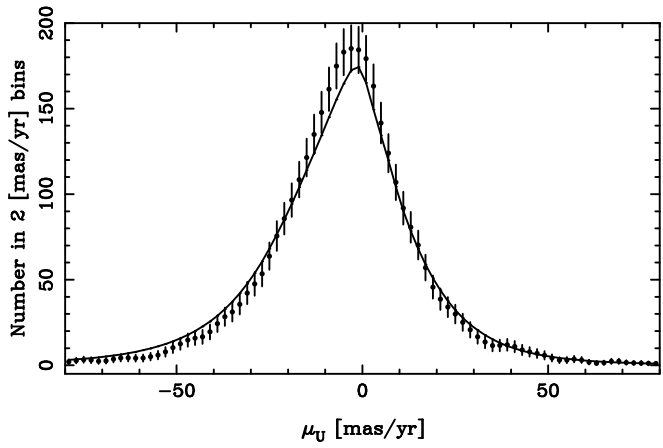
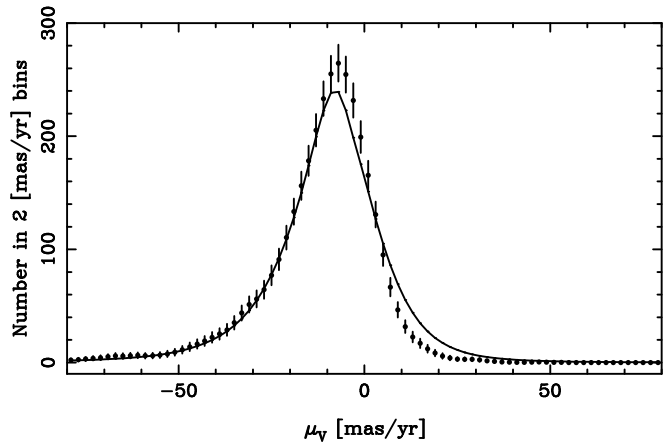
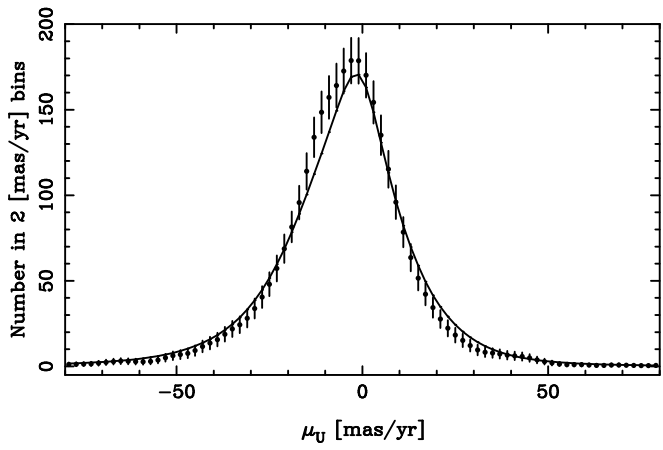
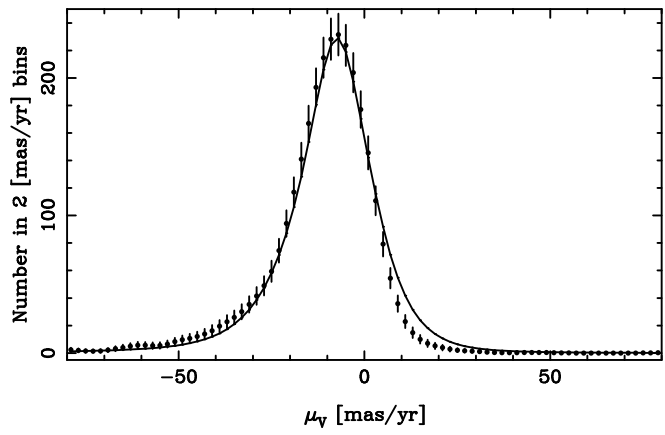
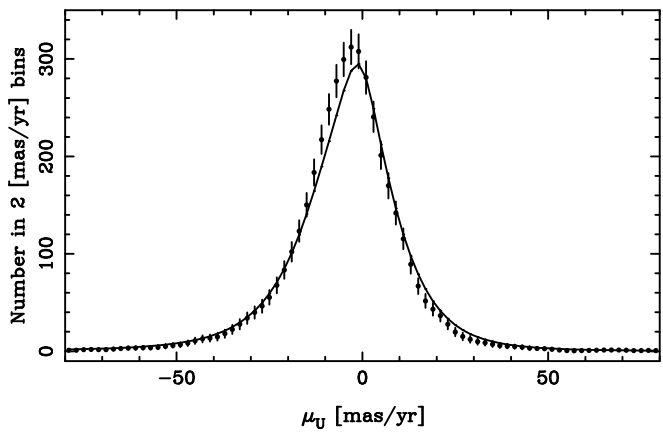
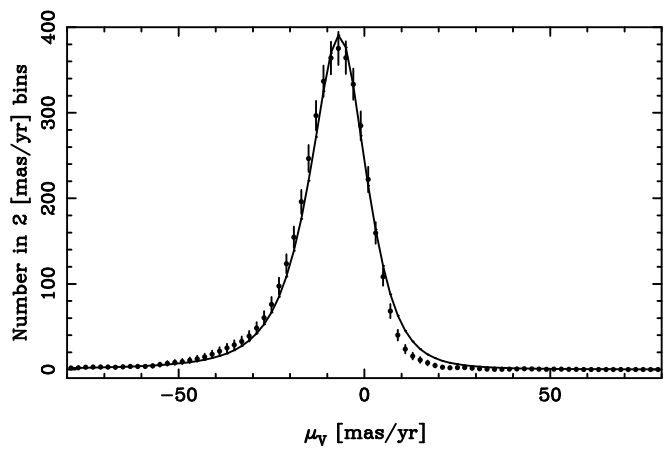
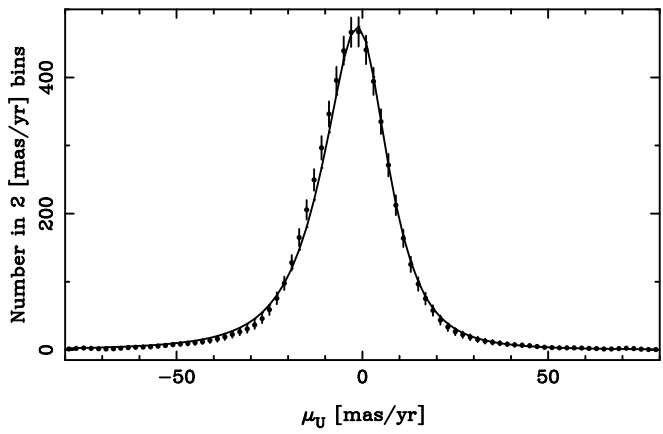
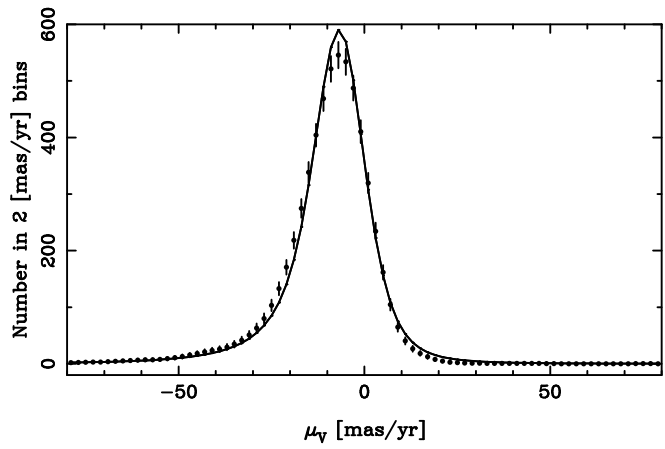

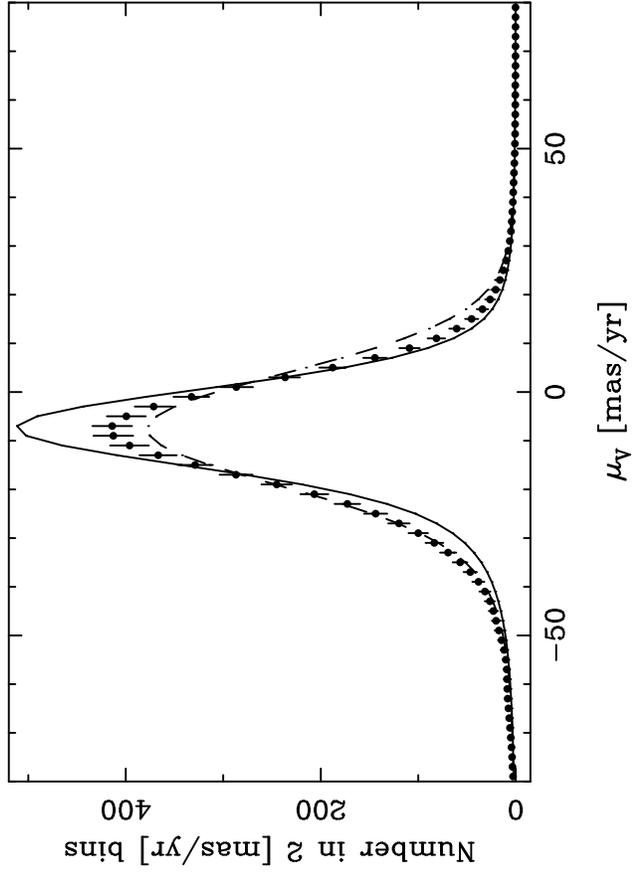
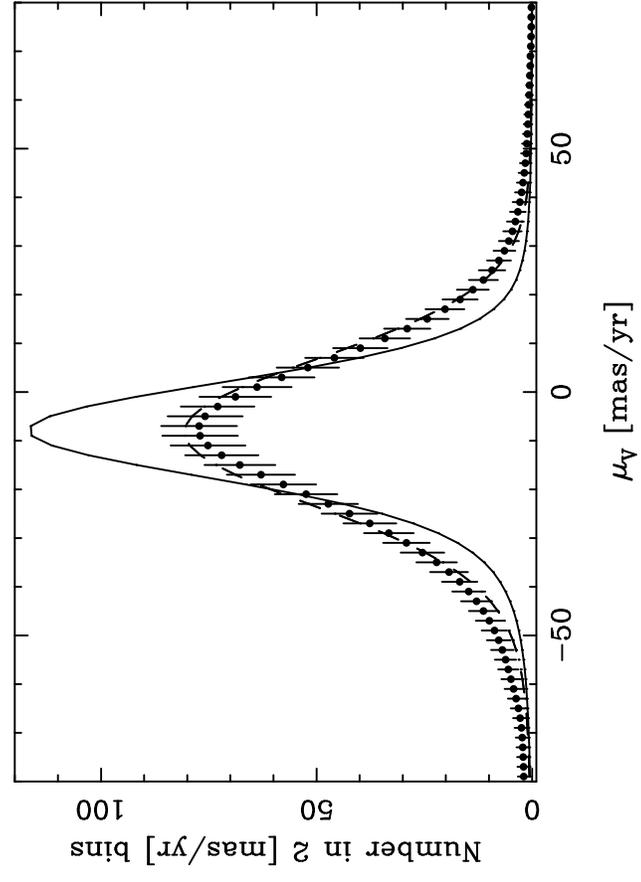
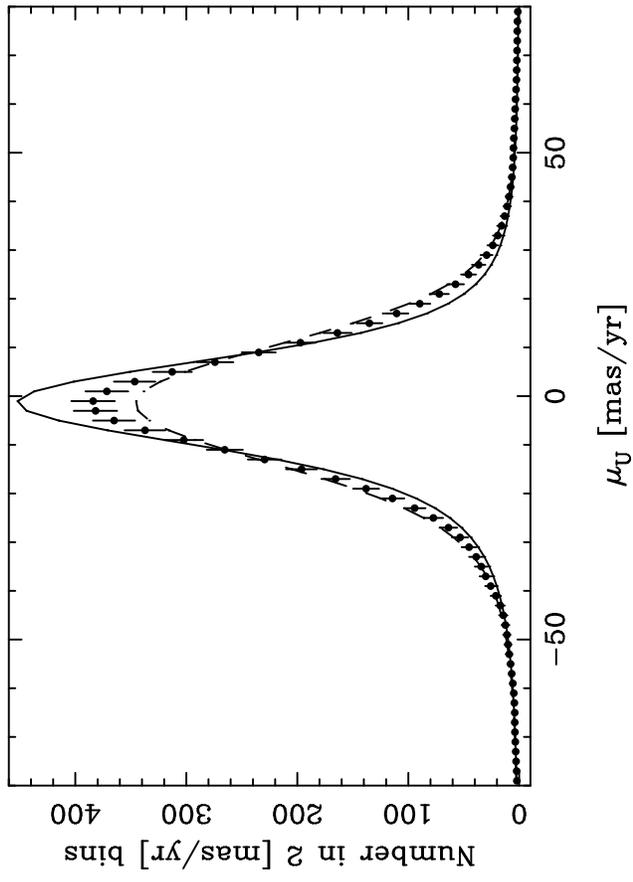
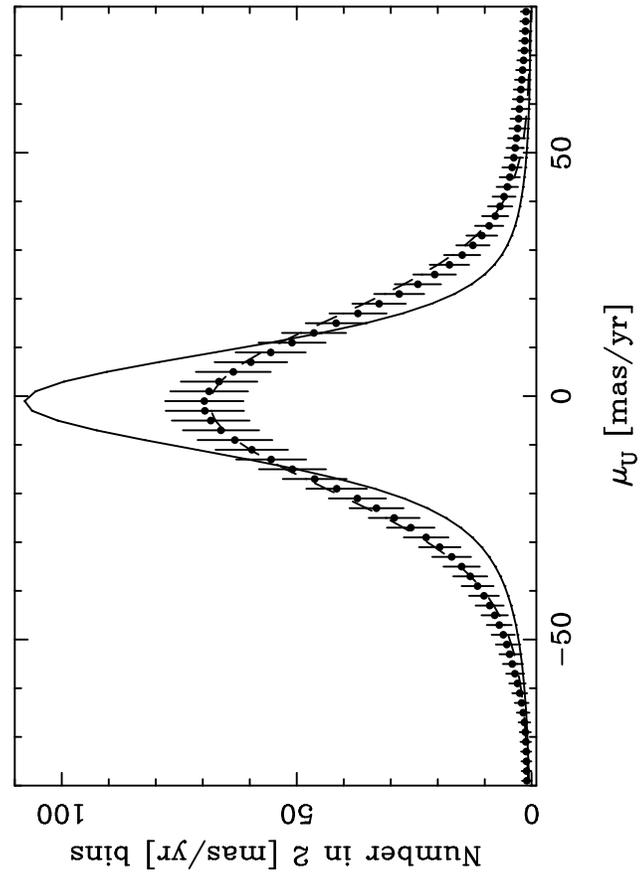